\pdfoutput=1
\documentclass[cernpreprint,atlasdraft=true,texlive=2020,UKenglish,texmf]{atlasdoc}
 
\usepackage{atlaspackage}
\usepackage[backref=false,block=none]{atlasbiblatex}
 
\usepackage[BSM]{atlasphysics}
 
\graphicspath{{logos/}{figures/}}
 
\usepackage{ANA-SUSY-2018-02-PAPER-defs}
\usepackage{multirow}
\usepackage{enumitem}

\AtlasTitle{Search for supersymmetry in events with four or more charged leptons in $139\,\mbox{fb\(^{-1}\)}$ of $\sqrt{s}=13 \,\ensuremath{\text{Te\kern -0.1em V}}$ $pp$ collisions with the ATLAS detector }
 
\AtlasAbstract{
A search for supersymmetry in events with four or more charged leptons (electrons, muons and $\tau$-leptons) is presented.
The analysis uses a data sample corresponding to $139\,\mbox{fb\(^{-1}\)}$  of proton-proton collisions delivered by the Large Hadron Collider at $\sqrt{s}=13\,\ensuremath{\text{Te\kern -0.1em V}}$ and recorded by the ATLAS detector.
Four-lepton signal regions with up to two hadronically decaying $\tau$-leptons are designed to target several supersymmetric models, while a general five-lepton signal region targets any new physics phenomena leading to a final state with five charged leptons.
Data yields are consistent with Standard Model expectations and results are used to set upper limits on contributions from processes beyond the Standard Model.
Exclusion limits are set at the 95\% confidence level in simplified models of general gauge-mediated supersymmetry, excluding higgsino masses up to $540\,\ensuremath{\text{Ge\kern -0.1em V}}$.
In $R$-parity-violating simplified models with decays of the lightest supersymmetric particle to charged leptons, lower limits of
$1.6\,\ensuremath{\text{Te\kern -0.1em V}}$,
$1.2\,\ensuremath{\text{Te\kern -0.1em V}}$, and
$2.5\,\ensuremath{\text{Te\kern -0.1em V}}$
are placed on wino, slepton and gluino masses, respectively.
}
 
\author{The ATLAS Collaboration}
 
\AtlasRefCode{SUSY-2018-02}

\AtlasJournal{JHEP}
 
\PreprintIdNumber{CERN-EP-2021-021}
 
\AtlasJournalRef{JHEP 07 (2021) 167}
\AtlasDOI{10.1007/JHEP07(2021)167}
\hypersetup{pdftitle={ATLAS document},pdfauthor={The ATLAS Collaboration}}
 
\begin{document}
 
\maketitle

\section{Introduction}
\label{sec:intro}
 
Standard Model (SM) processes rarely produce events with four or more charged leptons, while many new theories, such as supersymmetry (SUSY)~\cite{Golfand:1971iw,Volkov:1973ix,Wess:1974tw,Wess:1974jb,Ferrara:1974pu,Salam:1974ig}, predict events which would regularly decay to these multilepton final states.
This paper presents a search for new phenomena in final states with at least four isolated, charged leptons (electrons, muons or $\tau$-leptons) where up to two hadronically decaying $\tau$-leptons are considered.
Here, electrons and muons are referred to as `light leptons' and include those from leptonic $\tau$ decays.
The full proton-proton dataset delivered by the LHC and collected by the ATLAS experiment during the 2015--2018 data-taking period is used in the analysis,
corresponding to an integrated luminosity of $139~\ifb$~\cite{ATLAS-CONF-2019-021} at a centre-of-mass energy of $13\,\TeV$.
Several SUSY signal models are used to optimise the search, but the search itself is generally model-agnostic, using selections on either the presence of, or absence of, $Z$ bosons in the event, and loose requirements on either the effective mass or the missing transverse momentum.
Results are presented in terms of limits on SUSY models.
 
Previous searches for SUSY particles using signatures with three or more leptons were carried out
at the Tevatron collider~\cite{Abazov:2005ku,Abazov:2009zi,Abazov:2006nw,Aaltonen:2007an,Aaltonen:2008pv,Abulencia:2007mp},
and at the LHC by the ATLAS experiment~\cite{SUSY-2012-17,SUSY-2011-25,SUSY-2013-12,SUSY-2013-13,SUSY-2016-21,SUSY-2017-03}
and the CMS experiment~\cite{CMS-SUS-10-008,CMS-SUS-11-013,CMS-SUS-13-003,CMS-SUS-13-002,CMS-SUS-16-039, CMS-EXO-19-002}.
Searching for new physics using a four or more lepton final state may offer more sensitivity to some beyond the SM scenarios than using lower lepton multiplicities, as the very low SM background can allow for a looser selection and a more inclusive approach to be adopted.
This analysis closely follows the ATLAS analyses on the datasets at $7\,\TeV$~\cite{SUSY-2012-17} and $8\,\TeV$~\cite{SUSY-2013-13}, and on the partial dataset at $13\,\TeV$~\cite{SUSY-2016-21}.
Previous results are extended here by analysing the full ATLAS $13\,\TeV$ dataset, expanding the search with an additional channel selecting at least five leptons, and using data to constrain major sources of SM background.
 
\section{Targeted models}\label{sec:susymodels}
 
SUSY is a space-time symmetry that postulates the existence of a new superpartner for every SM particle, with spin differing by one half-unit from its SM partner:
each SM fermion (boson) is associated with a SUSY boson (fermion).
The new SUSY particles (sparticles) would have the same quantum numbers as their SM counterparts except for spin and provide a potential solution to the hierarchy problem~\cite{Sakai:1981gr,Dimopoulos:1981yj,Ibanez:1981yh,Dimopoulos:1981zb}.
The scalar superpartners of the SM fermions are the charged sleptons, $\slepton$, the sneutrinos, $\snu$, and the squarks, $\squark$, while the gluons have fermionic superpartners called gluinos ($\gluino$).
The bino, wino and higgsino fields are fermionic superpartners of the $\mathrm{SU(2)}\times\mathrm{U(1)}$ gauge fields of the SM, and the two complex scalar doublets of a minimally extended Higgs sector, respectively.
They mix to give mass eigenstates that are referred to as charginos $\tilde{\chi}_{i}^{\pm}$ $(i=1,2)$ and neutralinos $\tilde{\chi}_{j}^{0}$ $(j=1,2,3,4)$, numbered in order of increasing mass.

SUSY processes can result in proton decay at a rate that is in conflict with the stringent experimental constraints on the proton lifetime if they do not conserve both lepton number ($L$) and baryon number ($B$)~\cite{PhysRevD.99.032008}.
This conflict can be avoided by imposing the conservation of $R$-parity~\cite{Farrar:1978xj}, defined as $(-1)^{3(B-L)+2S}$, where $S$ is spin,
or by explicitly conserving either $B$ or $L$ in $R$-parity-violating (RPV) scenarios~\cite{Weinberg:1981wj,Sakai:1981pk}.
In $R$-parity-conserving (RPC) models, the lightest SUSY particle (LSP) is stable and a viable dark-matter candidate~\cite{Goldberg:1983nd,Ellis:1983ew}, and leptons can originate from unstable weakly interacting sparticles decaying to the LSP.
In RPV models, the LSP is unstable and decays into SM particles, including charged leptons and neutrinos when violating $L$ but not $B$.
Both the RPC and RPV SUSY scenarios can therefore result in signatures with high lepton multiplicities and substantial missing transverse momentum, selections on which can be used to suppress SM background processes effectively.

RPC and RPV SUSY models are used for signal region optimisation and to interpret the results of this analysis; each requires a different approach for signal selection, as discussed in Section~\ref{sec:eventreco}.
In all SUSY scenarios considered here, the light CP-even Higgs boson, $h$, of the minimal supersymmetric extension of the SM~\cite{Fayet:1976et,Fayet:1977yc} Higgs sector is assumed to be practically identical to the SM Higgs boson~\cite{Carena:2013ytb},
with mass and couplings compatible with the LHC measurements~\cite{HIGG-2016-33,CMS-HIG-19-004,HIGG-2018-57,CMS-HIG-17-031}.
In addition, the decoupling limit is used, which is defined by $m_A \gg m_Z$, while the CP-odd ($A$), the neutral CP-even ($H$), and the two charged ($H^\pm$) Higgs bosons are considered to be very heavy and thus considerably beyond the kinematic reach of the LHC.
 
\subsection{RPC SUSY scenarios}\label{sec:rpcsusysignal}
Naturalness~\cite{Barbieri:1987fn,deCarlos:1993yy} motivates light higgsino states ($\ninoone$, $\ninotwo$ and $\chinoonepm$); however, searching for higgsinos can be experimentally challenging.
The sparticles in the higgsino system are close in mass, thus decays of the $\ninotwo/\chinoonepm$ to a $\ninoone$ LSP result in low-momentum decay products that are difficult to reconstruct efficiently.
The LEP experiments searched for higgsino $\chinoonepm$ in approximately mass-degenerate scenarios and excluded chargino masses below $103.5\,\GeV$ (reduced to $92\,\GeV$ for small chargino--LSP mass differences between $0.1\,\GeV$ and $3\,\GeV$)~\cite{LEPSUSYWG:01-03.1}.
More recently, the ATLAS and CMS experiments have searched for higgsino production~\cite{SUSY-2018-16,CMS-SUS-16-048}, excluding higgsino $\ninotwo$ up to masses of $\sim$$240\,\GeV$ and down to $\ninotwo$--LSP mass differences of $1.5\,\GeV$.
 
General gauge-mediated (GGM) SUSY models~\cite{Meade:2008wd} offer an opportunity to study light higgsinos without relying on the reconstruction of experimentally challenging, low-momentum final states.
In the Planck-scale-mediated SUSY breaking scenario, the gravitino $\gravino$ is the fermionic superpartner of the graviton and its mass is comparable to the masses of the other SUSY particles, $m \sim 100\,\GeV$~\cite{Lahanas:1986uc,NILLES19841}.
In contrast, GGM models predict that the $\gravino$ is nearly massless and can be produced at the LHC via the decays of the higgsinos, e.g. $\ninoone \ra Z/h + \gravino$.
The leptonic decays of the $Z/h$ from the two $\ninoone$ decays can be reconstructed and are targeted in this analysis, giving an opportunity to study four-lepton signatures with one or more $Z$ boson candidates.
 
Simplified RPC SUSY models~\cite{Alwall:2008ve,Alwall:2008ag,Alves:2011wf} inspired by GGM are considered here, where an almost mass-degenerate higgsino system $\chinoonepm,\,\ninoone,\,\ninotwo$ and an LSP $\gravino$ with mass $1\,\MeV$ are the only SUSY particles within the reach of the LHC.
The $\chinoonepm$ and $\ninotwo$ masses are set to $1\,\GeV$ above the $\ninoone$ mass to ensure they decay promptly, and because they have only a weak coupling to the $\gravino$, the $\chinoonepm$ and $\ninotwo$ always decay to the $\ninoone$ via virtual $Z/W$ bosons.
The virtual $Z/W$ in turn decay to very soft final states that are not reconstructed, while the $\ninoone$ decays promptly to a gravitino plus a $Z$ or $h$ boson, $\ninoone \ra Z/h + \gravino$.
A higgsino system offers four production processes at the LHC: $\chinoonep\chinoonem$, $\chinoonepm\ninoone$, $\chinoonepm\ninotwo$ and $\ninoone\ninotwo$, all of which are considered in these GGM models, as shown in Figure~\ref{fig:HinoGGMdiagrams}.
The $\ninoone$ mass and $\ninoone \ra Z \gravino$ branching ratio are the two free parameters of the simplified GGM higgsino scenarios.
 
\begin{figure}[htbp]
\begin{center}
\includegraphics[width=0.4\textwidth]{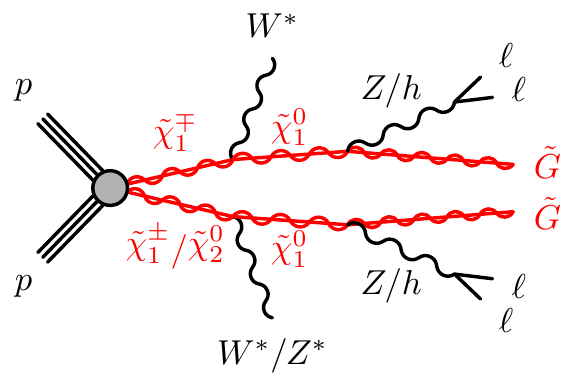}\hspace{0.03\textwidth}
\includegraphics[width=0.4\textwidth]{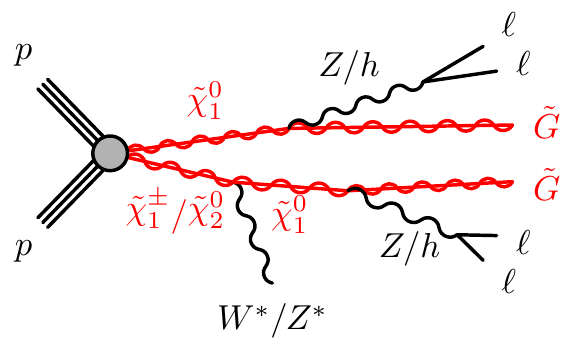}
\end{center}
\caption{Diagrams of the processes in the SUSY RPC GGM higgsino models.
The $W^{*}/\,Z^{*}$ produced in the $\chinoonepm/\ninotwo$ decays are off-shell ($m\sim 1\,\GeV$) and their decay products are usually not reconstructed.
The Higgs boson may decay into leptons and possible additional products via intermediate $\tau\tau$, $WW$ or $ZZ$ states.
\label{fig:HinoGGMdiagrams}}
\end{figure}

\subsection{RPV SUSY scenarios}\label{sec:rpvsusysignal}
 
In generic SUSY models with minimal particle content, the superpotential includes terms that violate conservation of $L$ and $B$:
 
\vspace{-0.4cm}
\begin{equation*}\label{eq:rpvpotential}
\frac{1}{2}\lambda_{ijk} L_iL_j\bar{E}_k +
\lambda_{ijk}' L_iQ_j\bar{D}_k +
\frac{1}{2}\lambda_{ijk}'' \bar{U_i}\bar{D_j}\bar{D_k} +
\kappa_iL_iH_2,
\end{equation*}
 
\vspace{-0.2cm}
\noindent
where $L_i$ and $Q_i$ indicate the lepton and quark $\mathrm{SU}(2)$-doublet superfields, respectively, and $\bar{E_i}$, $\bar{U_i}$ and $\bar{D_i}$ are the corresponding singlet superfields.
Quark and lepton generations are referred to by the indices $i$, $j$ and $k$, while the Higgs field that couples to up-type quarks is represented by the Higgs $\mathrm{SU(2)}$-doublet superfield $H_2$.
The $\lambda$, $\lambda'$ and $\lambda''$ parameters are three sets of new Yukawa couplings, while the $\kappa_i$ parameters have dimensions of mass.

Simplified models of RPV SUSY scenarios are considered here, with a bino neutralino ($\ninoone$) LSP which decays via an RPV interaction.
The lepton-number-violating superpotential term $\frac{1}{2}\lambda_{ijk} L_iL_j\bar{E}_k$ mediates the LSP decay into two charged leptons and a neutrino,
\vspace{-0.2cm}
\begin{equation}
\label{eq:LLERPV_N1decay}
\ninoone \to \ell^{\pm}_k \ell^{\mp}_{i/j} \nu_{j/i},
\end{equation}
 
\vspace{-0.2cm}
through a virtual slepton or sneutrino, with the allowed lepton flavours depending on the indices of the associated $\lambda_{ijk}$ couplings~\cite{Barbier:2004ez}.
The complex conjugate of the decay in Eq.~\eqref{eq:LLERPV_N1decay} is also allowed.
Thus, when two $\ninoone$ are present in a signal process, every signal event contains a minimum of four charged leptons and two neutrinos, giving an opportunity to study four-lepton SUSY signatures.
 
In principle, the nine\footnote{The 27 $\lambda_{ijk}$ RPV couplings are reduced to 9 by the antisymmetry requirement $\lambda_{ijk} = -\lambda_{jik}$ and the $i\neq j$ requirement for the generation of the $L$ terms in the superpotential.} $\lambda_{ijk}$ RPV couplings allow the $\ninoone$ to decay to every possible combination of charged-lepton pairs, where the branching ratio for each combination differs for each $\lambda_{ijk}$.
For example, for $\lambda_{121}\neq 0$ the branching ratios for $\ninoone \ra e\mu\nu$, $\ninoone \ra ee\nu$ and $\ninoone \ra \mu\mu\nu$ are 50\%, 50\% and 0\% respectively,
whereas for $\lambda_{122}\neq 0$ the corresponding branching ratios are 50\%, 0\% and 50\%.
It was shown in Ref.~\cite{SUSY-2013-13} that the four-charged-lepton search sensitivity is comparable in the cases of $\lambda_{121}\neq 0$ or $\lambda_{122}\neq 0$, and for $\lambda_{133}\neq 0$ or $\lambda_{233}\neq 0$.
Since the analysis reported here uses similar techniques for these cases, the number of $L$-violating RPV scenarios studied is reduced by making no distinction between the electron and muon decay modes of the $\ninoone$.
Two extremes of the $\lambda_{ijk}$ RPV couplings are considered:
\begin{itemize}
\item $\LLE 12k$ ($k \in {1,2}$) scenarios, where $\lambda_{12k}\neq 0$ and only decays to electrons and muons are included,
\item $\LLE i33$ ($i \in {1,2}$) scenarios, where $\lambda_{i33}\neq 0$ and only decays to $\tau$-leptons and either electrons or muons are included,
\end{itemize}
with all other RPV couplings assumed to be zero.
The branching ratios for the $\ninoone$ decay in the $\LLE 12k$ and $\LLE i33$ scenarios are shown in \Tab{\ref{tab:datasets_RPVsignalBR}}.
The sensitivity to $\lambda_{ijk}\neq 0$ couplings not considered here (e.g.\ $\lambda_{123}\neq 0$) is expected to be between that achieved in the $\LLE 12k$ and $\LLE i33$ scenarios.

\begin{table}[ht]
\centering
\caption{Decay modes and branching ratios for the $\ninoone$ LSP in the RPV models, where $\nu$ denotes neutrinos or antineutrinos of any lepton generation.
\label{tab:datasets_RPVsignalBR}}
\begin{tabular}{l ccc ccc}
\toprule
Scenario & \multicolumn{6}{c}{\ninoone branching ratios} \\
& $e^+e^-\nu$ & $e^\pm \mu^\mp \nu$ & $\mu^+\mu^-\nu$ & $e^\pm\tau^\mp\nu$ & $\tau^+\tau^-\nu$ & $\mu^\pm\tau^\mp\nu$ \\
\midrule
$\LLE 12k$ & $1/4$ & $1/2$ & $1/4$ & 0 & 0 & 0 \\
$\LLE i33$ & 0 & 0 & 0 & $1/4$ & $1/2$ & $1/4$ \\
\bottomrule
\end{tabular}
\end{table}
 
Pure-bino $\ninoone\ninoone$ production has a vanishingly small cross-section at the LHC, thus models that include one or more next-to-lightest SUSY particles (NLSP) are considered in order to obtain a reasonably large cross-section.
The choice of NLSP in the RPV SUSY scenarios determines the production cross-section, and can impact the signal acceptance to a lesser extent as intermediate decay products may also decay to leptons.
In all cases considered here, the NLSP is pair-produced in an RPC interaction and decays to the $\ninoone$ LSP (which itself undergoes an RPV decay).
Three different possibilities are considered for the NLSP in the $\LLE 12k$ and $\LLE i33$ scenarios:
\begin{itemize}
\item \textbf{wino NLSP}: mass-degenerate wino charginos and neutralinos are produced in association ($\chinoonep\chinoonem$ or $\chinoonepm\ninotwo$).
The charginos decay via $\chinoonepm\rightarrow W^{(*)} \ninoone$ with 100\% branching fraction, while the neutralinos decay via $\ninotwo \rightarrow Z^{(*)}\ninoone$ or $h\ninoone$ with 50\% branching fraction each, as shown in Figure~\ref{fig:RPVDiagrams_wino}.
\item $\pmb{\sleptonL}/\pmb{\snu}$~\textbf{NLSP}: mass-degenerate sleptons and sneutrinos of all three generations are produced in association ($\sleptonL\sleptonL$, $\snu\snu$, $\sleptonL\snu$, where the subscript L refers to the chirality of the partner lepton).
The sleptons decay via $\sleptonL \rightarrow \ell \ninoone$ and sneutrinos decay via $\snu \rightarrow \nu \ninoone$, both with 100\% branching fraction, as seen in Figure~\ref{fig:RPVDiagrams_LHSlepSnu}.
\item $\mathbf{\gluino}$~\textbf{NLSP}:   gluino pair-production, where the gluino decays with 100\% branching fraction via $\gluino \rightarrow q\bar{q} \ninoone$ ($q=u,d,s,c,b$ only, with equal branching fractions), as seen in Figure~\ref{fig:RPVDiagrams_gluino}. Decays to top quarks are not considered here, as this would introduce a significant change in signature for scenarios with mass difference $m(\gluino)-m(\ninoone)$ above and below $\sim 350\,\GeV$.
\end{itemize}
For the RPV models, the LSP mass is restricted to the range $10\,\GeV \leq m(\text{LSP}) \leq m(\text{NLSP}) - 10\,\GeV$ to ensure that both the RPC cascade decay and the RPV LSP decay are prompt.
Non-prompt decays of the $\ninoone$ in similar models were previously studied in Refs.~\cite{SUSY-2014-02,SUSY-2017-04}.

\begin{figure}[htbp]
\begin{center}
\subfloat[][wino NLSP]{\includegraphics[width=0.3\textwidth]{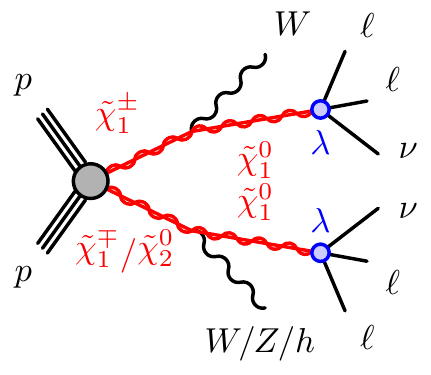}\label{fig:RPVDiagrams_wino}}\hspace{0.03\textwidth}
\subfloat[][$\sleptonL/\snu$ NLSP]{\includegraphics[width=0.3\textwidth]{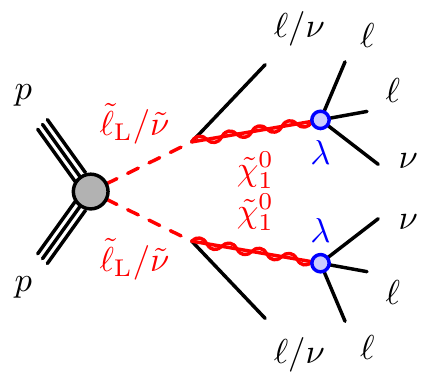}\label{fig:RPVDiagrams_LHSlepSnu}}\hspace{0.03\textwidth}
\subfloat[][$\gluino$ NLSP]{\includegraphics[width=0.3\textwidth]{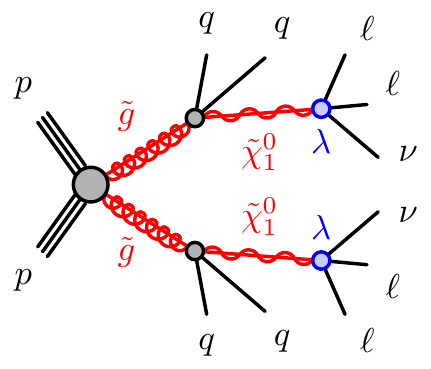}\label{fig:RPVDiagrams_gluino}}
\end{center}
\caption{Diagrams of the benchmark SUSY models of RPC NLSP pair production of a
\protect\subref{fig:RPVDiagrams_wino}~wino,
\protect\subref{fig:RPVDiagrams_LHSlepSnu}~slepton/sneutrino and
\protect\subref{fig:RPVDiagrams_gluino}~gluino, followed by the RPV decay of the $\ninoone$ LSP.
The LSP is assumed to decay as $\ninoone\ra \ell\ell\nu$ with 100\% branching ratio. }
\label{fig:RPVDiagrams}
\end{figure}

\section{ATLAS detector}
\label{sec:detector}
\newcommand{\AtlasCoordFootnote}{
ATLAS uses a right-handed coordinate system with its origin at the nominal interaction point (IP)
in the centre of the detector and the \(z\)-axis along the beam pipe.
The \(x\)-axis points from the IP to the centre of the LHC ring,
and the \(y\)-axis points upwards.
Cylindrical coordinates \((r,\phi)\) are used in the transverse plane,
\(\phi\) being the azimuthal angle around the \(z\)-axis.
The pseudorapidity is defined in terms of the polar angle \(\theta\) as \(\eta = -\ln \tan(\theta/2)\).
Angular distance is measured in units of \(\Delta R \equiv \sqrt{(\Delta\eta)^{2} + (\Delta\phi)^{2}}\).}
 
The ATLAS experiment~\cite{PERF-2007-01,ATLAS-TDR-2010-19,PIX-2018-001} at the LHC is a multipurpose particle detector
with a forward--backward symmetric cylindrical geometry and a near \(4\pi\) coverage in
solid angle.\footnote{\AtlasCoordFootnote}
It consists of an inner tracking detector (ID) surrounded by a thin superconducting solenoid
providing a \SI{2}{\tesla} axial magnetic field, electromagnetic and hadron calorimeters, and a muon spectrometer (MS).
The ID covers the pseudorapidity range \(|\eta| < 2.5\).
It consists of silicon pixel, silicon microstrip, and transition radiation tracking detectors.
Lead/liquid-argon (LAr) sampling calorimeters provide electromagnetic (EM) energy measurements with high granularity.
Hadronic calorimetry is provided by the steel/scintillator-tile calorimeter, which covers the central pseudorapidity range (\(|\eta| < 1.7\)).
The endcap and forward regions are instrumented with LAr calorimeters
for EM and hadronic energy measurements up to \(|\eta| = 4.9\).
The MS surrounds the calorimeters and is based on
three large air-core toroidal superconducting magnets with eight coils each.
The field integral of the toroids ranges between \num{2.0} and \SI{6.0}{\tesla\metre}
across most of the detector.
The MS includes a system of precision tracking chambers covering the region \(|\eta| < 2.7\) and fast detectors for triggering.
A two-level trigger system is used to select events.
The first-level trigger is implemented in hardware and uses a subset of the detector information
to accept events at a rate below \SI{100}{\kHz}.
This is followed by a software-based trigger that
reduces the accepted event rate to \SI{1}{\kHz} on average,
depending on the data-taking conditions.
 
\section{Data and simulated event samples}\label{sec:mcsim}
This analysis uses the full $\sqrt{s} = 13\,\TeV$ $pp$ dataset collected by the ATLAS experiment during the 2015--2018 data-taking period.
The average number of multiple $pp$ collisions in the same or nearby bunch crossings (pile-up) increased from 14 in 2015 to $\sim 38$ in 2018.
After the application of beam, detector and data-quality requirements~\cite{DAPR-2018-01}, the total integrated luminosity considered in this analysis corresponds to $139.0\pm2.4~\ifb$~\cite{ATLAS-CONF-2019-021}.
Events recorded during stable data-taking conditions are used in the analysis if the reconstructed primary vertex has at least two tracks with transverse momentum $\pt>500\,\MeV$ associated with it.
The primary vertex of an event is identified as the vertex with the highest $\Sigma \pt^2$ of associated tracks.

Events are selected using the single-lepton, dilepton, or trilepton triggers~\cite{TRIG-2018-05,Aad:2020uyd} listed in Table~\ref{tab:triggers},
where the trigger efficiencies are in the plateau region above the offline $\pt$ thresholds.
Dilepton (trilepton) triggers are used only when the leptons in the event fail $\pt$-threshold requirements for the single-lepton (single-lepton and dilepton) triggers.
The trigger efficiency for events with four (three) electrons/muons in signal SUSY scenarios is typically $>$99\% ($>$96\%).
For signal SUSY events with only two light leptons, the trigger efficiency is typically $>95\%$ for events with at least one electron and decreases to $\sim 90\%$ for events with only two muons.
 
\begin{table}[h]
\begin{center}
\caption{The triggers used in the analysis of 2015--2018 data.
The offline $\pt$ thresholds are required only for reconstructed charged leptons responsible for triggering the event.
Trigger thresholds increase across the years due to the increase in beam luminosity, and ``or'' denotes a move to a higher-threshold trigger during data-taking.
\label{tab:triggers} }
\vspace{0.2cm}
\small{
\begin{tabular}{ l lll }
\toprule
Trigger                            & \multicolumn{3}{l}{Offline $\pt$ threshold [$\,\GeV$] } \\
& 2015              & 2016     & 2017--2018                              \\
\midrule
Single $e$ (isolated)                & 25 & 27 & 27                                \\
Single $e$            & 61 & 61 & 61                                  \\
Single $\mu$ (isolated)              & 21 & 25 or 27 & 27                          \\
Single $\mu$          & 41 & 41 or 51 & 51                          \\
\midrule
Double $e$                         & 13, 13  & 18, 18 & (18, 18) or (25, 25)                \\
\midrule
Double $\mu$ { \textit (symmetric)}     & 11, 11  & (11, 11) or (15, 15)  & 15, 15                 \\
~~~~~~~~~~~~~~~~{\textit(asymmetric)}   & 19, 9   & (21, 9) or (23, 9)  & 23, 9           \\
\midrule
Double $e\mu$                      & 8($e$), 25($\mu$) & 8($e$), 25($\mu$)  & 8($e$), 25($\mu$)                 \\
& 18($e$), 15($\mu$)& 18($e$), 15($\mu$)& 18($e$), 15($\mu$)\\
&                      & 27($e$), 9($\mu$)& 27($e$), 9($\mu$)\\
\midrule
Triple $e\mu\mu$, $ee\mu$                      &  & 13($e$), 11($2\mu$) & 13($e$), 11($2\mu$) \\
&  & 13($2e$), 11($\mu$)& 13($2e$), 11($\mu$)\\
 
\bottomrule
\end{tabular}
}
\end{center}
\end{table}
 
Monte Carlo (MC) generators were used to simulate SM processes and new physics signals.
The SM processes considered are those that can lead to signatures with at least four reconstructed charged leptons.
Details of the signal and background MC simulation samples used in this analysis, as well as the order of the cross-section calculations in perturbative quantum chromodynamics used for yield normalisation, are shown in Table~\ref{tab:SMbkgMC}.

The SUSY signal processes were generated from leading-order (LO) matrix elements with up to two extra partons.
Jet--parton matching followed the CKKW-L prescription~\cite{Lonnblad:2012ix}, with a matching scale set to one quarter of the mass of the pair-produced SUSY particles.
Signal cross-sections were calculated to next-to-leading order in the strong coupling constant, adding the resummation of soft gluon emission at next-to-leading-logarithm accuracy (NLO+NLL)~\cite{Beenakker:1996ch,Kulesza:2008jb,Kulesza:2009kq,Beenakker:2009ha,Beenakker:2011fu,Fuks:2012qx,Fuks:2013vua,Fuks:2013lya}.
The nominal signal cross-section and its uncertainty were taken from an envelope of cross-section predictions using different parton distribution function (PDF) sets and factorisation and renormalisation scales, as described in Ref.~\cite{Borschensky:2014cia}.
 
The dominant irreducible background processes that can produce four prompt and isolated charged leptons are $ZZ$, $\ttbar Z$, $VVV$ and Higgs production (where $V=W,\,Z$, and includes off-shell $Z/\gamma$ contributions).
For simulated $ZZ$ production, the matrix elements contain all diagrams with four electroweak vertices, and they were calculated for up to one extra parton at NLO, and up to three extra partons at LO.
The production of top quark pairs with an additional $Z$ boson was simulated with matrix elements calculated at NLO precision.
Simulated triboson ($VVV$) production includes the processes $ZZZ$, $WZZ$ and $WWZ$ with four to six charged leptons, and was generated at NLO with additional LO matrix elements for up to two extra partons.
The simulation of Higgs processes includes Higgs production via gluon--gluon fusion (ggF) and vector-boson fusion (VBF), and associated production with a boson ($WH$, $ZH$) or a top--antitop pair ($\ttbar H$).
Other irreducible background processes with small cross-sections are grouped into a category labelled `Other', which contains the $tWZ$, $\ttbar WW$, $\ttbar ZZ$, $\ttbar WZ$, $\ttbar WH$, $\ttbar HH$, $\ttbar tW$ and $\ttbar \ttbar $ processes.
 
Top quark pair production and $Z$+jets are the dominant SM processes that may produce one or more non-prompt or misidentified leptons among the four charged leptons.
Processes such as $W$+jets, $WW$, $WZ$, and $\ttbar W$ also contribute to a four or more charged lepton signature, but at very small rates either due to a small production cross-section, or they require a higher multiplicity of non-prompt or misidentified leptons.
MC simulation of these processes is used as part of the estimation of the reducible background, as described in Section~\ref{sec:bgdet_red}.
Further information about the MC simulations of the reducible backgrounds can be found in Refs.~\cite{ATL-PHYS-PUB-2018-009,ATL-PHYS-PUB-2017-006}.

For all MC simulation samples, the propagation of particles through the ATLAS detector was modelled with $\GEANT4$~\cite{Agostinelli:2002hh} using the full ATLAS detector simulation~\cite{SOFT-2010-01}, except for the SUSY signal samples, which use a fast simulation based on a parameterisation of the response of the electromagnetic and hadronic calorimeters~\cite{ATL-PHYS-PUB-2010-013} and full simulation elsewhere.
The effect of pile-up is incorporated into the simulation by overlaying additional inelastic $pp$ events onto hard-scatter events.
These were generated with \PYTHIA8~\cite{Sjostrand:2014zea} with a set of tuned parameters called the A3 tune~\cite{ATL-PHYS-PUB-2016-017} and the MSTW2008LO PDF set~\cite{Martin:2009iq}.
Simulated events are reconstructed in the same manner as data, and are weighted to match the distribution of the expected mean number of interactions per bunch crossing in data.
The simulated MC samples are corrected to account for differences from the data in the triggering efficiencies,
lepton reconstruction efficiencies,
$b$-quark jet identification efficiencies,
and the energy and momentum measurements of leptons and jets.
 
\begin{table}[h]
\centering
\caption{Summary of the simulated SM background and signal samples used in this analysis, where $V=W,Z$, and includes off-shell contributions. ``Tune'' refers to the set of parameter values used by the generator.
\label{tab:SMbkgMC}}
\vspace{0.2cm}
\scriptsize
\resizebox{\linewidth}{!}{
\begin{tabular}{ l cccc }
\toprule
Process   &   Generator(s)  & Cross-section  &  Tune  &             PDF set             \\
&         &                     calculation  &        &  \\ \midrule
 
$ZZ$, $WZ$, $WW$  & \SHERPA 2.2.2~\cite{Bothmann:2019yzt}  &  NLO~\cite{ATL-PHYS-PUB-2016-002} & \SHERPA default & NNPDF30NNLO~\cite{Ball:2014uwa} \\
\midrule
 
$VVV$            & \SHERPA 2.2.1                          &  NLO~\cite{ATL-PHYS-PUB-2016-002} & \SHERPA default & NNPDF30NNLO     \\
\midrule
 
$H$ via ggF, VBF, $VH$           & \POWHEGBOX v2~\cite{Nason:2004rx, Frixione:2007vw,Alioli:2010xd} + \PYTHIA 8.212~\cite{Sjostrand:2014zea}  &  NNNLO+NNLL~\cite{deFlorian:2016spz,Anastasiou:2016cez,Anastasiou:2015ema,Dulat:2018rbf,Aglietti:2004nj,Actis:2008ug,Bonetti:2018ukf} &    AZNLO~\cite{STDM-2012-23}     &     CTEQ6L1~\cite{Pumplin:2002vw} \\
$\ttbar H$      & \POWHEGBOX v2 + \PYTHIA 8.230  &   NLO~\cite{deFlorian:2016spz}   & A14~\cite{ATL-PHYS-PUB-2014-021}    &  NNPDF23LO~\cite{Ball:2012cx}   \\
\midrule
 
$\ttbar Z$, $\ttbar W$ & \MGMCatNLO 2.3.3~\cite{Alwall:2014hca} + \PYTHIA 8.210  &  NLO~\cite{ATL-PHYS-PUB-2016-005}   &   A14  &   NNPDF23LO            \\
\midrule
 
$\ttbar WW$                 & \MGMCatNLO~2.2.2 + \PYTHIA 8.186 &  NLO~\cite{ATL-PHYS-PUB-2016-005} & A14  & NNPDF23LO    \\
$\ttbar WZ$, $tWZ$          & \MGMCatNLO~2.3.3 + \PYTHIA 8.212 &  NLO~\cite{ATL-PHYS-PUB-2016-005} & A14  & NNPDF23LO    \\
$\ttbar ZZ$, $\ttbar WZ$, $\ttbar WH$, $\ttbar HH$  & \MGMCatNLO~2.6.7 + \PYTHIA 8.240 &  NLO~\cite{ATL-PHYS-PUB-2016-005} & A14  & NNPDF23LO    \\
$\ttbar tW$, $\ttbar\ttbar$ & \MGMCatNLO~2.2.2 + \PYTHIA 8.186 &  NLO~\cite{Alwall:2014hca}  & A14  & NNPDF23LO    \\
\midrule
 
\ttbar  &  \POWHEGBOX v2 + \PYTHIA 8.230   &  NNLO+NNLL~\cite{Beneke:2011mq,Cacciari:2011hy,Baernreuther:2012ws,Czakon:2012zr,Czakon:2012pz,Czakon:2013goa,Czakon:2011xx} & A14 &  NNPDF23LO               \\
\midrule
 
$Z$+jets, $W$+jets &  \POWHEGBOX v1 + \PYTHIA 8.186  &  NNLO~\cite{Anastasiou:2003ds}  & AZNLO  &  CTEQ6L1            \\
\midrule
 
SUSY signal  &  \MGMCatNLO 2.2.2 + \PYTHIA 8.230  & NLO+NLL~\cite{Beenakker:1996ch,Kulesza:2008jb,Kulesza:2009kq,Beenakker:2009ha,Beenakker:2011fu,Fuks:2012qx,Fuks:2013vua,Fuks:2013lya} &  A14  & NNPDF23LO            \\
 
\bottomrule
\end{tabular}
}
\end{table}
 
\section{Event reconstruction}\label{sec:eventreco}
 
This analysis uses reconstructed electrons, muons, $\tau$-leptons, and jets, which are classified as `preselected' or `signal' using various kinematic and quality criteria.
Preselected objects must satisfy a loose set of criteria and pass the overlap removal procedure, which resolves ambiguities among reconstructed objects.
Signal leptons are those preselected leptons that satisfy a more stringent set of criteria; those failing the signal lepton requirements are used as part of the background estimation in Section~\ref{sec:bgdet_red}.
The $\pt$ thresholds for leptons are nominally low; however, $\pt$ thresholds are higher for the one, two, or three leptons responsible for triggering the event via the single-lepton, dilepton, or trilepton triggers listed in Table~\ref{tab:triggers}.

The missing transverse momentum, $\met$, is the magnitude of the negative vector sum of the transverse momenta of all preselected objects (electrons, photons, muons, and jets, including all jets with $|\eta|<4.5$) and an additional soft term~\cite{ATLAS-CONF-2018-023}.
Hadronically decaying $\tau$-leptons are included in the $\met$ as jets.
The soft term is constructed from the tracks matched to the primary vertex, but not associated with identified physics objects.
By using tracks, it cannot account for the neutral component of calorimeter energy deposits; however, this allows the soft term to be nearly independent of pile-up~\cite{PERF-2016-07}.
 
Preselected electrons are reconstructed using calibrated clusters of energy deposits in the electromagnetic calorimeter that are matched to a track in the ID, and must have $\pt>4.5\,\GeV$ and $|\eta|<2.47$.
They must also satisfy the tracking- and calorimeter-based `loose and B-layer' criteria of the likelihood-based identification algorithm~\cite{EGAM-2018-01}.
Preselected muons are reconstructed by combining tracks in the ID with tracks in the MS, and must have $\pt>3\,\GeV$ and $|\eta|<2.7$.
They must also satisfy `medium' identification requirements~\cite{MUON-2018-03}, which are based on the number of hits in the different ID and MS subsystems, and on the significance of the charge-to-momentum ratio.
The cosmic-ray muon background is suppressed by rejecting events containing one or more muons that have a transverse impact parameter $|d_0|>0.2\,$mm, or a longitudinal impact parameter $|z_0|>1\,$mm, both relative to the primary vertex.
Preselected electrons and muons must point back to the primary vertex, with $|z_{0}\sin{\theta}|$ required to be less than $0.5\,$mm.

Jets are reconstructed from three-dimensional calorimeter energy clusters using the anti-$k_t$ algorithm~\cite{Cacciari:2008gp} with a radius parameter of $R=0.4$.
The jets are calibrated following Ref.~\cite{PERF-2016-04} and must have $\pt>20\,\GeV$ and $|\eta|<2.8$.
Events with large calorimeter noise or non-collision backgrounds are suppressed by rejecting events with jets that fail to satisfy the quality criteria described in Ref.~\cite{ATLAS-CONF-2015-029}.
A multivariate technique based on quantities related to reconstructed secondary vertices is used to identify jets with $|\eta|<2.5$ that originate from $b$-quarks (referred to as `$b$-tagging').
The $b$-tagging algorithm~\cite{FTAG-2018-01} used here correctly identifies $b$-quark jets in simulated $\ttbar$ samples with an efficiency of 85\% and a rejection factor of 25 for light-flavour jets.

Leptonically decaying $\tau$-leptons are reconstructed as electrons and muons as described above.
Hadronically decaying $\tau$-leptons are denoted by $\tauhad$, and their visible decay products are reconstructed as jets, as described above, with $\pt>10\,\GeV$ and $|\eta|<2.47$.
In this analysis, kinematic variables built with $\tauhad$ candidates use only their visible decay products.
The $\tauhad$ reconstruction algorithm~\cite{ATLAS-CONF-2017-029} uses the electromagnetic and hadronic shower shapes in the calorimeters, as well as information about the tracks within $\Delta R=0.2$ of the jet direction.
Since $\tau$-leptons mostly decay into either one or three charged hadrons together with a neutrino (and often additional neutral hadrons), $\tauhad$ candidates are required to have one or three associated tracks, referred to as `prongs'.
The preselected $\tauhad$ candidates must have $\pt>20\,\GeV$, $|\eta|<1.37$ or $1.52<|\eta|<2.47$, total charge of their constituent tracks equal to $\pm1$, and the $\tauhad$ energy scale is corrected using an $\eta$- and $\pt$-dependent calibration.
A recurrent neural network (RNN) uses discriminating track and cluster variables to optimise $\tauhad$ identification,
where `loose', `medium' and `tight' working points are defined~\cite{ATL-PHYS-PUB-2019-033}.
The RNN-based identification is used to define signal $\tauhad$ candidates, but not preselected $\tauhad$ candidates.
Transition radiation tracker and calorimeter information is used to suppress electrons misidentified as preselected $\tauhad$ candidates.

To avoid double counting of identified physics objects, preselected charged leptons and jets must survive an overlap removal procedure, applied in the following order:
\vspace{-5pt}
\begin{enumerate}[nolistsep]
\item Any $\tauhad$ within $\Delta R=0.2$ of an electron or muon is removed.
\item Any electron sharing an ID track with a muon is removed.
\item Any jet within $\Delta R=0.2$ of an electron is removed.
\item Any electron within $\Delta R=0.4$ of a jet is removed (to suppress electrons from semileptonic decays of $c$- and $b$-hadrons).
\item Any jet with fewer than three associated tracks is removed either if a muon is within $\Delta R=0.2$ or if the muon can be matched to a track associated with the jet.
\item Any muon within $\Delta R=0.4$ of a jet is removed (to suppress muons from semileptonic decays of $c$- and $b$-hadrons).
\item Any jet within $\Delta R=0.4$ of a preselected $\tauhad$ passing `medium' RNN-based identification requirements is removed.
\end{enumerate}
To suppress low-mass particle decays,
if surviving electrons and muons form an opposite-sign (OS) pair with $m_{\textrm OS} < 4\,\GeV$,
or form a same-flavour, opposite-sign (SFOS) pair in the $\Upsilon(1S)$--$\Upsilon(3S)$ mass range $8.4<\msfos<10.4\,\GeV$, both leptons are discarded.
Finally, to suppress leptons from a decay chain with multiple heavy flavour quarks undergoing leptonic decay, e.g. $b \rightarrow  Wc (\rightarrow Ws)$ where $W\rightarrow \ell\bar{\nu}$, if two leptons are found within $\Delta R = 0.6$ of one another and one of them has $\pt <30\,\GeV$, both leptons are discarded.

Reconstructed charged leptons may be `real', defined to be prompt and genuinely isolated leptons (including those from leptonic $\tau$ decays), or `fake/non-prompt', defined to be non-prompt or non-isolated leptons that could originate from semileptonic decays of $b$- and $c$-hadrons, from in-flight decays of light mesons, from misidentification of particles within light-flavour or gluon-initiated jets, or from photon conversions.
To suppress fake/non-prompt leptons, preselected objects surviving overlap removal are required to satisfy additional identification criteria and are referred to as \textit{signal leptons/jets}.
Signal electrons must have $\pt>7\,\GeV$ and signal muons must have $\pt>5\,\GeV$.
Signal electrons must also satisfy `medium' likelihood-based identification criteria~\cite{EGAM-2018-01},
while signal $\tauhad$ must satisfy the `medium' RNN-based identification criteria~\cite{ATL-PHYS-PUB-2019-033}.
Signal electrons and muons must pass $\pt$-dependent isolation requirements imposed to reduce the contributions from semileptonic decays of hadrons and jets misidentified as prompt leptons.
The `Loose' isolation working point is used for electrons and muons, as described in Refs.~\cite{EGAM-2018-01} and \cite{MUON-2018-03}, including updates to improve the performance under conditions with higher pile-up encountered during 2017 and 2018 data-taking.
To improve the identification of closely spaced charged leptons (e.g.\ from boosted decays), contributions to the isolation energy and \pt sums from nearby electrons and muons passing all other signal lepton requirements are removed.
To further suppress electrons and muons originating from secondary vertices, the transverse impact parameter normalised to its uncertainty must be small, $|d_{0}|/\sigma_{d_{0}}<5\,(3)$ for electrons (muons).
To reduce pile-up effects, signal jets with $\pt<120\,\GeV$ and $|\eta|<2.5$ must satisfy additional criteria using the `medium' working point of the jet-vertex-tagging algorithm described in Ref.~\cite{PERF-2014-03}.

\section{Signal regions}\label{sec:signalregions}
The search strategy for the SUSY scenarios considered here selects events with at least four signal leptons ($e$, $\mu$, $\tauhad$) and the events are classified according to the number of light signal leptons ($L$ = $e$, $\mu$) and signal $\tauhad$ ($T$) required as follows:
$4L0T$, with at least four light leptons and no $\tauhad$ multiplicity requirement;
$3L1T$, with exactly three light leptons and at least one $\tauhad$;
or $2L2T$, with exactly two light leptons and at least two $\tauhad$.
A general region, $5L0T$, with at least five light leptons and no $\tauhad$ multiplicity requirement is also considered.
The signal region (SR) definitions are summarised in Table~\ref{tab:SR_defs}.

To target the RPC GGM scenarios, events with $4L0T$ are selected and these must have two pairs of SFOS leptons that are both consistent with a leptonic $Z$ boson decay.
The SFOS pair with mass closer to the $Z$ boson mass is labelled as the first $Z$ candidate, while the other SFOS pair is labelled as the second $Z$ candidate.
The first (second) $Z$ candidate must have an invariant mass $m(LL)$ in the range $81.2$--$101.2\,\GeV$ ($61.2$--$101.2\,\GeV$).
The peak of the first $Z$ candidate is narrower due to the ordering of the $Z$ candidates, so that widening the low-mass side of the $m(LL)$ window used for the selection of a second $Z$ candidate increases the GGM signal acceptance.
GGM scenarios with branching ratio ${\mathcal{B}}(\ninoone \rightarrow \gravino h)>0$ will have a significant $h\rightarrow b\bar{b}$ component, but the four-lepton analysis is not sensitive to these decays, so $b$-tagged jets are vetoed to suppress the $\ttbar Z$ and $\ttbar$ SM backgrounds.
Two SRs are defined with $4L0T$, no $b$-tagged jets, a first- and second-$Z$ requirement, and different selections on $\met$: a loose signal region (\SRLLLLE) with $\met>100\,\GeV$, and a tighter signal region (\SRLLLLF) with $\met>200\,\GeV$, optimised for the low-mass and high-mass higgsino GGM scenarios, respectively.
Two further SRs that showed an excess in the $13\,\TeV$ partial dataset analysis in Ref.~\cite{SUSY-2016-21} are also examined here, and are defined with $4L0T$, no requirement on $b$-tagged jets, with a first- and second-$Z$ requirement, and with different selections on $\met$: a loose signal region (\SRLLLLC, labelled \texttt{SR0C} in Ref.~\cite{SUSY-2016-21}) with $\met>50\,\GeV$, and a tighter signal region (\SRLLLLD, labelled \texttt{SR0D} in Ref.~\cite{SUSY-2016-21}) with $\met>100\,\GeV$.
The two newly defined regions, \SRLLLLE\ and \SRLLLLF, are subsets of these two regions, \SRLLLLC\ and \SRLLLLD.

For the RPV scenarios, events with $4L0T$ are used to target the $\LLE 12k$ models, and events with $4L0T$, $3L1T$, and $2L2T$ are used to target the $\LLE i33$ models.
To suppress SM backgrounds with a $Z$ boson, a $Z$ veto is required, which rejects events where any SFOS lepton pair combination has an invariant mass close to the $Z$ boson mass, in the range $81.2$--$101.2\,\GeV$.
The $Z$ veto is extended to three- and four-lepton invariant mass combinations to suppress events where a photon radiated from a $Z\ra\ell\ell$ decay converts to a second SFOS lepton pair; any event with an $\ell^{+}\ell^{-}\ell'^{\pm}$ or $\ell^{+}\ell^{-}\ell'^{+}\ell'^{-}$ system with invariant mass in the range $81.2$--$101.2\,\GeV$ is rejected (the flavour of $\ell$ and $\ell'$ may be different).
A small number of four-lepton events will satisfy neither the $Z$ requirement described above for the GGM scenarios nor the $Z$ veto; however, these are assumed to come from $ Z \ra \ell^+\ell^-\gamma$ and $Z \ra \ell^+\ell^-\ell^+\ell^-$ decays, which are not considered to be signal-like.
 
The gluino and wino RPV models can produce $b$-quarks ($\gluino \rightarrow b\bar{b} \ninoone$, or $\ninotwo \rightarrow \ninoone h, h\rightarrow b\bar{b}$) and these decay chains are an important component of the signal for high $\Delta m(\mathrm{NLSP},\ninoone) = m(\mathrm{NLSP})-m(\ninoone)$.
A veto on the presence of $b$-tagged jets is required for some signal regions to minimise heavy-flavour SM backgrounds, and at least one $b$-tagged jet is required for other signal regions to improve sensitivity to high $\Delta m(\mathrm{NLSP},\ninoone)$ gluino and wino RPV scenarios.
 
In order to separate the RPV SUSY signal from the SM background, the effective mass of the event, $\meff$, is used, defined as the scalar sum of the $\met$,
the $\pt$ of signal leptons and the $\pt$ of all jets with $\pt>40\,\GeV$.
The $\pt>40\,\GeV$ requirement for jets aims to suppress contributions from pile-up and the underlying event.
A selection using the $\meff$ rather than the $\met$ is particularly effective for the RPV SUSY scenarios, which produce multiple high-energy leptons (and in some cases jets), but only low to moderate $\met$ from neutrinos in the final state.
The chosen $\meff$ thresholds are found to be broadly optimal for the wide range of RPV scenarios with different NLSPs considered in this paper.
 
Three general signal regions are defined with a $Z$ veto, no $b$-tagged jets, and $\meff>600\,\GeV$: \SRLLLLA\ with $4L0T$, \SRLLLTA\ with $3L1T$, and \SRLLTTA\ with $2L2T$.
These signal regions are non-optimal for the SUSY scenarios considered here and select regions with low levels of SM background to target new phenomena decaying to four-lepton final states.
Two further signal regions are defined with $4L0T$ and a $Z$ veto: a high-$\meff$ signal region (\SRLLLLB) with no $b$-tagged jets and $\meff>1250\,\GeV$, and a signal region (\SRLLLLG) with one or more $b$-tagged jets and $\meff>1300\,\GeV$, both optimised for RPV $\LLE 12k$ scenarios.
Similarly, two further signal regions are defined with $3L1T$ and a $Z$ veto: a high-$\meff$ signal region (\SRLLLTB) with no $b$-tagged jets and $\meff>1000\,\GeV$, and a signal region (\SRLLLTC) with one or more $b$-tagged jets and $\meff>1300\,\GeV$, both optimised for RPV $\LLE i33$ scenarios.
Finally, two signal regions are defined with $2L2T$ and a $Z$ veto: a high-$\meff$ signal region (\SRLLTTB) with no $b$-tagged jets and $\meff>1000\,\GeV$, and a signal region (\SRLLTTC) with one or more $b$-tagged jets and $\meff>1100\,\GeV$, both optimised for RPV $\LLE i33$ scenarios.
 
A general signal region, \SRLLLLL, with at least five light leptons is also defined, with no further selection applied.
 
\begin{table}[h]
\newcolumntype{R}{>{$}r<{$}}
\newcolumntype{L}{>{$}l<{$}}
\newcolumntype{M}{R@{${}>{}$}L}
\renewcommand\arraystretch{1.3}
\centering
\caption{Signal region definitions. The $Z$ boson column refers to the $Z$ veto or selection of a first and second $Z$ candidate as described in the text. \label{tab:SR_defs}}
\vspace{0.2cm}
\scriptsize{
\resizebox{\linewidth}{!}{
\begin{tabular}{ llcccc M c }
\toprule
Name & Signal Region & $N(e,\mu)$ & $N(\tauhad)$ & $N(b\textrm{-tagged jets})$ &     $Z$ boson      & \multicolumn{2}{c}{Selection} & Target         \\ \midrule
 
$4L0T$ & \SRLLLLE   &  $\geq4$   &     $\geq0$     &  $=0$                 & require 1st \& 2nd & $\met$  & $100\,\GeV$         & higgsino GGM   \\
& \SRLLLLF   &  $\geq4$   &     $\geq0$     &  $=0$                 & require 1st \& 2nd & $\met$  & $200\,\GeV$         & higgsino GGM   \rule{0pt}{3ex} \\
\cline{2-9}
\rule{0pt}{3ex} & \SRLLLLC   &  $\geq4$   &     $\geq0$     &  $\geq0$              & require 1st \& 2nd & $\met$  & $50\,\GeV$          & Excess from Ref.~\cite{SUSY-2016-21}   \\
& \SRLLLLD   &  $\geq4$   &     $\geq0$     &  $\geq0$              & require 1st \& 2nd & $\met$  & $100\,\GeV$         & Excess from Ref.~\cite{SUSY-2016-21}  \rule{0pt}{3ex}  \\
\cline{2-9}
\rule{0pt}{3ex} & \SRLLLLA   &  $\geq4$   &     $\geq0$     &  $=0$                 &        veto        & $\meff$ & $600\,\GeV$         & General        \\
& \SRLLLLB   &  $\geq4$   &     $\geq0$     &  $=0$                 &        veto        & $\meff$ & $1250\,\GeV$        & RPV $\LLE12k$  \\
& \SRLLLLG   &  $\geq4$   &     $\geq0$     &  $\geq1$              &        veto        & $\meff$ & $1300\,\GeV$        & RPV $\LLE12k$  \\
\midrule
$3L1T$ & \SRLLLTA   &    $=3$    &   $\geq1$    &  $=0$                 &        veto        & $\meff$ & $600\,\GeV$         & General \\
& \SRLLLTB   &    $=3$    &   $\geq1$    &  $=0$                 &        veto        & $\meff$ & $1000\,\GeV$        & RPV $\LLE i33$ \\
& \SRLLLTC   &    $=3$    &   $\geq1$    &  $\geq1$              &        veto        & $\meff$ & $1300\,\GeV$        & RPV $\LLE i33$ \\
\midrule
$2L2T$ & \SRLLTTA   &    $=2$    &   $\geq2$    &  $=0$                 &        veto        & $\meff$ & $600\,\GeV$         & General \\
& \SRLLTTB   &    $=2$    &   $\geq2$    &  $=0$                 &        veto        & $\meff$ & $1000\,\GeV$        & RPV $\LLE i33$ \\
& \SRLLTTC   &    $=2$    &   $\geq2$    &  $\geq1$              &        veto        & $\meff$ & $1100\,\GeV$        & RPV $\LLE i33$ \\
\midrule
 
$5L0T$ & \SRLLLLL   &  $\geq5$   &     $\geq0$     &  $\geq0$              & -- & \multicolumn{2}{c}{--}         & General   \\

\bottomrule
\end{tabular}
}       }
\end{table}

\section{Background determination}\label{sec:bgdet}
 
The SM background is composed of processes that can give rise to four real or fake/non-prompt leptons and these are classified into two categories:
\begin{description}
\item[Irreducible background:] hard-scattering processes giving rise to events with four or more real leptons, $ZZ$, $\ttbar Z$, $\ttbar WW$, $\ttbar ZZ$, $\ttbar WZ$, $\ttbar WH$, $\ttbar HH$, $tWZ$, $VVZ$ ($ZZZ$, $WZZ$, $WWZ$), Higgs ($H$ via ggF, $WH$, $ZH$, $H$ via VBF, $\ttbar H$), $\ttbar \ttbar $, $\ttbar t W$.
\item[Reducible background:] processes leading to events with at least one fake/non-prompt lepton, \ttbar, $Z$+jets, $WZ$, $WW$, $WWW$, $\ttbar W$, $\ttbar t$. Processes listed under irreducible that do not undergo a decay to four real leptons (e.g.\ $ZZ\ra q\bar{q}\ell\ell$) are also included in the reducible background.
\end{description}
Backgrounds with three or more fake/non-prompt leptons (e.g.\ $W$+jets) are found to be $<1\%$ of the total SM background in four-lepton regions using the method outlined in Section~\ref{sec:bgdet_red} and are neglected.
The systematic uncertainty of the reducible background is increased to cover any effect from neglected backgrounds (discussed in Section~\ref{sec:systematics}).
 
In the four-lepton signal regions, the main irreducible backgrounds are $ZZ$, and $\ttbar Z$, while the reducible background is dominated by the two-fake/non-prompt-lepton backgrounds $\ttbar$ and $Z$+jets.
The $ZZ$ and $\ttbar Z$ backgrounds are estimated using MC simulation normalised to data in control regions (CR), while the other irreducible backgrounds are estimated from MC simulation.
The reducible backgrounds are derived from data using a fake-factor method.
Signal regions with $4L0T$ are dominated by irreducible background processes, whereas the reducible background processes dominate the $3L1T$ and $2L2T$ regions.
The predictions for irreducible and reducible backgrounds are tested in validation regions (Section~\ref{sec:bgdet_valid}).
 
For \SRLLLLL, the main irreducible background processes are $VZZ$ and Higgs, followed by small contributions from $ZZ\ra 6\ell$ and $\ttbar Z \ra 5\ell$, where virtual photons convert into lepton pairs (internal conversions).
However, reducible background processes are the leading source of events in the $5L0T$ signal region, and are dominated by $ZZ\ra 4\ell$ and $\ttbar Z \ra 4\ell$.
 
The \HistFitter~\cite{Baak:2014wma} software framework is used when constraining the $ZZ$ and $\ttbar Z$ background normalisations and a `background-only fit' of observations in the CRs is used to estimate the expected background in the SRs, without considering any CR signal contamination.
A likelihood function is built as a product of Poisson probability functions, describing the observed and expected number of events in the CRs and SRs.
The observed numbers of events in various CRs and SRs are used in a combined profile likelihood fit to determine the expected SM background yields in each of the SRs.
The systematic uncertainties in the expected SM background yields described in Section~\ref{sec:systematics} are included as nuisance parameters, constrained to be Gaussian with a width determined by the size of the uncertainty.
Common nuisance parameters take into account the correlations between CRs and SRs, and background processes.
The fit parameters are determined by maximising the product of the Poisson probability functions and the Gaussian constraints on the nuisance parameters.
 
\subsection{Irreducible background determination }\label{sec:bgdet_irr}
The irreducible background processes $ZZ$ and  $\ttbar Z$ are estimated using MC simulation normalised to data yields in CRs which are orthogonal to the SRs and minimise potential signal contamination.
By normalising the MC simulation to data, the estimation of $ZZ$ and $\ttbar Z$ is improved in the SRs.
A simultaneous fit to the CRs and SRs (see in Section~\ref{sec:results}) provides the final estimate of the yields and their uncertainties.
 
The $ZZ$ and  $\ttbar Z$ control region definitions are shown in Table~\ref{tab:CRZZttZ_defs}.
The $ZZ$ CR, \CRZZ, is defined with at least four light leptons, no $b$-tagged jets, a first- and second-$Z$ requirement, and $\met<50\,\GeV$,
while the $\ttbar Z$ CR, \CRttZ, is defined with $4L0T$, at least one $b$-tagged jet, only one $Z$ boson candidate, and $\met>100\,\GeV$.
The background-only fit is used to obtain normalisation factors for the $ZZ$ and $\ttbar Z$ MC simulation in their CRs of $1.15 \pm 0.09$ and $1.06 \pm 0.24$, respectively.
The uncertainties quoted for the normalisation factors include the statistical uncertainty of the data and MC simulation in the CR, as well as the experimental and theory uncertainties from the subtraction of contaminating SM processes (see Section~\ref{sec:systematics}).
The $\meff$ distributions for \CRZZ\ and \CRttZ\ after the simultaneous fit is performed are shown in Figure~\ref{fig:CRmeff}.
 
Since the regions \CRZZ\ and \CRttZ\ include five-light-lepton events, they are both restricted to exactly four light leptons when estimating the backgrounds for \SRLLLLL.
In these restricted CRs, normalisation factors of $1.14 \pm 0.09$ and $1.07 \pm 0.25$ are obtained for the $ZZ$ and $\ttbar Z$ backgrounds, respectively.
 
\begin{table}[h]
\newcolumntype{R}{>{$}r<{$}}
\newcolumntype{L}{>{$}l<{$}}
\renewcommand\arraystretch{1.3}
\centering
\caption{Irreducible background control region definitions. Both \CRZZ\ and \CRttZ\ are restricted to $N(e,\mu)=4$ when considering \SRLLLLL. The $Z$ boson column refers to the $Z$ veto or selection of a first/second $Z$ candidate as described in the text. \label{tab:CRZZttZ_defs}}
\vspace{0.2cm}
\small{
\begin{tabular}{ lccccc }
\toprule
Region & $N(e,\mu)$ & $N(\tauhad)$ & $N(b\textrm{-tagged jets})$ &     $Z$ boson      & Selection \\ \midrule
\CRZZ   &  $\geq4$   &     $\geq0$     &  $=0$                 & require 1st \& 2nd~~~~~~~\, & $\met < 50\,\GeV$~         \\
\CRttZ  &  $\geq4$   &     $\geq0$     &  $\geq1$              & require 1st \& veto 2nd & ~$\met > 100\,\GeV$   \\
\bottomrule
\end{tabular}
}
\end{table}

\begin{figure}[h]
\centering
\subfloat[][$\meff$ in \CRZZ]{\includegraphics[width=0.475\textwidth]{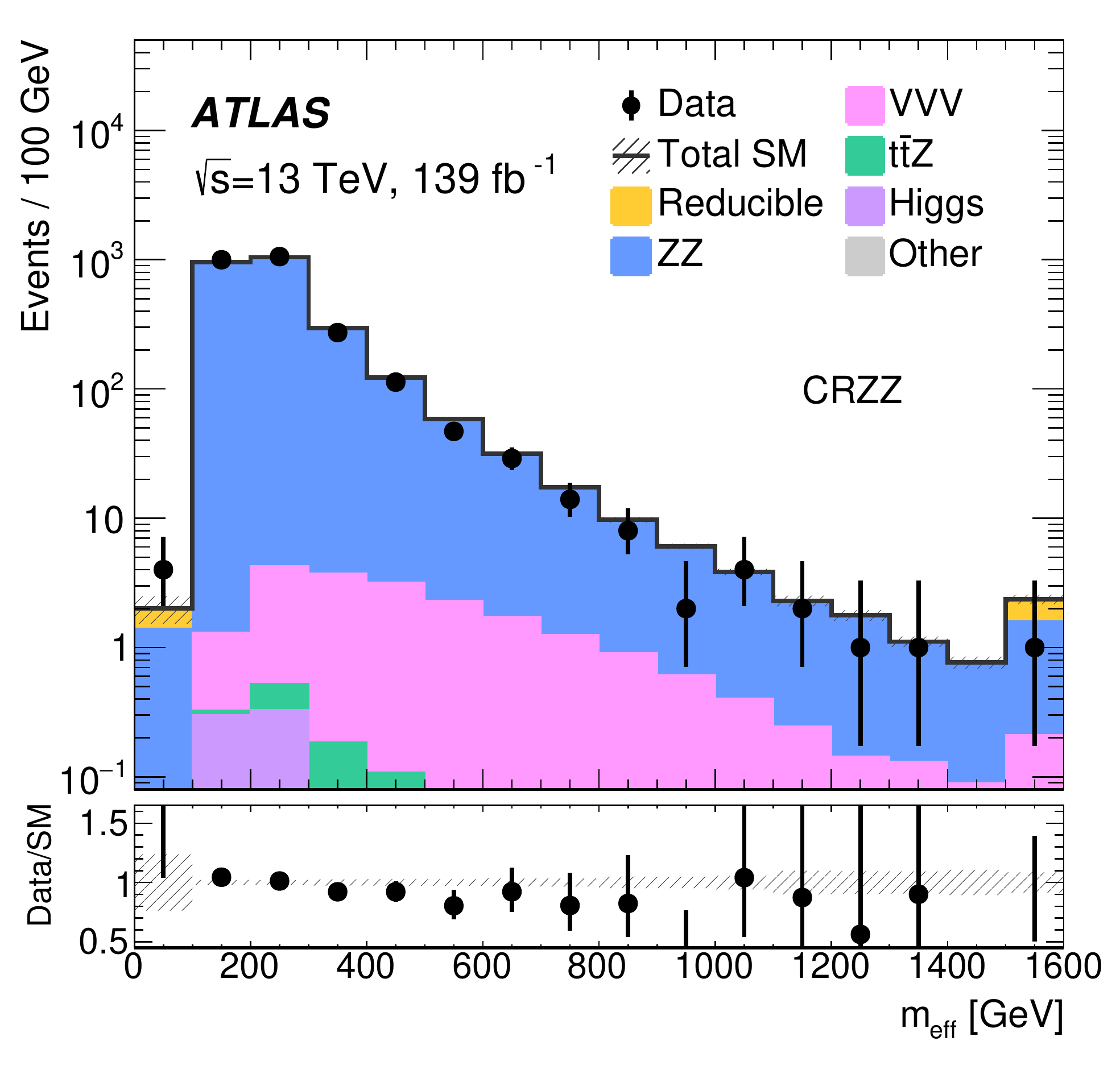}\label{fig:CRZZmeff}}
\subfloat[][$\meff$ in \CRttZ]{\includegraphics[width=0.475\textwidth]{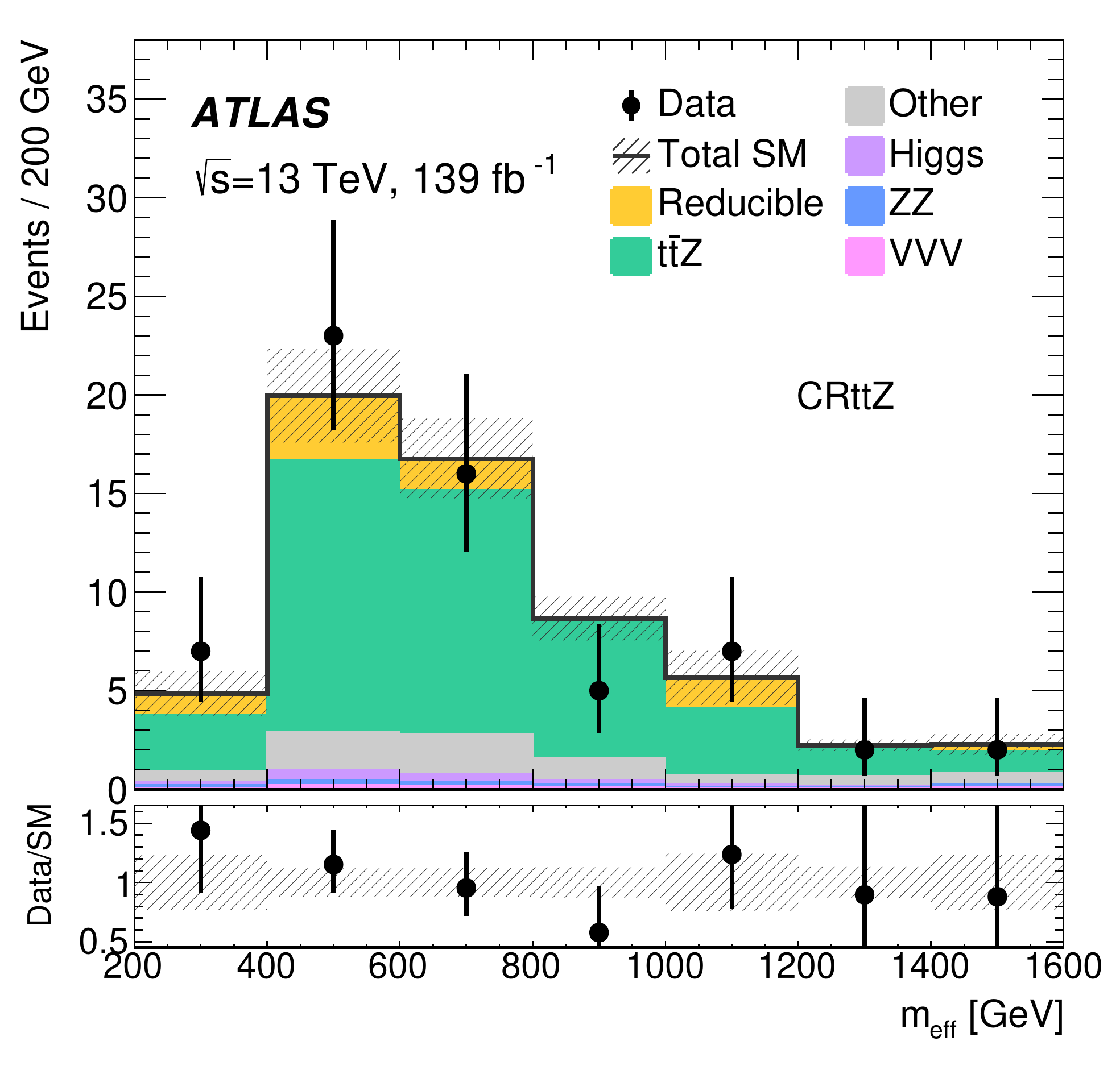}\label{fig:CRttZmeff}}\\
\caption{ The $\meff$ distributions for data and the SM backgrounds in \CRZZ\ and \CRttZ\ after the background-only fit.
``Other'' is the sum of the $tWZ$, $\ttbar WW$, $\ttbar ZZ$, $\ttbar WZ$, $\ttbar WH$, $\ttbar HH$, $\ttbar tW$, and $\ttbar \ttbar $ backgrounds.
The last bin includes the overflow.
The lower panel shows the ratio of the observed data to the expected SM background yield in each bin.
Both the statistical and systematic uncertainties in the SM background are included in the shaded band.
\label{fig:CRmeff}
}
\end{figure}

\subsection{Reducible background determination }\label{sec:bgdet_red}
The number of reducible background events in a given region is estimated from data with a hybrid fake-factor method that uses a combination of data and MC simulation.
Preselected leptons surviving overlap removal are classified as `signal' or `loose' depending on whether they pass or fail the signal lepton requirements, respectively.
A very loose selection on the identification RNN of $>0.05$ is also applied to the preselected $\tauhad$, as those with very low RNN scores are typically gluon-induced jets and jets arising from pile-up, which is not the case for the signal $\tauhad$ candidates.
Probabilities for a fake/non-prompt lepton to be identified as a signal or loose lepton are calculated from simulation and corrected to data where possible.
The ratio $F=f/{\bar{ f}}$ for fake/non-prompt leptons is then defined as the ``fake factor'', where $f$ ($\bar f$) is the probability that a fake/non-prompt lepton is identified as a signal (loose) lepton.

The reducible background prediction is extracted by applying fake factors to control regions in data.
The CR definition only differs from that of the associated SR in the quality of the required leptons; here exactly one (\texttt{CR1}) or two (\texttt{CR2}) of the four leptons must be identified as a loose lepton, as shown in Table~\ref{tab:CR_defs}.
In $3L1T$ and $5L0T$ events, the contribution from events with two fake/non-prompt light leptons is negligible, as is the contribution from one and two fake/non-prompt light leptons in $2L2T$ events.
 
\begin{table}[h]
\centering
\caption{Reducible background control region definitions where ``\texttt{L}'' and ``\texttt{T}'' denote signal light leptons and $\tauhad$, while ``\texttt{{\scriptsize l}}'' and ``\texttt{{\scriptsize t}}'' denote loose light leptons and $\tauhad$.
Loose leptons are preselected leptons surviving overlap removal that do not pass signal lepton requirements.
Additional selections for $b$-tagged jets,  $Z$ veto/requirement, $\met$, and $\meff$ are applied to match a given signal or validation region.
\label{tab:CR_defs}}
\vspace{0.2cm}
\small{
\begin{tabular}{ ll cccc }
\toprule
Reducible bkg.     & Control Region & $N(e,\mu)$ & $N(e,\mu)$ & $N(\tauhad)$ & $N(\tauhad)$ \\
estimation for &                & signal     &   loose    &    signal    &    loose     \\ \midrule
$4L0T$         & \texttt{CR1\_LLL{\scriptsize l}}      & $=3$       &  $\geq1$   &     $\geq0$     &   $\geq0$    \\
& \texttt{CR2\_LL{\scriptsize l}{\scriptsize l}}      & $=2$       &  $\geq2$   &     $\geq0$     &   $\geq0$    \\ \midrule
$3L1T$         & \texttt{CR1\_LLL{\scriptsize t}}      & $=3$       &    $=0$    &     $=0$     &   $\geq1$    \\
& \texttt{CR1\_LLT{\scriptsize l}}      & $=2$       &    $=1$    &   $\geq1$    &   $\geq0$    \\
& \texttt{CR2\_LL{\scriptsize l}{\scriptsize t}}      & $=2$       &    $=1$    &     $=0$     &   $\geq1$    \\ \midrule
$2L2T$         & \texttt{CR1\_LLT{\scriptsize t}}      & $=2$       &    $=0$    &     $=1$     &   $\geq1$    \\
& \texttt{CR2\_LL{\scriptsize t}{\scriptsize t}}      & $=2$       &    $=0$    &     $=0$     &   $\geq2$    \\ \midrule
$5L0T$           & \texttt{CR1\_LLLL{\scriptsize l}}     & $=4$       &  $\geq1$   &     $\geq0$     &   $\geq0$    \\ \bottomrule
\end{tabular}
}
\end{table}
 
The fake factors depend on the lepton flavour, the source of the fake/non-prompt lepton, and the production process.
Fake factors are calculated separately for each fake/non-prompt-lepton flavour ($e$, $\mu$, $\tauhad$) and source (light-flavour jets, heavy-flavour jets, gluon-initiated jets for $\tauhad$ only, and photon conversions for electrons and $\tauhad$ only), where these categories are referred to as fake/non-prompt-lepton `types'.
The fake factor per fake/non-prompt-lepton type for each production process ($\ttbar$ and $Z$+jets, or $ZZ$ for $5L$) is binned in lepton $\pt$, $\eta$, proximity to other leptons ($\Delta R$) for electrons and muons, and number of prongs for $\tauhad$.
The statistical uncertainties on the fake factors for the dominant types of fake/non-prompt-lepton are very small.
To correctly account for the relative abundances of fake/non-prompt-lepton types and production processes, a weighted average $F_{w}$ of fake factors is computed in each CR, as
 
\vspace{-0.4cm}
\begin{equation*}
\label{eq:fr}
F_{w}=\sum_{i,j} \left( F^{ij} \times R^{ij} \times s^{i} \right).
\end{equation*}
 
\vspace{-0.3cm}
\noindent
The term $F^{ij}$ is the corresponding fake factor for fake/non-prompt leptons of type $i$ from process $j$ calculated using MC simulation.
The fake factors are weighted by the `process fractions', $R^{ij}$, that describe the fraction of fake/non-prompt leptons of type $i$ from process $j$ in that region.
The process fractions are determined from MC simulation in the corresponding \texttt{CR2}, and are similar to the process fractions obtained in the signal regions from MC simulation, which suffer from having few events.
For the five lepton regions, no corresponding CR2 is used for the reducible estimation and the process fractions are determined from MC simulation in the corresponding CR1.
To account for possible differences between data and MC simulation, the fake factors obtained from simulation are corrected to data using `scale factors', $s^{i}$.
The scale factors are assumed to be independent of the physical process (e.g.\ $\ttbar$, $Z$+jets) and depend on the fake/non-prompt-lepton type only.
They are determined from data in regions enriched in objects of a given fake/non-prompt-lepton type, where MC simulation is used to remove any small contamination from leptons not from the fake/non-prompt-lepton type under study.
 
For fake/non-prompt leptons from heavy-flavour jets, the scale factor is measured in a $\ttbar$-dominated control sample.
The heavy-flavour scale factors are seen to have a modest $\pt$-dependence,
decreasing for electrons from $ 1.18 \pm 0.10$ to $1.08 \pm 0.08$ as the electron $\pt$ increases from $7\,\GeV$ to $20\,\GeV$.
For muons, the heavy-flavour scale factor is seen to be less dependent on $\pt$, changing from $ 1.00 \pm 0.04$ to $0.94 \pm 0.10$ as the muon $\pt$ increases from $5\,\GeV$ to $20\,\GeV$.
For 1-prong (3-prong) $\tauhad$, the heavy-flavour scale factor decreases from $1.26 \pm 0.07$ to $0.93 \pm 0.11$ ($ 1.15 \pm 0.06$ to $0.97 \pm 0.12$) as the $\tauhad$ $\pt$ increases from $20\,\GeV$ to $50\,\GeV$.
Uncertainties quoted for the scale factors include the statistical uncertainties of the data and MC simulation.
 
The scale factor for fake/non-prompt $\tauhad$ originating from light-flavour jets is measured separately for one- and three-prong $\tauhad$ in a control sample dominated by $Z$+jets events.
The scale factors are seen to be $\pt$-dependent,
decreasing from $ 1.115 \pm 0.009$ to $0.919 \pm 0.017$ ($1.340 \pm 0.023$ to $ 1.04 \pm 0.05$) as the 1-prong (3-prong) $\tauhad$ $\pt$ increases from $20\,\GeV$ to $50\,\GeV$.
The scale factor for $\tauhad$ originating from gluon-initiated jets is assumed to be the same as the one obtained from light-flavour jets.
The scale factor for fake/non-prompt electrons originating from light-flavour jets is measured in a $W$+jets-dominated control sample, where the light-flavour scale factor increases from $ 1.05 \pm 0.29$ to $1.38 \pm 0.09$ as the electron $\pt$ increases from $7\,\GeV$ to $20\,\GeV$.
The contribution to the signal regions from fake/non-prompt muons originating from light-flavour jets or leptons from photon conversions is very small and the scale factor cannot be measured reliably using data.
Therefore, values of $1.00 \pm 0.10$ are used instead, motivated by similar uncertainties in the other scale factor measurements.
 
The final estimate of the number $N^\textrm{SR}_\textrm{red}$ of background events with one or two fake/non-prompt leptons in each SR is determined from the number of events in data in the corresponding CRs, $N^\textrm{CR1}_\textrm{data}$ and $N^\textrm{CR2}_\textrm{data}$, according to
 
\vspace{-0.4cm}
\begin{equation}
\label{eq:fakes}
N^\textrm{SR}_\textrm{red}  = [N^\textrm{CR1}_\textrm{data}-N^\textrm{CR1}_\textrm{irr}]\times F_{w,1} - [N^\textrm{CR2}_\textrm{data}-N^\textrm{CR2}_\textrm{irr}]\times F_{w,1}\times F_{w,2},
\end{equation}
 
\vspace{-0.2cm}
\noindent where $F_{w,1}$ and $F_{w,2}$ are the two weighted fake factors constructed using the highest- and second-highest-$\pt$ loose leptons in the CRs, respectively.
Only the first term is used to estimate the reducible background for five lepton events.
The small contributions from irreducible background processes in the CRs, $N^\textrm{CR1,CR2}_\textrm{irr}$, are evaluated using MC simulation and subtracted from the corresponding number of events seen in data.
The second term removes the double-counting of events with two fake/non-prompt leptons in the first term.
Processes with one fake/non-prompt-lepton are included in the reducible background estimation, but have very low contributions to CR1 and CR2 because of their small cross-sections e.g.\ $WZ$ and $\ttbar W$.
Both CR1 and CR2 are dominated by the two-fake/non-prompt-lepton processes $\ttbar$ and $Z$+jets, so the first term is roughly twice the second term.
Higher-order terms in $F_w$ describing three- and four-fake/non-prompt-lepton backgrounds are neglected, as are some terms with a very small contribution; e.g.\ in $3L1T$ events, the contribution from events with two fake/non-prompt light leptons is negligible.
A systematic uncertainty is applied to account for these neglected terms, as described in the following section.

\section{Systematic uncertainties \label{sec:systematics} }
 
The uncertainties affecting the SM and signal simulation-based estimates can be divided into three components: statistical uncertainty of the MC simulation, experimental uncertainty in event reconstruction ($e$, $\mu$, $\tauhad$ and jets, $\met$), and theoretical uncertainty.
The reducible background estimation is affected by different sources of uncertainty associated with data counts in control regions and uncertainties in the weighted fake factors.
The leading SM background in the signal regions targeting the higgsino GGM models (\texttt{SR0-ZZ}) is generally $ZZ$ production, and the jet experimental and theoretical uncertainties are seen to dominate the total uncertainty.
The exception is \SRLLLLF, where the reducible background and its associated uncertainties are the leading contribution to the total SM background and its uncertainty.
The reducible background is also the leading component of the SM background in most of the high-$\meff$ signal regions targeting the RPV models, as well as the five-lepton signal region, and the uncertainty in the reducible background dominates the total uncertainty.
The exceptions to this are \SRLLLLA\ and \SRLLLLG, where the irreducible processes and theory uncertainties also contribute significantly to the total SM background and uncertainty.
The primary sources of systematic uncertainty, described below, are summarised in Figure~\ref{fig:systUncert}.

\begin{figure}[h]
\centering
\includegraphics[width=0.95\textwidth]{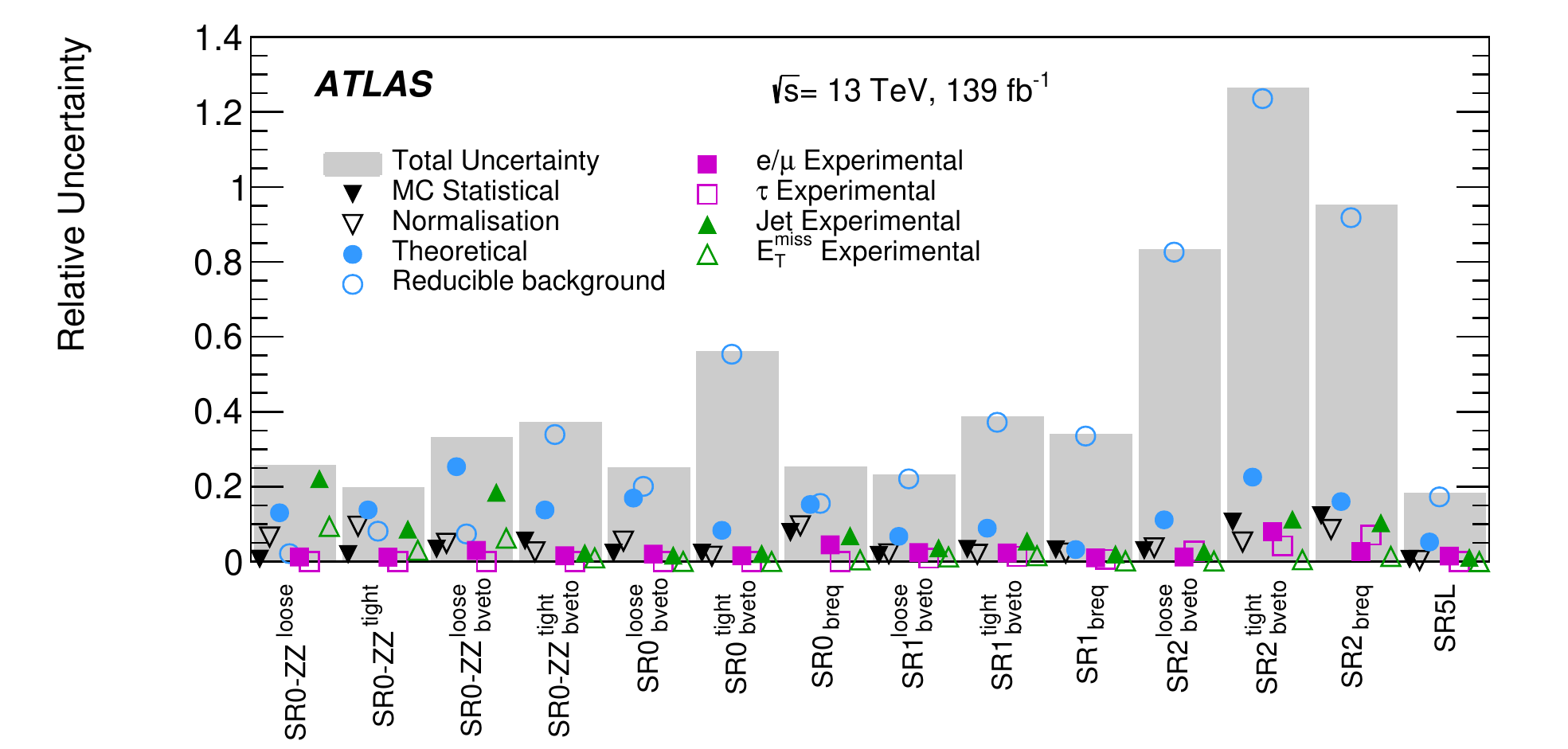}
\caption{Breakdown of the dominant systematic uncertainties in the background estimates for the signal regions. Theoretical uncertainties in the simulation-based estimates, including those in the acceptance for events in $ZZ$ and $\ttbar Z$ simulations, are grouped under the ``Theoretical'' category, while the ``Normalisation'' category includes the statistical uncertainties of the data counts in \protect\CRZZ\ and \protect\CRttZ\ and the uncertainty from the fitted normalisation factors. For simplicity, the individual uncertainties are added in quadrature for each category, without accounting for correlations. The total background uncertainty is taken from the background-only fit where correlations of individual uncertainties are accounted for.
\label{fig:systUncert}
}
\end{figure}

The statistical uncertainty of the MC simulation-based background estimate is small and typically less than $5\%$ of the total background estimate in the signal regions. However, this rises to $10$--$15\%$ in \SRLLLLG, \SRLLTTB, and \SRLLTTC\ where tight selections on $\meff$ are made.
The experimental uncertainties include
the uncertainties associated with electrons, muons, $\tauhad$, jets, and $\met$,
as well as the uncertainty associated with the simulation of pile-up,
and uncertainty in the luminosity.
The uncertainty in the combined 2015--2018 integrated luminosity is 1.7\%~\cite{ATLAS-CONF-2019-021}, obtained using the LUCID-2 detector~\cite{LUCID2} for the primary luminosity measurements.
The uncertainties associated with pile-up and luminosity are included in the total uncertainty in Figure~\ref{fig:systUncert}.
The experimental uncertainties pertaining to electrons, muons and $\tauhad$ include the uncertainties due to the
lepton identification efficiencies, lepton energy scale and energy resolution, and isolation and trigger efficiencies.
Systematic uncertainties from electron, muon, and $\tauhad$ sources are generally small in all signal regions, at a few percent relative to the total expected background.
The uncertainties associated with jets are due to the jet energy scale, jet energy resolution, jet vertex tagging, and $b$-tagging.
The jet energy resolution is the dominant source of uncertainty for jets in the signal regions.
Uncertainties in the object momenta are propagated to the $\met$ measurement, and additional uncertainties in $\met$
arising from energy deposits not associated with any reconstructed objects are also considered.
The jet and $\met$ uncertainties are generally of the order of a few percent in the signal regions,
but approach $10$--$20\%$ in \SRLLLLC, \SRLLLLD\ and \SRLLLLE, where selections on $\met$ are made.

Theoretical uncertainties in the simulation-based estimates include
the theoretical cross-section uncertainties due to the choice of renormalisation and factorisation scales and PDFs,
the acceptance uncertainty due to PDF and scale variations, and the choice of MC generator.
The theoretical cross-section uncertainties for the irreducible backgrounds used in this analysis are
20\% for $VVV$~\cite{ATL-PHYS-PUB-2016-002},
10\% for $\ttbar H$~\cite{Dittmaier:2012vm},
5\% for $H$ via ggF~\cite{Dittmaier:2012vm},
and 20\% for $\ttbar WZ/WW/WH/HH$, $tWZ$, and Higgs via VBF/$VH$,
where the order of each cross-section calculation is shown in Table~\ref{tab:SMbkgMC}.
 
Uncertainties in the acceptance are evaluated for the irreducible processes $ZZ$, $\ttbar Z$, $VVV$, $\ttbar H$ and Higgs, where for each background the uncertainties are correlated across all regions.
The uncertainty in the acceptance from varying the strong coupling constant \alphas within its uncertainty is included for the $ZZ$, $VVV$, Higgs, and $\ttbar H$ backgrounds.
Uncertainties arising from varying the renormalisation and factorisation scales (\muR and \muF) by factors of $2$ and $1/2$ are considered -- the \muR and \muF variations are taken as uncorrelated and the additional coherent up/down variation of \muR and \muF is also considered.
Resummation and merging-scale variations are also considered for the $ZZ$ sample.
The uncertainty related to the choice of CKKW merging scale ($20\,\GeV$) was evaluated by considering alternative values of $15\,\GeV$ and $30\,\GeV$.
The uncertainty related to the resummation scale \muQ ($2\,\GeV$, the upper cut-off in perturbative calculations of parton shower evolution), was evaluated by varying the scale by factors of $1/4$ and $4$.
Finally, for $\SHERPA$ $ZZ$ samples the impact of using an alternative recoil scheme for single-particle emission in the $\SHERPA$ parton shower is studied by comparing the nominal samples with alternative ones in which the CSS scheme is used (CSSKIN)~\cite{Schumann:2007mg}.
The theoretical uncertainties in the $ZZ$ acceptance dominate the total uncertainties in signal regions where the irreducible backgrounds are a large fraction of the total SM background: \SRLLLLD, \SRLLLLE, \SRLLLLA, and \SRLLLLG.

The uncertainty in the reducible background is dominated by the statistical uncertainty of the data events in the corresponding CR1 and CR2, which is the leading source of uncertainty in many of the signal regions where the reducible background is a large fraction of the total SM background, as seen in Figure~\ref{fig:systUncert}.
The uncertainty in the weighted fake factors includes the statistical uncertainty of the MC simulation in the process fractions, the uncertainty in the fake/non-prompt-lepton scale factors, and the statistical uncertainty from the fake factors determined from simulation.
The uncertainties in the fake factors from each fake/non-prompt-lepton type are treated as correlated across processes.
Thus, since both CR1 and CR2 are dominated by two-fake/non-prompt-lepton processes with the same source of fake/non-prompt lepton, correlations in the fake factors applied to CR1 and CR2 cause the uncertainties from the weighted fake factors to largely cancel out between the first and second terms in Eq.~\eqref{eq:fakes}.
Finally, a conservative uncertainty is applied to account for the neglected terms in Eq.~\eqref{eq:fakes}.
For example, for $4L0T$ events the three- and four-fake/non-prompt-lepton terms are neglected.
Weighted fake factors are applied to data events with one signal light lepton and three loose light leptons to estimate an upper limit on this neglected contribution for each $4L0T$ region. The calculated upper limit plus its $1\sigma$ statistical uncertainty is then added to the reducible background uncertainty, adding an absolute uncertainty of $0.17$ events in \SRLLLLA.
This is repeated for the $3L1T$ and $2L2T$ regions, accounting for the neglected terms with one or two fake/non-prompt light leptons as necessary, adding an absolute uncertainty of $0.59$ events in \SRLLLTA, and $2.5$ events in \SRLLTTA.
Finally, for \SRLLLLL, the neglected terms with two or three fake/non-prompt light leptons are evaluated in the same way, adding an absolute uncertainty of $0.5$ events.

Systematic uncertainties in the SUSY signal yields from experimental sources are typically of the order of 10\%, while typical uncertainties in the signal acceptance and distribution shapes due to scale and parton shower variations are found to range from $\sim5\%$ in $4L0T$ regions to $\sim20\%$ in $2L2T$ regions.
The systematic uncertainty in the signal cross-section is described in Section~\ref{sec:mcsim}.

\section{Background modelling validation}\label{sec:bgdet_valid}
 
To validate the modelling of the SM backgrounds, the yields and shapes of key kinematic variable distributions are compared with data in validation regions (VR).
The VRs are defined to be close to, yet disjoint from the SRs and CRs, and be dominated by the background process under consideration, as shown in Table~\ref{tab:VR_defs}.
The $ZZ^{(*)}$ modelling is validated in a region (\VRZZ) defined with $4L0T$, no $b$-tagged jets, and only one $Z$ boson candidate, while the $\ttbar Z$ modelling is validated in a region (\VRttZ) with $4L0T$, a $Z$ veto, at least one $b$-tagged jet, and high $\meff$.
For signal regions that veto $Z$ boson candidates, three general VRs are defined by reversing the $\meff$ requirement: \VRLLLL, \VRLLLT, and \VRLLTT.
Two further VRs with $3L1T$ and $2L2T$ are defined with one $Z$ boson requirement to check the reducible background modelling across the full range of $\meff$: \VRLLLTZ\ and \VRLLTTZ.
The background model adopted in the VRs is the same as in the SRs, with the $ZZ$ and $\ttbar$ backgrounds obtained from MC simulation normalised to data, the other irreducible processes from MC simulation, and the reducible background estimated from data using the fake-factor method with process fractions and loose lepton control regions corresponding to the VRs.
The systematic uncertainties of the SM backgrounds in the VRs are evaluated as in Section~\ref{sec:systematics}.
 
\begin{table}[h]
\centering
\caption{Validation region definitions.  The $Z$ boson column refers to the $Z$ veto or selection of a first/second $Z$ candidate as described in the text.  \label{tab:VR_defs}}
\vspace{0.2cm}
\footnotesize{
\resizebox{\linewidth}{!}{
\begin{tabular}{ ll ccc c cc }
\toprule
Name & Validation & $N(e,\mu)$ & $N(\tauhad)$ & $N(b{\textrm{-jets}})$ & $Z$ boson               & Selection         &         Target         \\
& Region     &            &              &                      &                         &                   &  \\ \midrule
$4L0T$ & \VRZZ       & $\geq4$    &     $=0$     &  $=0$                & require 1st \& veto 2nd & --                & $ZZ^{(*)}$          \\
& \VRttZ      & $\geq4$    &     $=0$     &  $\geq1$             & veto                    & $400<\meff<1300\,\GeV$ & $\ttbar Z^{(*)}$          \\
& \VRLLLL        & $\geq4$    &     $=0$     &  $=0$                & veto                    & $\meff<600\,\GeV$ & $\ttbar$, $Z$+jets, $Z^{(*)}Z^{(*)}$ \\
\midrule
$3L1T$ & \VRLLLT        & $=3$       &   $\geq1$    &  $=0$                & veto                    & $\meff<600\,\GeV$ & $\ttbar$, $Z$+jets    \\
& \VRLLLTZ       & $=3$       &   $\geq1$    &  $=0$                & require 1st             & --                & $Z$+jets    \\
\midrule
$2L2T$ & \VRLLTT        & $=2$       &   $\geq2$    &  $=0$                & veto                    & $\meff<600\,\GeV$ & $\ttbar$, $Z$+jets    \\
& \VRLLTTZ       & $=2$       &   $\geq2$    &  $=0$                & require 1st             & --                & $Z$+jets    \\
\bottomrule
\end{tabular}
}}
\end{table}

Observed and expected event yields in the VRs are shown in Table~\ref{tab:VRyields}, where good agreement is seen in general within statistical and systematic uncertainties.
No significant deviations from SM expectations are observed in any VR.
\begin{table}[h]
\renewcommand\arraystretch{1.5}
\newcolumntype{R}{>{$}r<{$}}
\newcolumntype{L}{>{$}l<{$}}
\newcolumntype{M}{R@{${}\pm{}$}L}
\centering
\caption{Expected and observed yields for $139\,\ifb$ in the validation regions after the background-only fit. ``Other'' is the sum of the $tWZ$, $\ttbar WW$, $\ttbar ZZ$, $\ttbar WZ$, $\ttbar WH$, $\ttbar HH$, $\ttbar tW$, and $\ttbar \ttbar $ backgrounds.
Both the statistical and systematic uncertainties in the SM background are included in the uncertainties shown.
\label{tab:VRyields}}
\vspace{0.2cm}
\small{
\resizebox{\linewidth}{!}{
\begin{tabular}{ l MM MM MM MM MM MM MM}
\toprule
& \multicolumn{2}{c}{\VRZZ} & \multicolumn{2}{c}{\VRttZ} & \multicolumn{2}{c}{\VRLLLL} & \multicolumn{2}{c}{\VRLLLT} & \multicolumn{2}{c}{\VRLLLTZ} & \multicolumn{2}{c}{\VRLLTT} & \multicolumn{2}{c}{\VRLLTTZ} \\ \midrule
Observed  & \multicolumn{2}{c}{$874$} & \multicolumn{2}{c}{$42$} & \multicolumn{2}{c}{$216$} & \multicolumn{2}{c}{$192$} & \multicolumn{2}{c}{$620$} & \multicolumn{2}{c}{$156$} & \multicolumn{2}{c}{$505$} \\ \midrule
Total SM & \multicolumn{2}{c}{$929_{-34}^{+43}$} & $40$ & $5$ & $230$ & $12$ & $199$ & $19$ & \multicolumn{2}{c}{$624_{-47}^{+46}$} & $164$ & $30$ & $479$ & $20$ \\ \midrule
$ZZ$ & $737$ & $20$ & \multicolumn{2}{c}{$2.7_{-1.1}^{+1.0}$~} & $160$ & $7$ & $36.8$ & $2.3$ & \multicolumn{2}{c}{$200_{-11}^{+9}$} & \multicolumn{2}{c}{$23.3_{-1.3}^{+1.2}$~~} & $119$ & $7$ \\
$\ttbar Z$ & \multicolumn{2}{c}{~$5.2_{-1.4}^{+1.3}$} & $16$ & $4$ & $2.0$ & $0.5$ & $0.9$ & $0.4$ & \multicolumn{2}{c}{$2.3_{-1.1}^{+1.2}$} & $0.36$ & $0.1$ & $0.14$ & $0.04$ \\
Higgs & \multicolumn{2}{c}{~$77_{-26}^{+36}$} & $5.5$ & $0.8$ & \multicolumn{2}{c}{$3.1_{-0.7}^{+1.0}$} & $2.4$ & $0.5$ & $13.6$ & $3.0$ & $1.87$ & $0.31$ & $8.8$ & $1.8$ \\
$VVV$ & $26$ & $5$ & \multicolumn{2}{c}{$0.43_{-0.11}^{+0.10}$~~~} & $5.4$ & $1.1$ & $2.8$ & $0.8$ & \multicolumn{2}{c}{$8.7_{-2.2}^{+2.1}$} & $1.23$ & $0.26$ & $1.52$ & $0.33$ \\
Other & \multicolumn{2}{c}{~$2.2_{-0.5}^{+0.4}$} & $3.1$ & $0.4$ & \multicolumn{2}{c}{$0.45_{-0.11}^{+0.10}$} & \multicolumn{2}{c}{$0.24_{-0.20}^{+0.19}$} & $0.9$ & $0.5$ & $0.15$ & $0.04$ & \multicolumn{2}{c}{$0.082_{-0.028}^{+0.025}$~~} \\
Reducible & \multicolumn{2}{c}{~$81_{-11}^{+10}$} & \multicolumn{2}{c}{$12.1_{-3.6}^{+3.5}$~~~~} & $59$ & $9$ & $156$ & $19$ & $397$ & $45$ & $137$ & $30$ & $350$ & $19$ \\
\bottomrule
\end{tabular}
}}
\end{table}
 
The $\met$, $\meff$, and lepton $\pt$ distributions in the VRs are shown in Figure~\ref{fig:VRMeff} and Figure~\ref{fig:VR012LepPtAll}.
The $\meff$ distributions in \VRLLLL, \VRLLLT, \VRLLTT, and \VRttZ\ can be seen in the lower $\meff$ bins in Figure~\ref{fig:SRLLLLABmeff}, Figure~\ref{fig:SRLLLTABmeff}, Figure~\ref{fig:SRLLTTABmeff}, and Figure~\ref{fig:SRLLLLGmeff}, respectively.
 
\begin{figure}[h!]
\centering
\subfloat[][$\met$ in \VRZZ]{\includegraphics[width=0.475\textwidth]{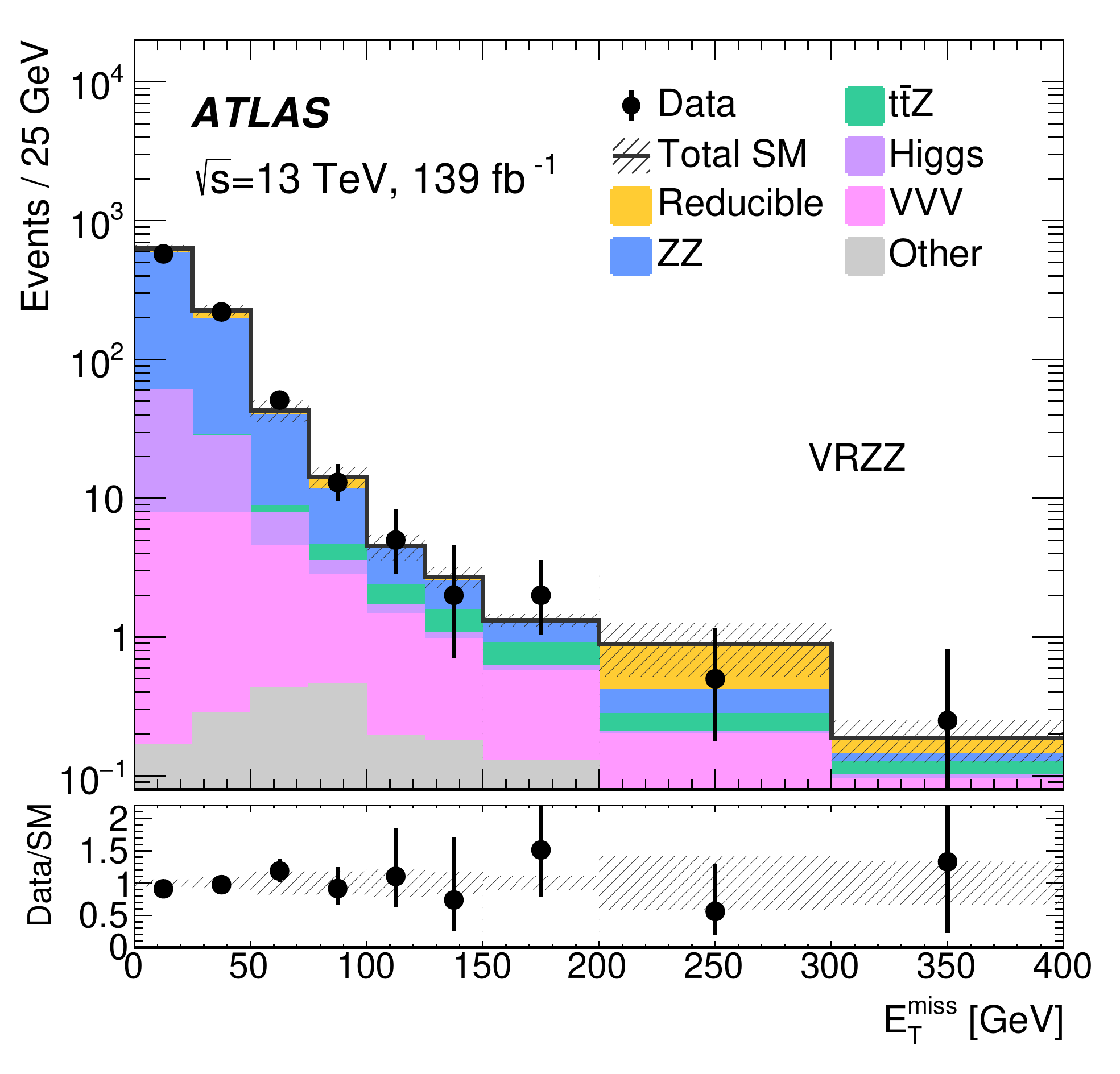}\label{fig:VRZZMet}}
\subfloat[][$\met$ in \VRttZ]{\includegraphics[width=0.475\textwidth]{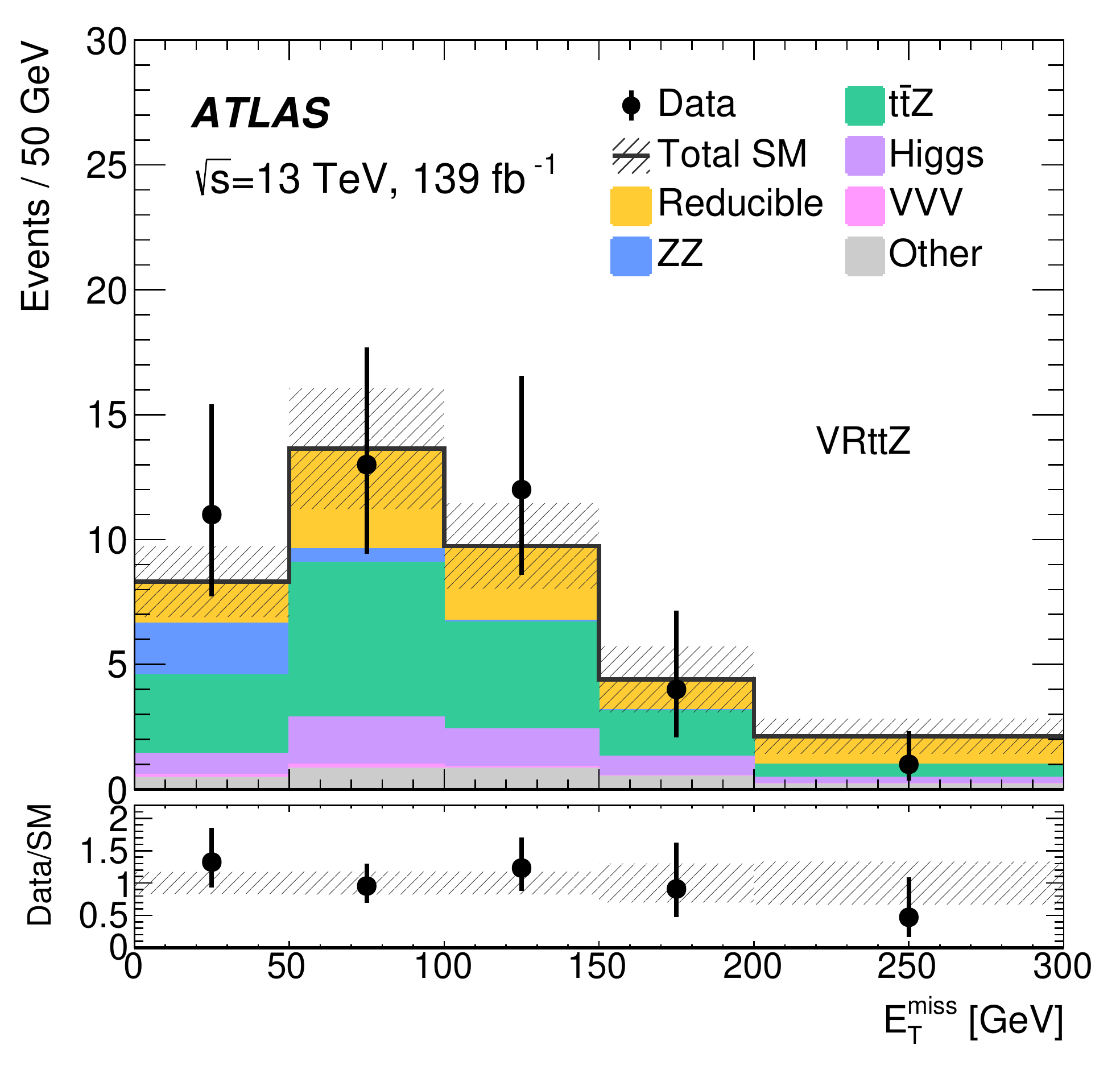}\label{fig:VRttZMet}}\\
\subfloat[][$\meff$ in \VRLLLTZ]{\includegraphics[width=0.475\textwidth]{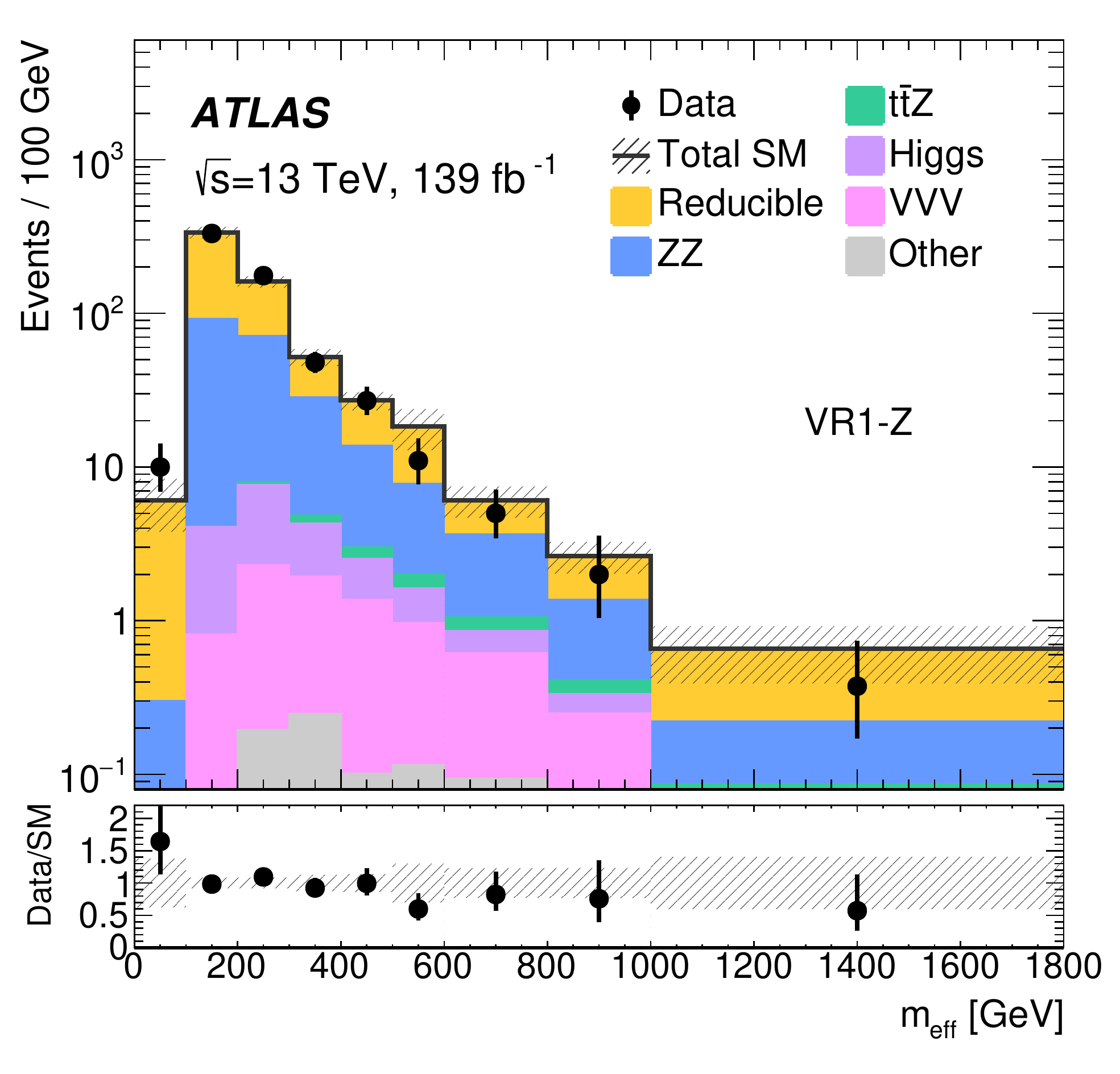}\label{fig:VR1ZMeff}}
\subfloat[][$\meff$ in \VRLLTTZ]{\includegraphics[width=0.475\textwidth]{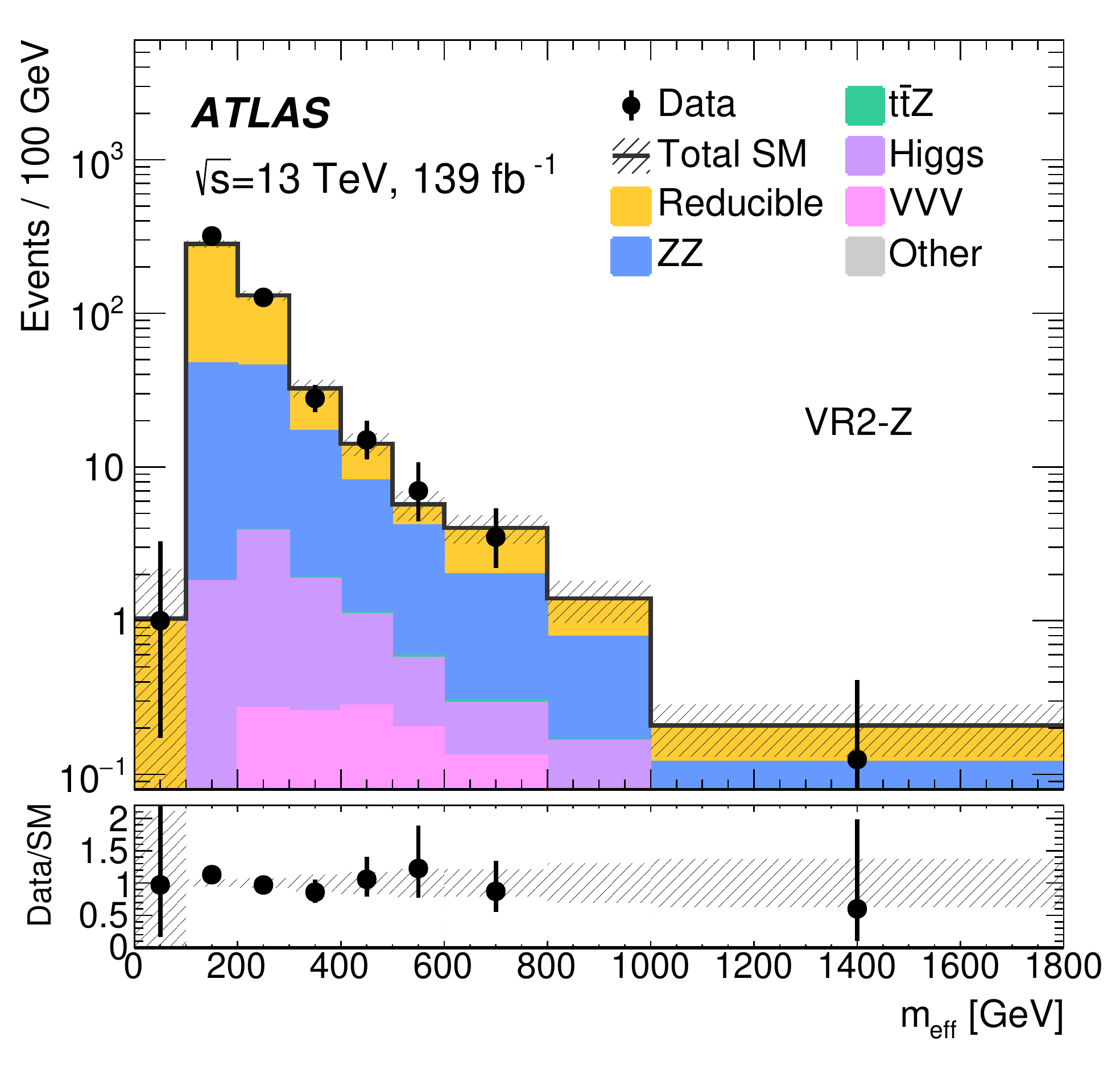}\label{fig:VR2ZMeff}}\\
\caption{ The distributions for data and the estimated SM backgrounds in the validation regions after the background-only fit, for the $\met$ in (a) \VRZZ\ and (b) in \VRttZ, and the $\meff$ in (c) \VRLLLTZ\ and in (d) \VRLLTTZ.
``Other'' is the sum of the $tWZ$, $\ttbar WW$, $\ttbar ZZ$, $\ttbar WZ$, $\ttbar WH$, $\ttbar HH$, $\ttbar tW$, and $\ttbar \ttbar $ backgrounds.
The last bin includes the overflow.
The lower panel shows the ratio of the observed data to the expected SM background yield in each bin.
Both the statistical and systematic uncertainties in the SM background are included in the shaded band.
\label{fig:VRMeff}
}
\end{figure}

\begin{figure}[h!]
\centering
\subfloat[][$\pt\left({e,\mu}\right)$ in \VRLLLL]{\includegraphics[width=0.475\textwidth]{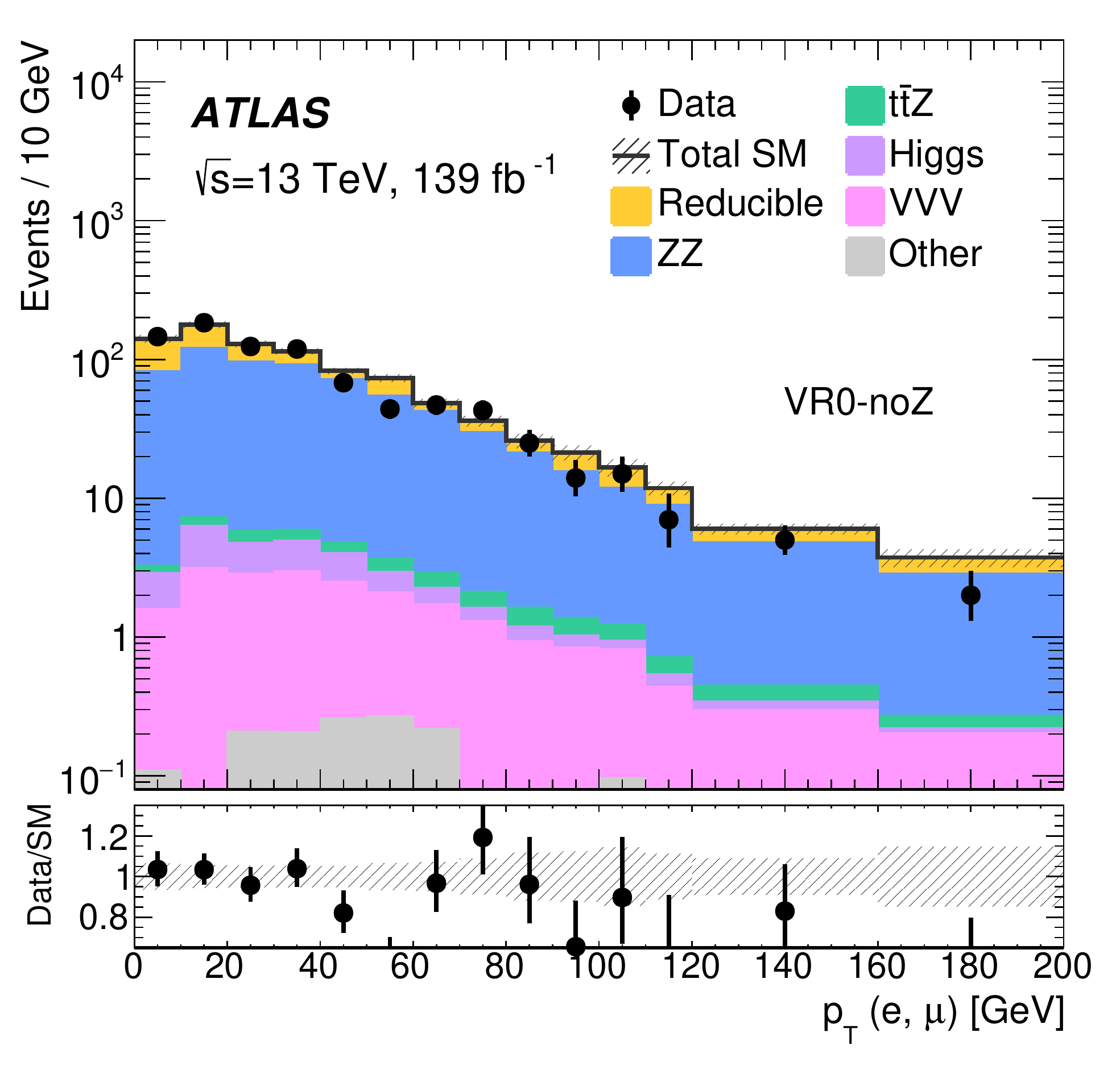}\label{fig:VR0LepPt}}
\subfloat[][$\pt\left({e,\mu}\right)$ in \VRLLLT]{\includegraphics[width=0.475\textwidth]{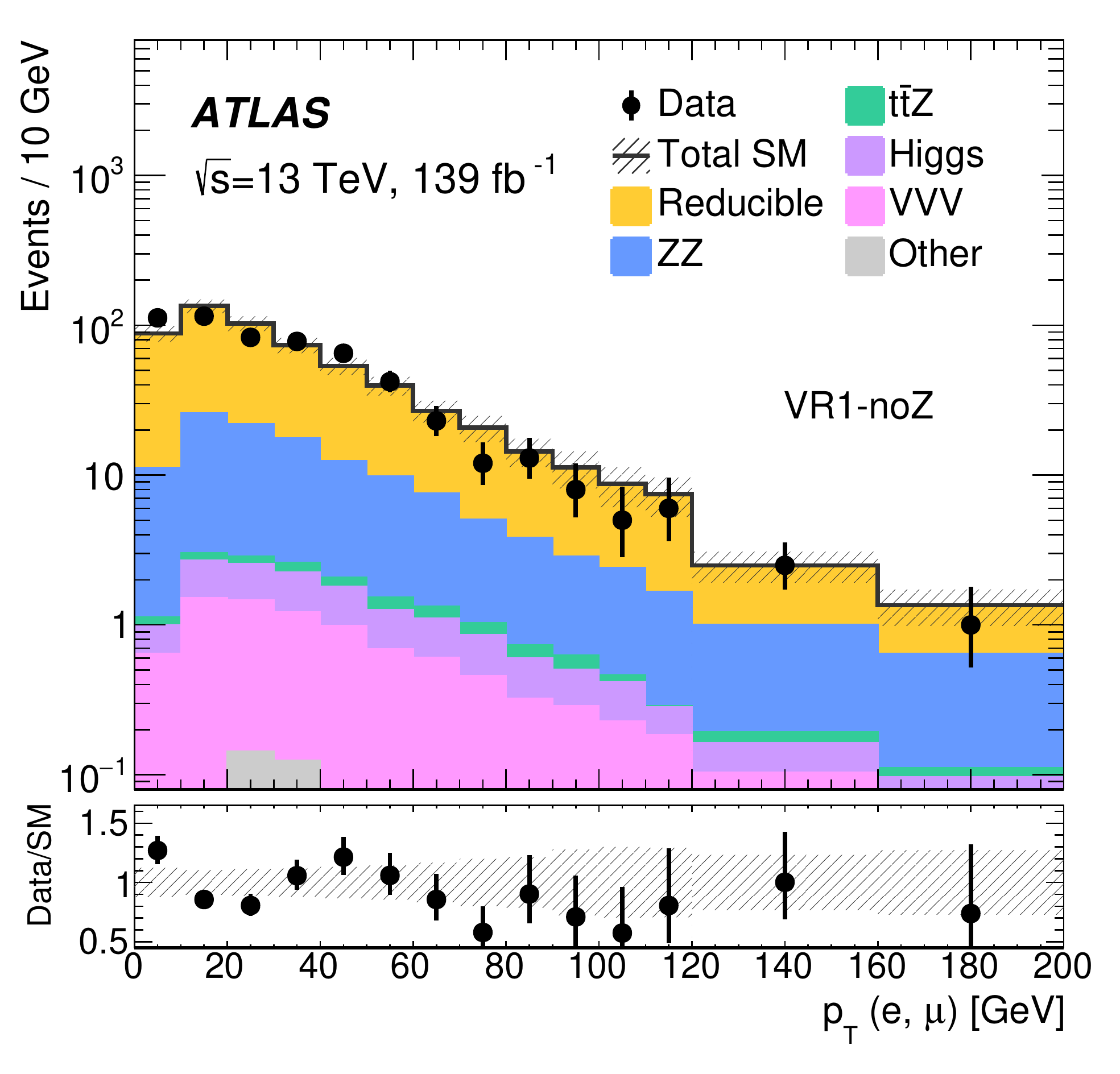}\label{fig:VR1LepPt}}\\
\subfloat[][$\pt\left({\tauhad}\right)$ in \VRLLLT]{\includegraphics[width=0.475\textwidth]{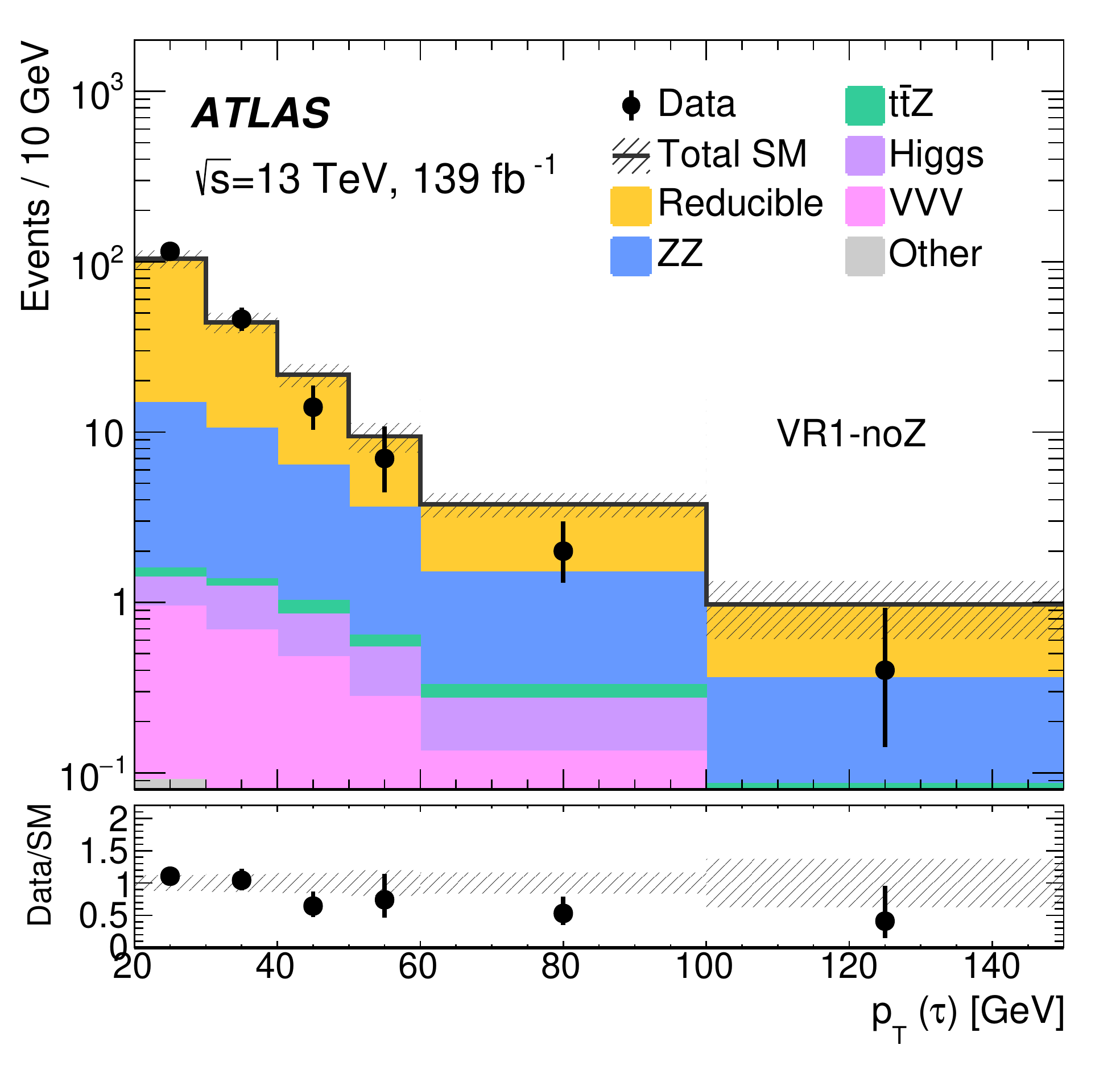}\label{fig:VR1TauPt}}
\subfloat[][$\pt\left({\tauhad}\right)$ in \VRLLTT]{\includegraphics[width=0.475\textwidth]{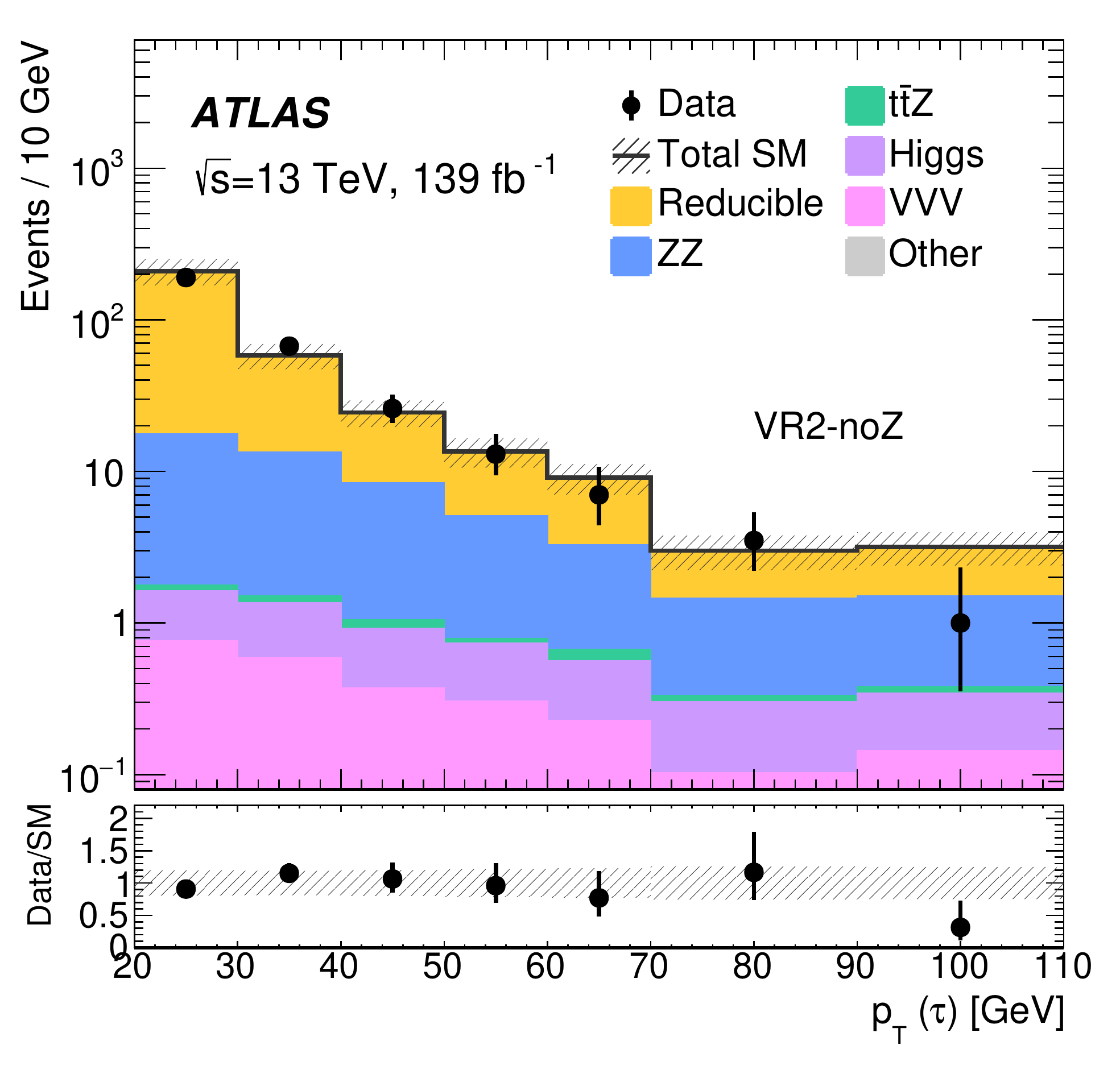}\label{fig:VR2TauPt}}\\
\caption{ The distributions for data and the estimated SM backgrounds in the validation regions after the background-only fit, for the light lepton $\pt$ in (a) \VRLLLL\ and (b) \VRLLLT, and the $\tauhad$ $\pt$ in (c) \VRLLLT\ and (d) \VRLLTT.
``Other'' is the sum of the $tWZ$, $\ttbar WW$, $\ttbar ZZ$, $\ttbar WZ$, $\ttbar WH$, $\ttbar HH$, $\ttbar tW$, and $\ttbar \ttbar $ backgrounds.
The last bin includes the overflow.
The lower panel shows the ratio of the observed data to the expected SM background yield in each bin.
Both the statistical and systematic uncertainties in the SM background are included in the shaded band.
\label{fig:VR012LepPtAll}
}
\end{figure}

\section{Results}\label{sec:results}
 
The observed number of events in each signal region is reported in Table~\ref{tab:SRyields} and Figure~\ref{fig:signifplot}, along with the background expectations and uncertainties.
The $\met$ and $\meff$ distributions for all events passing signal region requirements, except the $\met$ or $\meff$ requirement itself, are shown in Figures~\ref{fig:SRmet}--\ref{fig:SRmeffB}.
 
\begin{table}[h]
\renewcommand\arraystretch{1.5}
\newcolumntype{R}{>{$}r<{$}}
\newcolumntype{L}{>{$}l<{$}}
\newcolumntype{M}{R@{${}\pm{}$}L}
\centering
\caption{Expected and observed yields for $139~\ifb$ in the signal regions after the background-only fit.
``Other'' is the sum of the $tWZ$, $\ttbar WW$, $\ttbar ZZ$, $\ttbar WZ$, $\ttbar WH$, $\ttbar HH$, $\ttbar tW$, and $\ttbar \ttbar $ backgrounds.
Both the statistical and systematic uncertainties in the SM background are included in the uncertainties shown.
\label{tab:SRyields}}
\scriptsize{
\resizebox{\linewidth}{!}{
\begin{tabular}{ l MM MM MM MM MM MM MM}
\toprule
& \multicolumn{2}{c}{\SRLLLLC} & \multicolumn{2}{c}{\SRLLLLD} & \multicolumn{2}{c}{\SRLLLLE} & \multicolumn{2}{c}{\SRLLLLF} & \multicolumn{2}{c}{\SRLLLLA} & \multicolumn{2}{c}{\SRLLLLB} & \multicolumn{2}{c}{\SRLLLLG} \\ \midrule
 
Observed & \multicolumn{2}{c}{$157$} & \multicolumn{2}{c}{$17$} & \multicolumn{2}{c}{$5$} & \multicolumn{2}{c}{$1$} & \multicolumn{2}{c}{$11$} & \multicolumn{2}{c}{$1$} & \multicolumn{2}{c}{$3$} \\ \midrule
Total SM  & \multicolumn{2}{c}{$161_{-43}^{+41}$} & \multicolumn{2}{c}{$18.4_{-3.3}^{+3.6}$~~~} & \multicolumn{2}{c}{$7.3_{-1.9}^{+2.4}$} & $1.1$ & $0.4$ & \multicolumn{2}{c}{$11.5_{-2.2}^{+2.9}$} & \multicolumn{2}{c}{$3.5_{-2.2}^{+2.0}$} & \multicolumn{2}{c}{$1.19_{-0.28}^{+0.30}$} \\ \midrule
$ZZ$ & \multicolumn{2}{c}{$125_{-42}^{+40}$} & \multicolumn{2}{c}{$4.5_{-2.1}^{+2.6}$} & \multicolumn{2}{c}{$3.7_{-1.7}^{+2.2}$} & \multicolumn{2}{c}{$0.05_{-0.04}^{+0.11}$} & \multicolumn{2}{c}{$7.6_{-1.7}^{+1.6}$} & \multicolumn{2}{c}{$0.64_{-0.29}^{+0.28}$} & \multicolumn{2}{c}{$0.19_{-0.19}^{+0.15}$} \\
$\ttbar Z$  & $15$ & $4$ & $7.4$ & $1.8$ & $0.87$ & $0.24$ & \multicolumn{2}{c}{$0.12_{-0.04}^{+0.05}$} & \multicolumn{2}{c}{$0.7_{-0.19}^{+0.18}$} & \multicolumn{2}{c}{$0.02_{-0.015}^{+0.014}$} & $0.49$ & $0.13$ \\
Higgs  & $0.79$ & $0.1$ & $0.29$ & $0.05$ & \multicolumn{2}{c}{$0.09_{-0.027}^{+0.028}$} & \multicolumn{2}{c}{$0.0046_{-0.0018}^{+0.0019}$~~} & $0.24$ & $0.04$ & \multicolumn{2}{c}{$0.02_{-0.006}^{+0.007}$} & \multicolumn{2}{c}{$0.16_{-0.06}^{+0.05}$} \\
$VVV$  & \multicolumn{2}{c}{$7.9_{-2.0}^{+1.9}$} & $2.4$ & $0.6$ & $2.2$ & $0.5$ & $0.44$ & $0.12$ & $1.6$ & $0.4$ & $0.21$ & $0.06$ & \multicolumn{2}{c}{$0.083_{-0.029}^{+0.027}$} \\
Other  & $3.3$ & $0.7$ & $1.7$ & $0.4$ & $0.32$ & $0.09$ & \multicolumn{2}{c}{$0.04_{-0.014}^{+0.013}$} & \multicolumn{2}{c}{$0.142_{-0.032}^{+0.029}$} & \multicolumn{2}{c}{$0.032_{-0.022}^{+0.019}$~~~} & \multicolumn{2}{c}{$0.27_{-0.05}^{+0.06}$} \\
Reducible  & \multicolumn{2}{c}{$9.1_{-4.4}^{+3.4}$} & \multicolumn{2}{c}{$2.0_{-1.7}^{+1.5}$} & \multicolumn{2}{c}{$0.15_{-0.15}^{+0.54}$} & $0.4$ & $0.4$ & \multicolumn{2}{c}{$1.2_{-1.2}^{+2.3}$} & \multicolumn{2}{c}{$2.6_{-2.2}^{+1.9}$~} & \multicolumn{2}{c}{$0.00_{-0.00}^{+0.19}$} \\
 
\bottomrule
\toprule
 
& \multicolumn{2}{c}{\SRLLLTA} & \multicolumn{2}{c}{\SRLLLTB} & \multicolumn{2}{c}{\SRLLLTC} & \multicolumn{2}{c}{\SRLLTTA} & \multicolumn{2}{c}{\SRLLTTB} & \multicolumn{2}{c}{\SRLLTTC} & \multicolumn{2}{c}{\SRLLLLL} \\ \midrule
Observed & \multicolumn{2}{c}{$7$} & \multicolumn{2}{c}{$2$} & \multicolumn{2}{c}{$2$} & \multicolumn{2}{c}{$5$} & \multicolumn{2}{c}{$2$} & \multicolumn{2}{c}{$1$} & \multicolumn{2}{c}{$21$} \\ \midrule
Total SM  & \multicolumn{2}{c}{$7.7_{-1.9}^{+1.8}$} & \multicolumn{2}{c}{$1.6_{-0.7}^{+0.6}$} & $2.2$ & $0.7$ & \multicolumn{2}{c}{$3.4_{-1.6}^{+2.8}$} & \multicolumn{2}{c}{$0.35_{-0.13}^{+0.44}$} & \multicolumn{2}{c}{$0.52_{-0.13}^{+0.50}$~} & $12.4$ & $2.3$ \\ \midrule
$ZZ$ & \multicolumn{2}{c}{$2.0_{-0.6}^{+0.4}$} & \multicolumn{2}{c}{$0.39_{-0.19}^{+0.13}$} & $0.04$ & $0.04$ & \multicolumn{2}{c}{$1.54_{-0.4}^{+0.3}$~~~} & \multicolumn{2}{c}{$0.23_{-0.13}^{+0.08}$} & $0.06$ & $0.06$ & $0.49$ & $0.04$ \\
$\ttbar Z$ & $0.19$ & $0.1$ & \multicolumn{2}{c}{$0.029_{-0.029}^{+0.047}$} & $0.22$ & $0.06$ & \multicolumn{2}{c}{$0.058_{-0.025}^{+0.024}$~~} & $0.0$ & $0.0$ & $0.19$ & $0.07$ & $0.034$ & $0.009$ \\
Higgs  & $0.24$ & $0.07$ & \multicolumn{2}{c}{$0.033_{-0.020}^{+0.019}$} & \multicolumn{2}{c}{$0.14_{-0.06}^{+0.05}$} & $0.2$ & $0.04$ & \multicolumn{2}{c}{$0.033_{-0.011}^{+0.010}$} & $0.2$ & $0.07$ & $0.96$ & $0.14$ \\
$VVV$  & \multicolumn{2}{c}{$0.66_{-0.17}^{+0.16}$} & \multicolumn{2}{c}{$0.16_{-0.05}^{+0.04}$} & $0.021$ & $0.008$ & \multicolumn{2}{c}{$0.38_{-0.10}^{+0.09}$} & \multicolumn{2}{c}{$0.084_{-0.027}^{+0.025}$} & \multicolumn{2}{c}{$0.024_{-0.010}^{+0.009}$~~} & \multicolumn{2}{c}{$3.0_{-0.7}^{+0.6}$} \\
Other  & $0.009$ & $0.005$ & \multicolumn{2}{c}{$0.02_{-0.018}^{+0.013}$} & \multicolumn{2}{c}{$0.183_{-0.034}^{+0.039}$} & $0.005$ & $0.005$ & $0.0014$ & $0.0012$ & \multicolumn{2}{c}{$0.054_{-0.015}^{+0.019}$~~} & $0.22$ & $0.19$ \\
Reducible & $4.7$ & $1.7$ & $1.0$ & $0.6$ & $1.6$ & $0.7$ & \multicolumn{2}{c}{$1.2_{-1.2}^{+2.8}$} & \multicolumn{2}{c}{$0.0_{-0.0}^{+0.4}$} & \multicolumn{2}{c}{$0.0_{-0.0}^{+0.5}$~~} & $7.7$ & $2.1$ \\
 
\bottomrule
\end{tabular}
}}
\end{table}

\begin{figure}[h]
\centering
\includegraphics[width=0.8\textwidth]{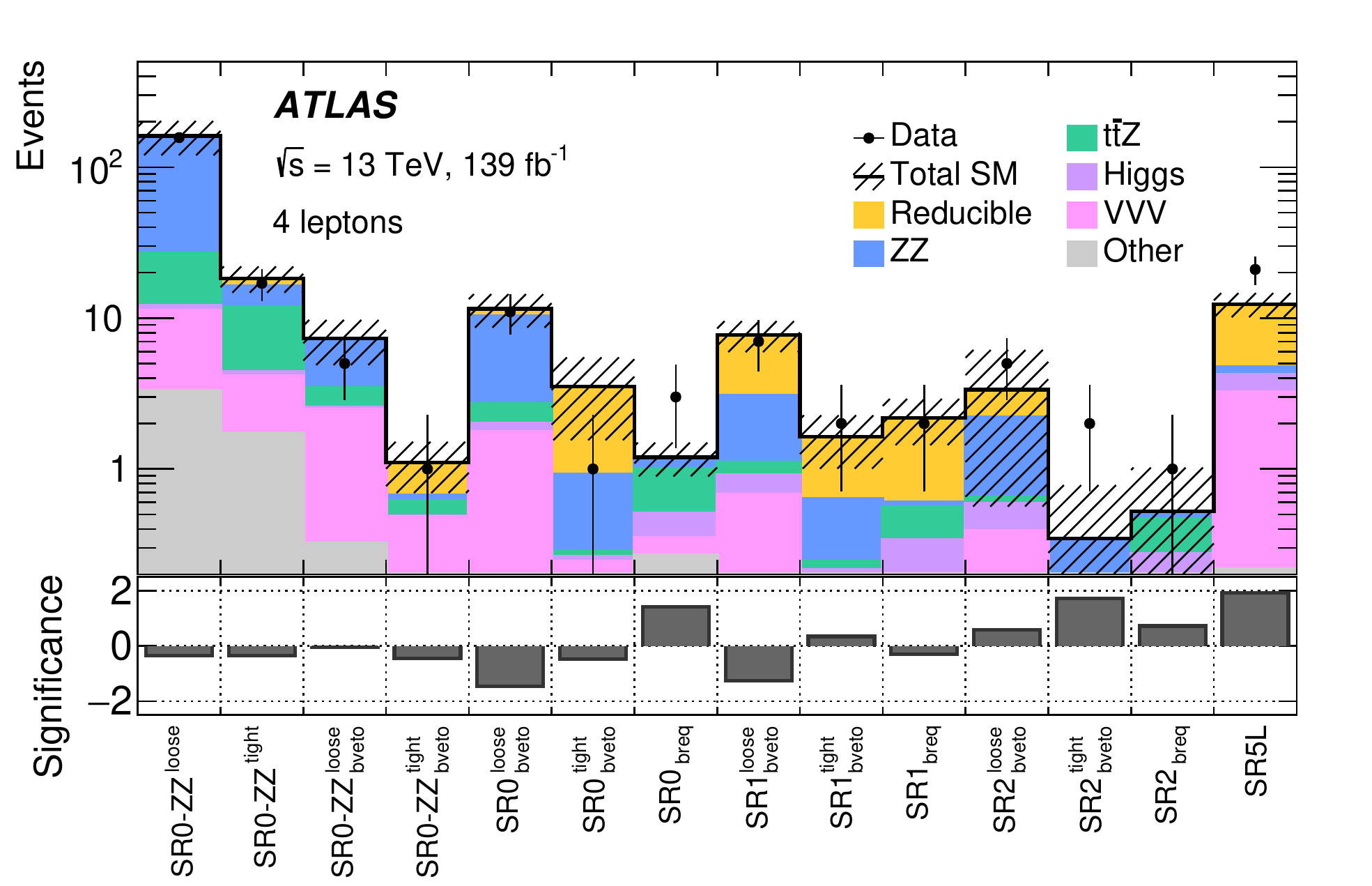}
\caption{
Expected and observed yields in the signal regions after the background-only fit.
``Other'' is the sum of the $tWZ$, $\ttbar WW$, $\ttbar ZZ$, $\ttbar WZ$, $\ttbar WH$, $\ttbar HH$, $\ttbar tW$, and $\ttbar \ttbar $ backgrounds.
Both the statistical and systematic uncertainties in the SM background are included in the uncertainties shown.
The significance of any difference between observed and expected yields is shown in the bottom panel, calculated with the profile likelihood method from Ref.~\protect\cite{Cousins:2007bmb}.
\label{fig:signifplot} }
\end{figure}
 
\begin{figure}[h]
\centering
\subfloat[][$\met$ in \SRLLLLC\ and \SRLLLLD]{\includegraphics[width=0.475\textwidth]{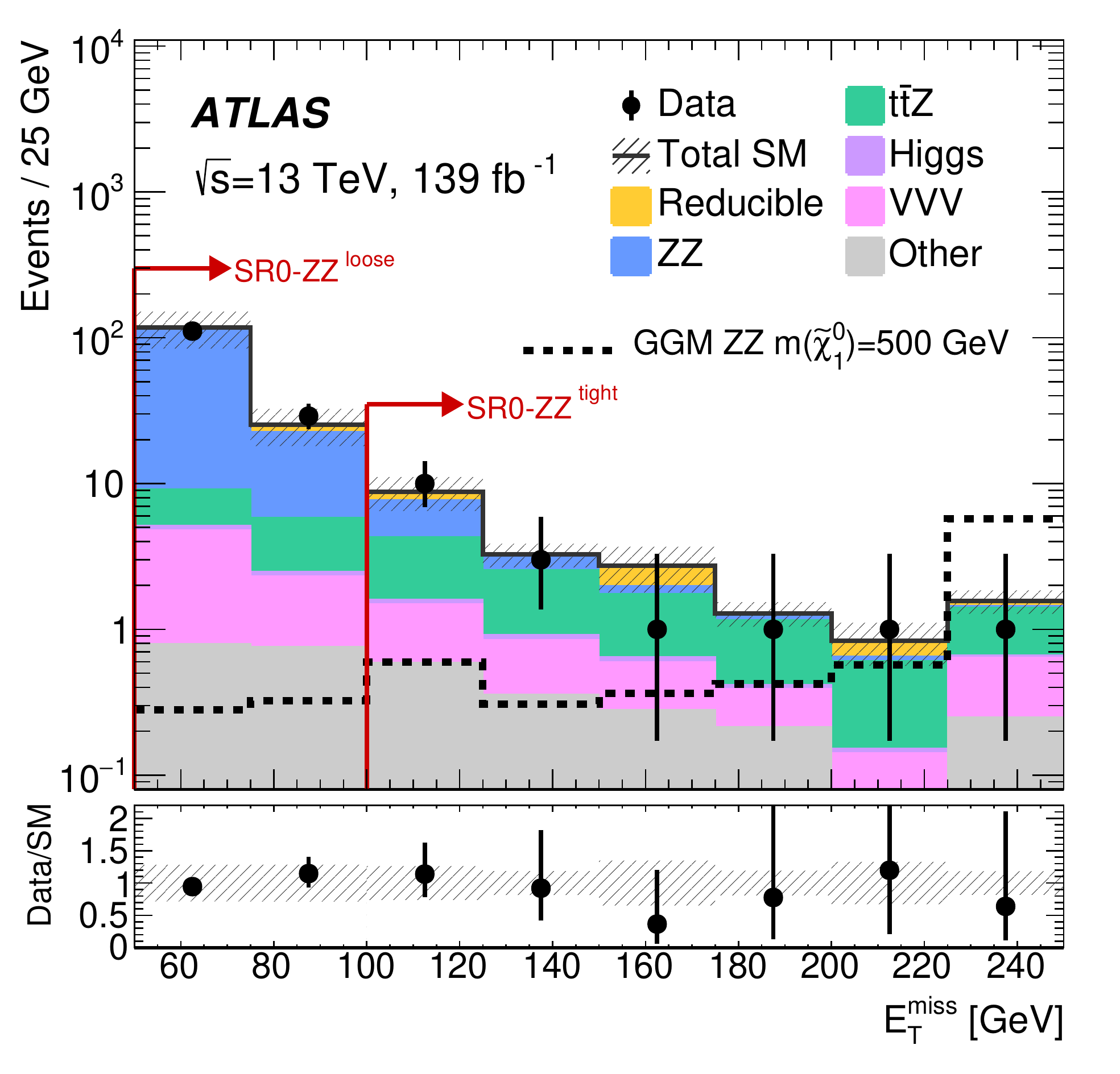}\label{fig:SRLLLLCDmet}}
\subfloat[][$\met$ in \SRLLLLE\ and \SRLLLLF]{\includegraphics[width=0.475\textwidth]{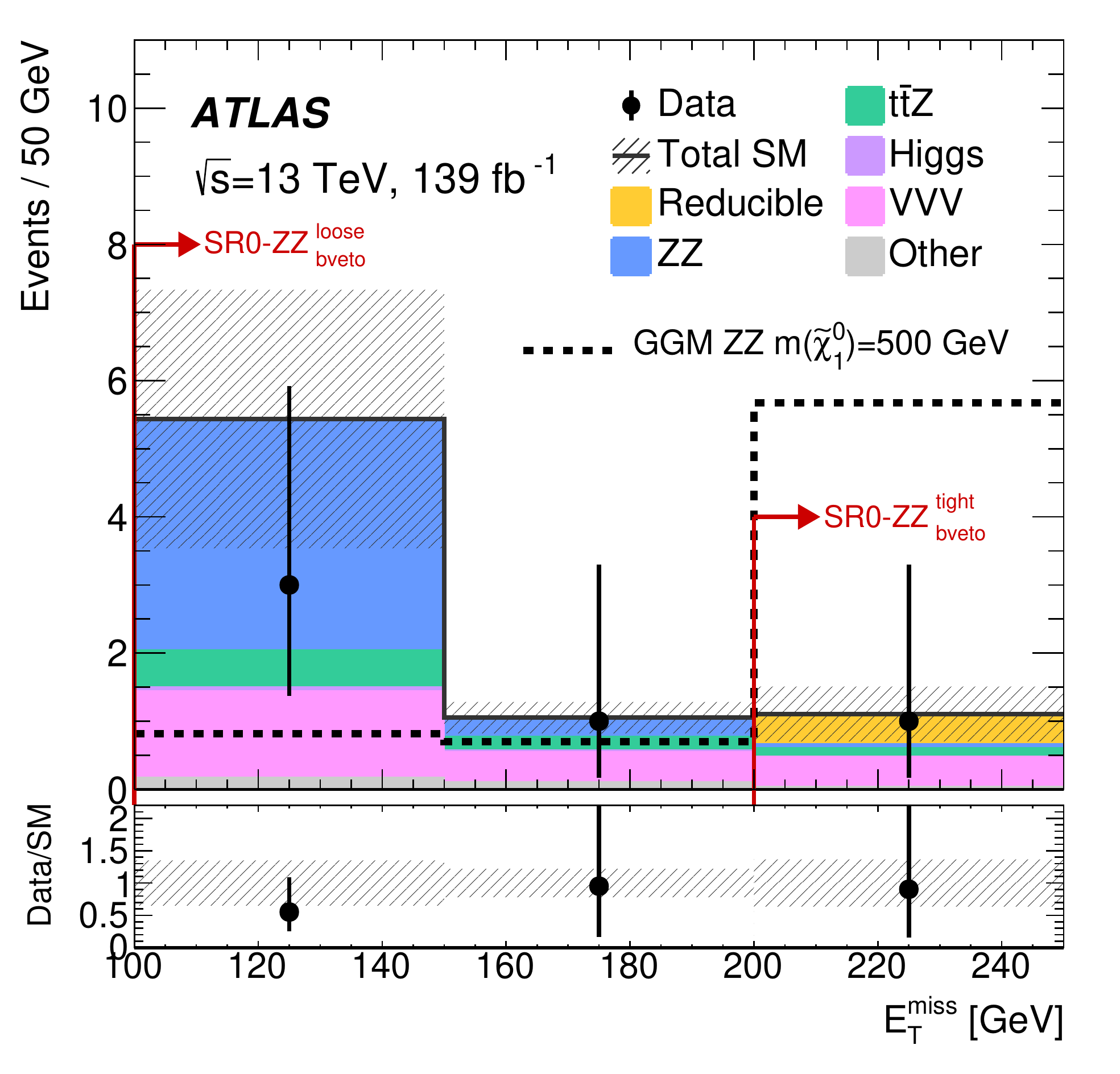}\label{fig:SRLLLLEFmet}}\\
\subfloat[][$\met$ in \SRLLLLL]{\includegraphics[width=0.475\textwidth]{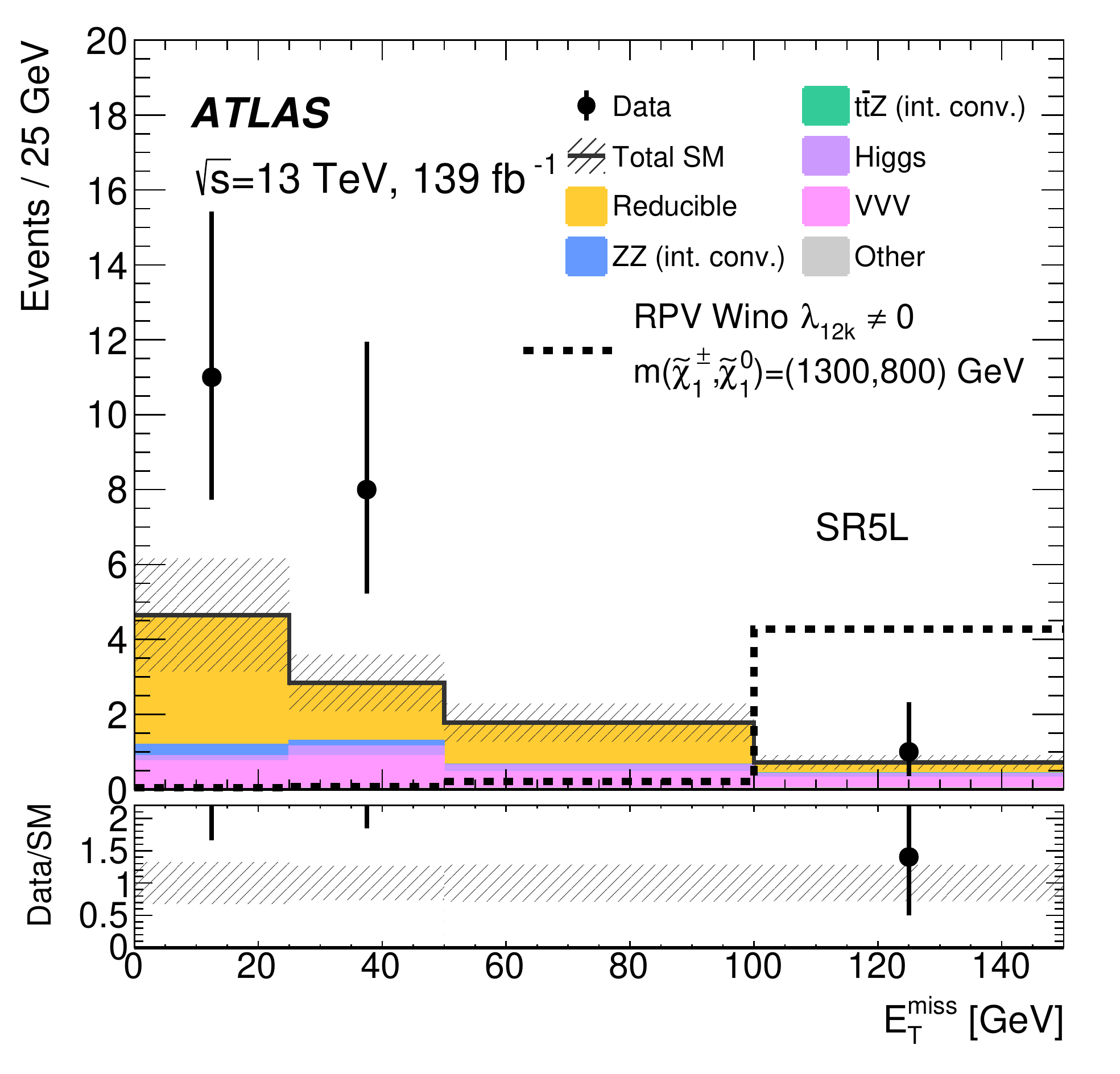}\label{fig:SRLLLLLmeff}}\\
\caption{
The $\met$ distribution in (a) \protect\SRLLLLC\ and \protect\SRLLLLD, (b) \protect\SRLLLLE\ and \protect\SRLLLLF, and (c) \protect\SRLLLLL\ for events passing the signal region requirements except the $\met$ requirement.
Distributions for data, the estimated SM backgrounds after the background-only fit, and an example SUSY scenario are shown.
``Other'' is the sum of the $tWZ$, $\ttbar WW$, $\ttbar ZZ$, $\ttbar WZ$, $\ttbar WH$, $\ttbar HH$, $\ttbar tW$, and $\ttbar \ttbar $ backgrounds.
The last bin captures the overflow events.
The lower panel shows the ratio of the observed data to the expected SM background yield in each bin.
Both the statistical and systematic uncertainties in the SM background are included in the shaded band.
The red arrows indicate the $\met$ selections in the signal regions.
\label{fig:SRmet} }
\end{figure}
 
\begin{figure}[h]
\centering
\subfloat[][$\meff$ in \SRLLLLA\ and \SRLLLLB]{\includegraphics[width=0.475\textwidth]{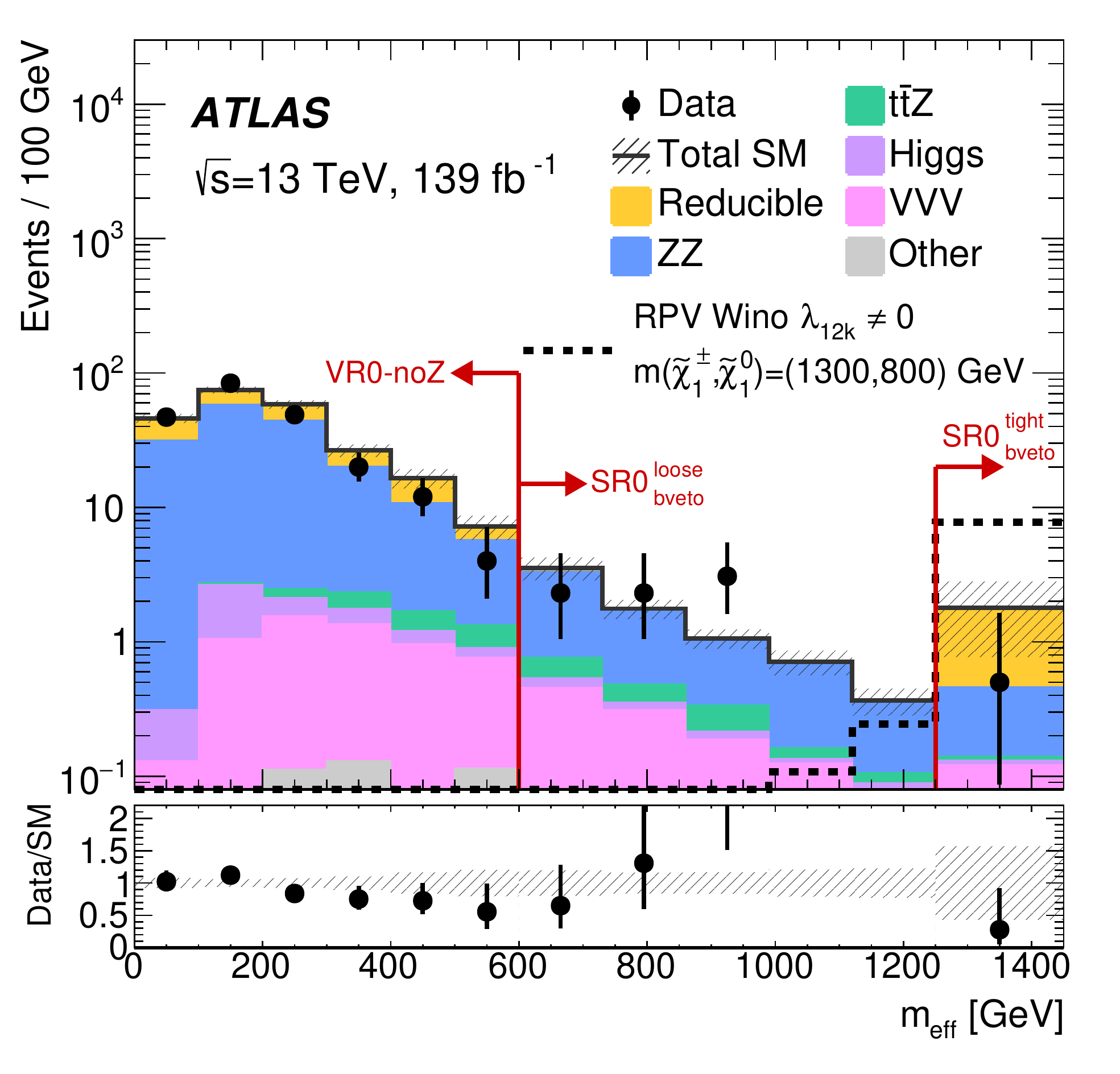}\label{fig:SRLLLLABmeff}}
\subfloat[][$\meff$ in \SRLLLTA\ and \SRLLLTB]{\includegraphics[width=0.475\textwidth]{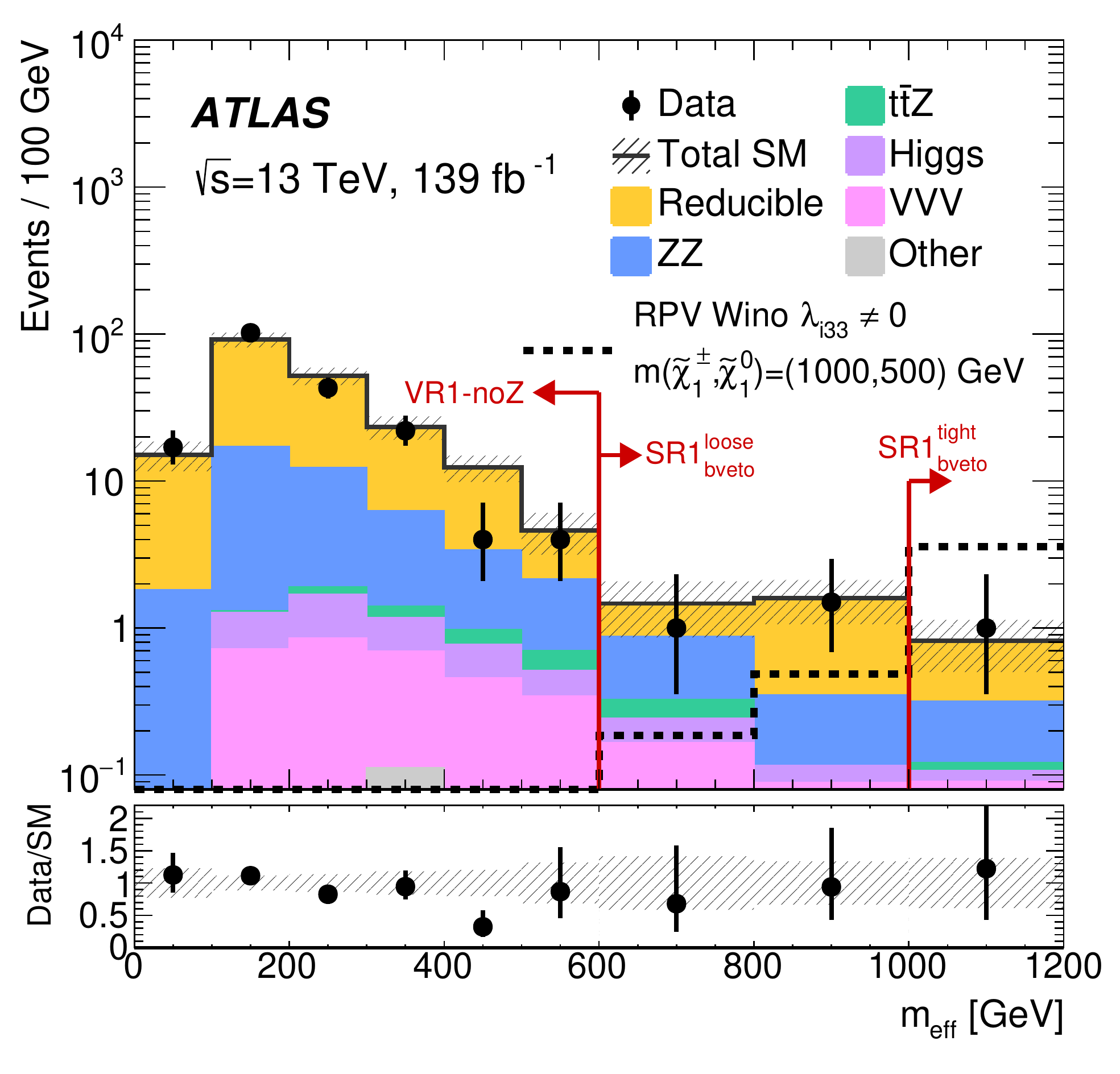}\label{fig:SRLLLTABmeff}}\\
\subfloat[][$\meff$ in \SRLLTTA\ and \SRLLTTB]{\includegraphics[width=0.475\textwidth]{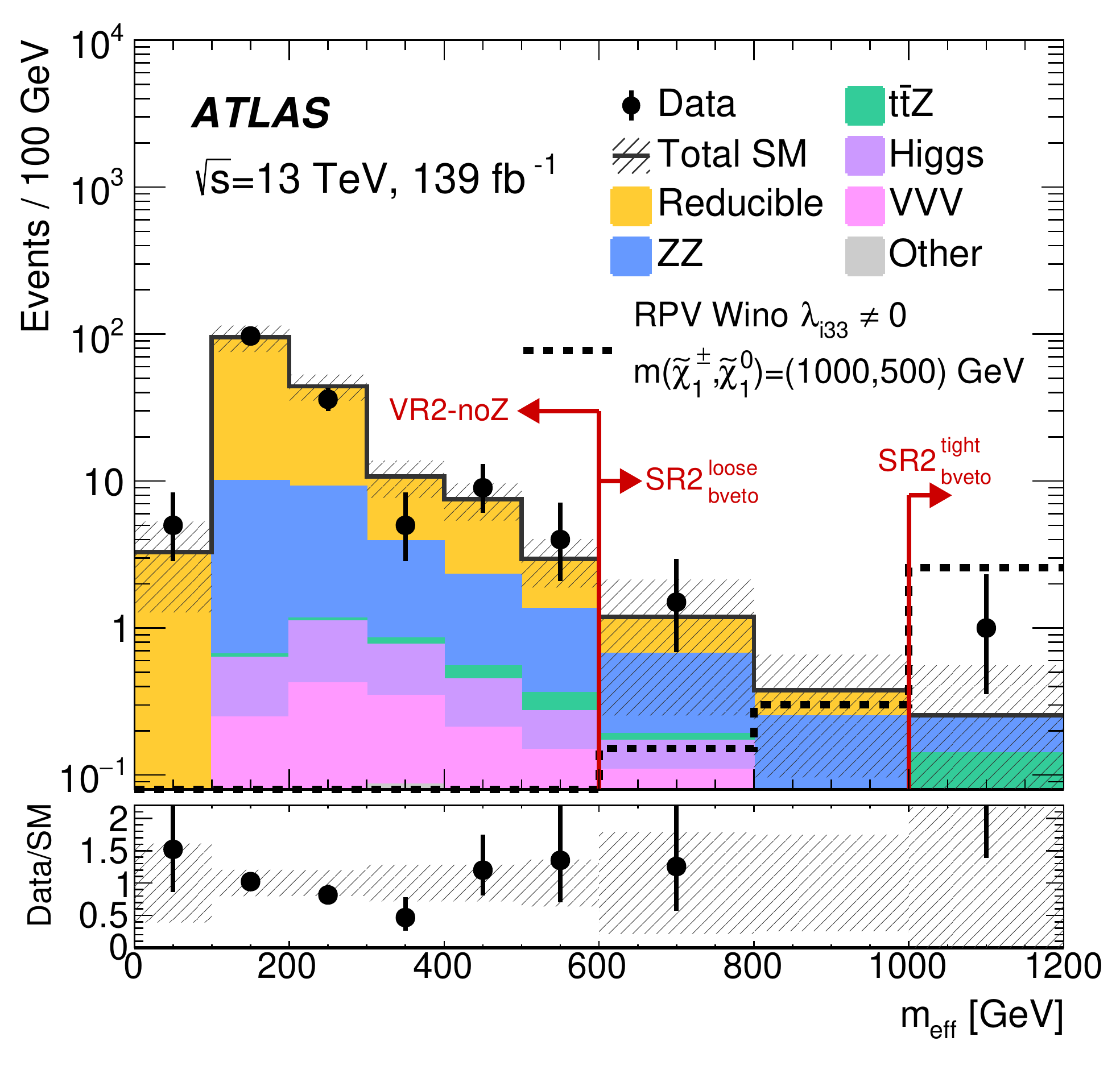}\label{fig:SRLLTTABmeff}}\\
\caption{
The $\meff$ distribution in (a) \protect\SRLLLLA\ and \protect\SRLLLLB, (b) \protect\SRLLLTA\ and \protect\SRLLLTB, and (c) \protect\SRLLTTA\ and \protect\SRLLTTB\ for events passing the signal region requirements except the $\meff$ requirement.
Distributions for data, the estimated SM backgrounds after the background-only fit, and an example SUSY scenario are shown. ``Other'' is the sum of the $tWZ$, $\ttbar WW$, $\ttbar ZZ$, $\ttbar WZ$, $\ttbar WH$, $\ttbar HH$, $\ttbar tW$, and $\ttbar \ttbar $ backgrounds.
The last bin captures the overflow events.
The lower panel shows the ratio of the observed data to the expected SM background yield in each bin.
Both the statistical and systematic uncertainties in the SM background are included in the shaded band.
The red arrows indicate the $\meff$ selections in the signal and validation regions.
\label{fig:SRmeff} }
\end{figure}
 
\begin{figure}[h]
\centering
\subfloat[][$\meff$ in \SRLLLLG]{\includegraphics[width=0.475\textwidth]{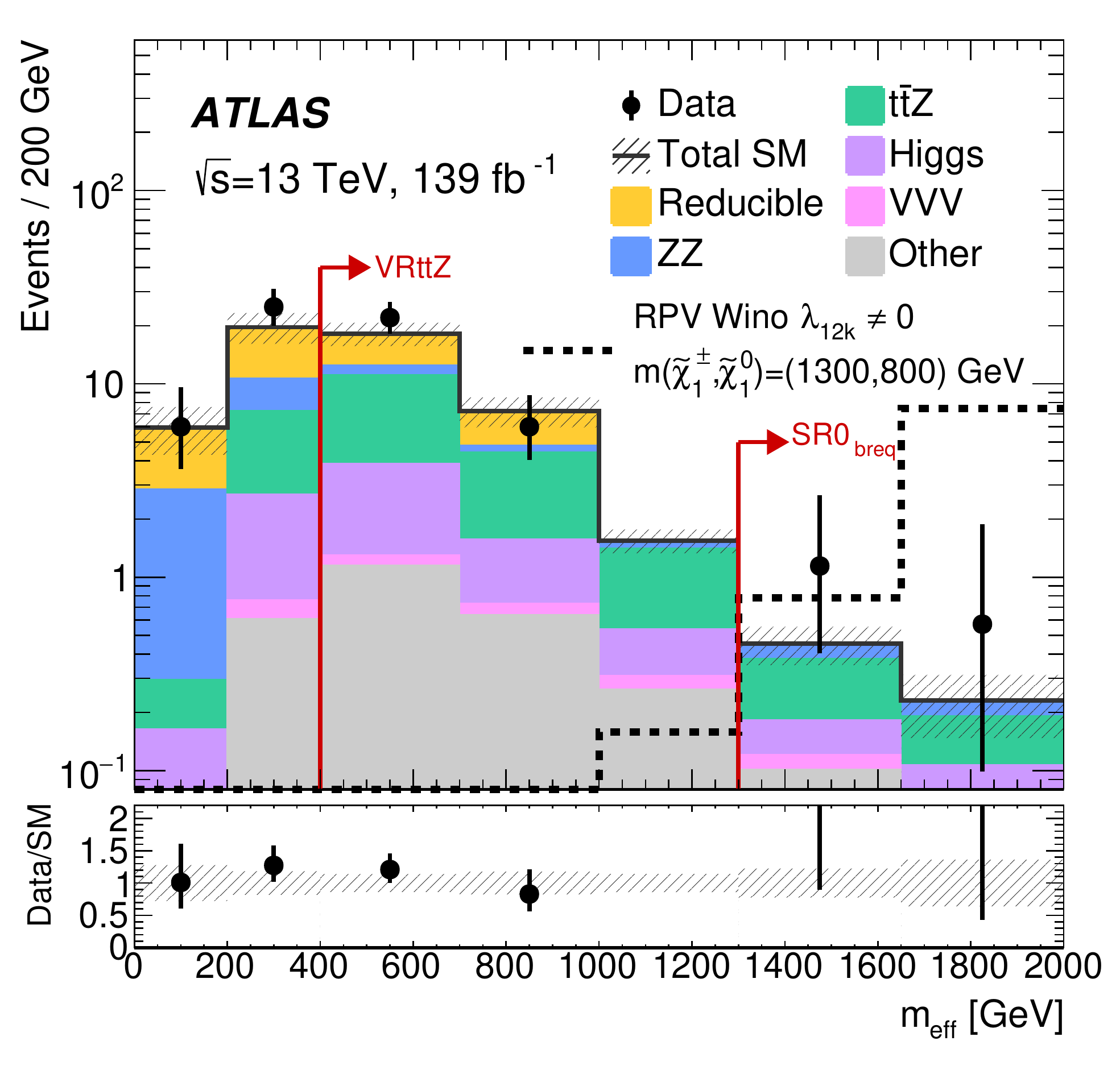}\label{fig:SRLLLLGmeff}}
\subfloat[][$\meff$ in \SRLLLTC]{\includegraphics[width=0.475\textwidth]{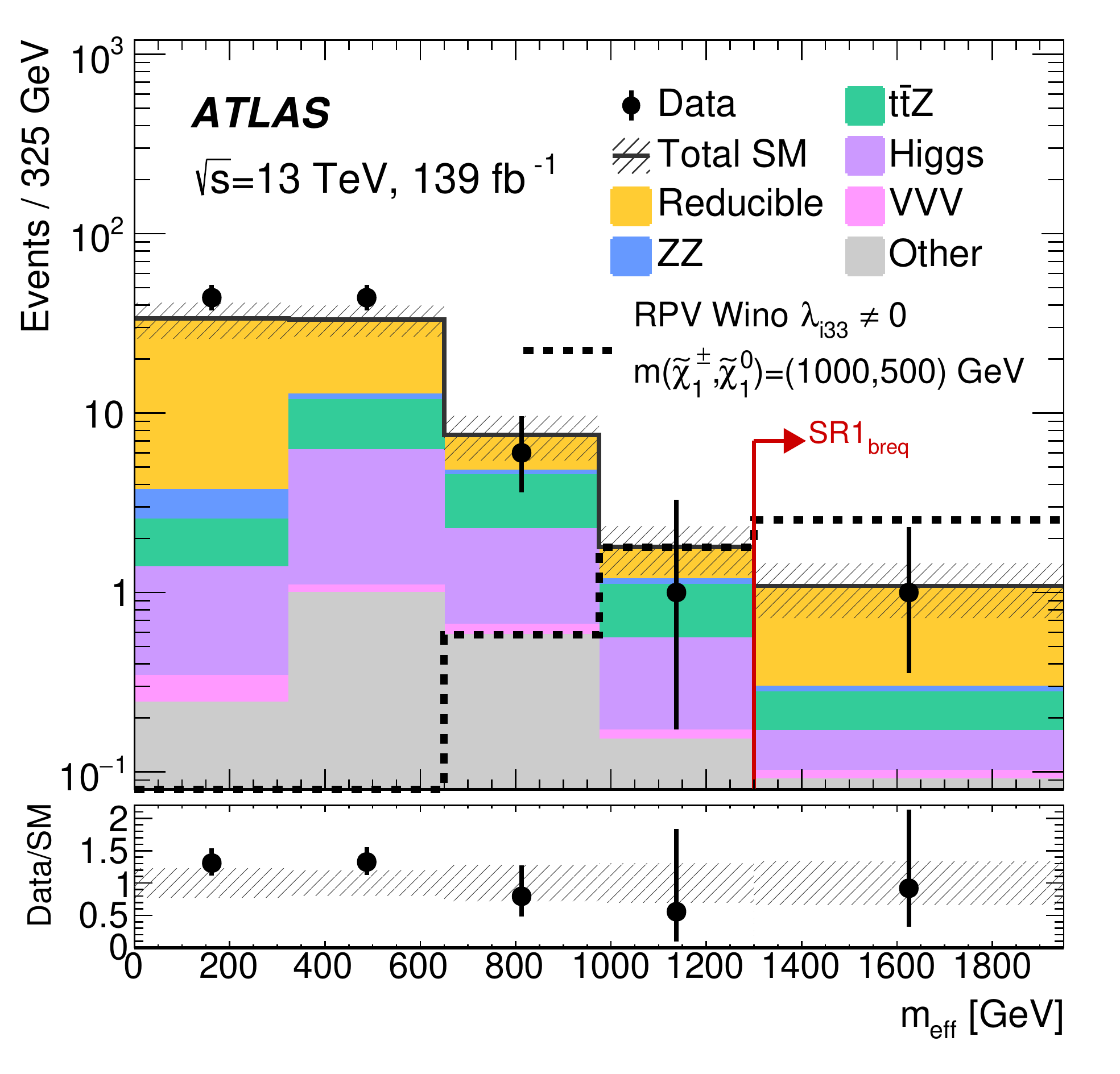}\label{fig:SRLLLTCmeff}}\\
\subfloat[][$\meff$ in \SRLLTTC]{\includegraphics[width=0.475\textwidth]{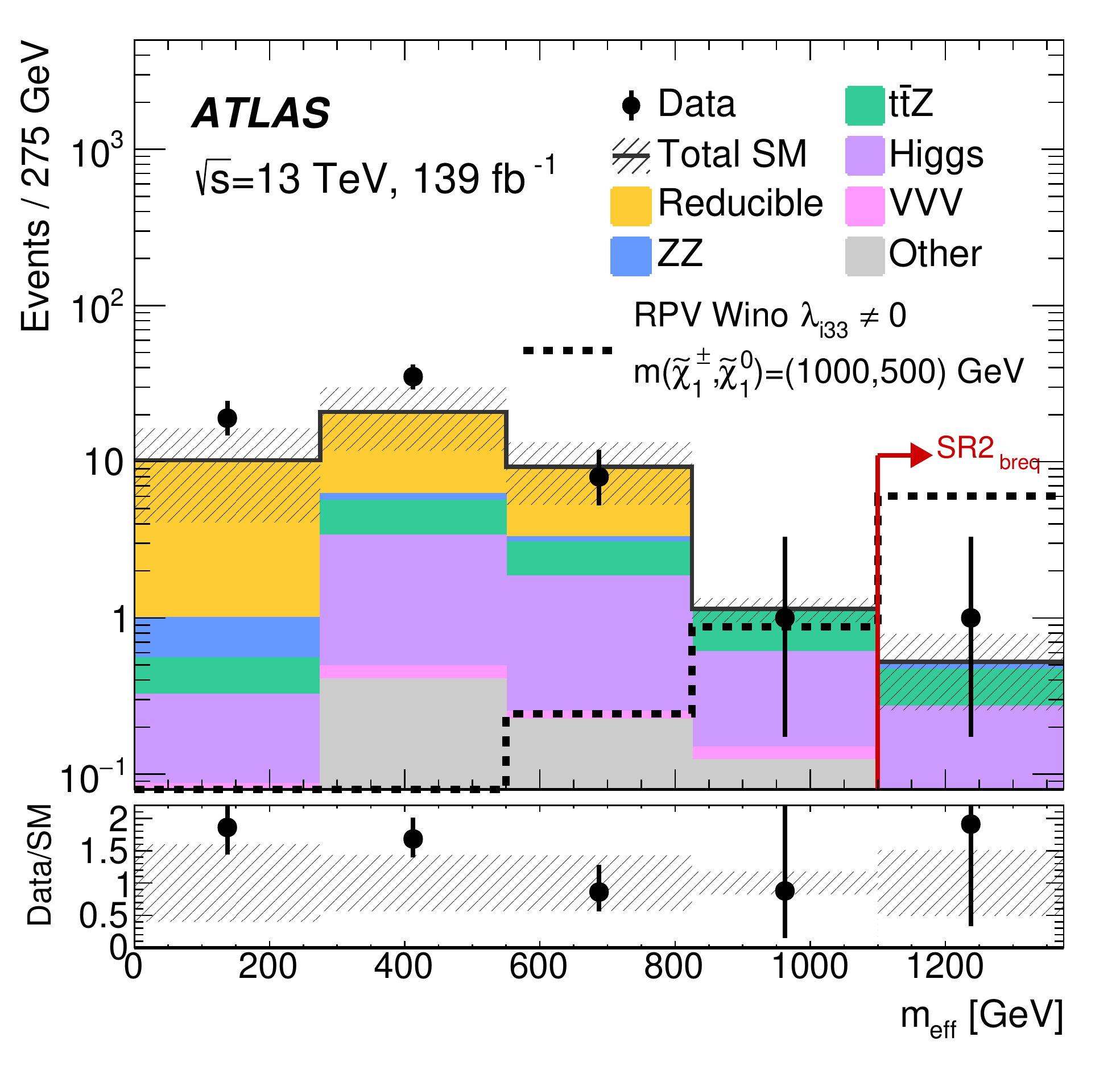}\label{fig:SRLLTTCmeff}}\\
\caption{
The $\meff$ distribution in (a) \protect\SRLLLLG, (b) \protect\SRLLLTC,  and (c) \protect\SRLLTTC\ for events passing the signal region requirements except the $\meff$ requirement.
Distributions for data, the estimated SM backgrounds after the background-only fit, and an example SUSY scenario are shown. ``Other'' is the sum of the $tWZ$, $\ttbar WW$, $\ttbar ZZ$, $\ttbar WZ$, $\ttbar WH$, $\ttbar HH$, $\ttbar tW$, and $\ttbar \ttbar $ backgrounds.
The last bin captures the overflow events.
The lower panel shows the ratio of the observed data to the expected SM background yield in each bin.
Both the statistical and systematic uncertainties in the SM background are included in the shaded band.
The red arrows indicate the $\meff$ selections in the signal and validation regions.
\label{fig:SRmeffB} }
\end{figure}

The \HistFitter software framework is used for the statistical interpretation of the results.
In order to quantify the probability for the background-only hypothesis to fluctuate to the observed number of events or higher, a one-sided $p_0$-value is calculated using pseudo-experiments, where the profile likelihood ratio is used as a test statistic~\cite{Cowan:2010js} to exclude the signal-plus-background hypothesis.
Signal contamination in the CRs is assumed to be zero in this model-independent fit.
The observations are consistent with the SM expectations, with \SRLLLLL\ giving rise to the highest local significance of $1.9\sigma$.
 
A signal model can be excluded at 95\% confidence level (CL) if the CL$_{\mathrm{s}}$~\cite{Read:2002hq} of the signal-plus-background hypothesis is below $0.05$.
For each signal region, the expected and observed upper limits at 95\% CL on the number of beyond-the-SM events ($S^{95}_{\mathrm{exp}}$ and $S^{95}_{\mathrm{obs}}$) are calculated using the model-independent signal fit.
The 95\% CL upper limits on the signal cross-section times efficiency ($\langle\epsilon{\mathrm \sigma}\rangle_{\mathrm{obs}}^{95}$) also calculated for each signal region and shown in Table~\ref{tab:SRModInd}.

\begin{table}[h]
\renewcommand\arraystretch{1.3} \centering
\caption{The model-independent limits calculated from the signal region observations;
the 95\% CL upper limit on
the visible cross-section times efficiency ($\langle\epsilon{\mathrm \sigma}\rangle_{\mathrm{obs}}^{95}$),
the observed number of signal events($S^{95}_{\mathrm{obs}}$), and
the signal events given the expected number of background events ($S^{95}_{\mathrm{exp}}$, $\pm1\sigma$ variations of the expected number) calculated by  performing  pseudo-experiments for each signal region.
\label{tab:SRModInd}}
\vspace{0.2cm}
\small{
\begin{tabular}{ lccc }
\toprule
& $\langle\epsilon{\mathrm \sigma}\rangle_{\mathrm{obs}}^{95}$ [fb]  &  $S_{\mathrm{obs}}^{95}$  & $S_{\mathrm{exp}}^{95}$ \\
\midrule
\SRLLLLC    & $0.481$ &  $66.86$~~ & $ { 67.43 }^{ +20.43 }_{ -15.71 }$\, \\
\SRLLLLD    & $0.081$ &  $11.28$~~ & $ { 11.52 }^{ +4.81 }_{ -3.34 }$~~ \\
\SRLLLLE    & $0.043$ &  $6.01$ & $ { 7.10 }^{ +2.82 }_{ -1.90 }$ \\
\SRLLLLF    & $0.028$ &  $3.87$ & $ { 3.63 }^{ +1.44 }_{ -0.63 }$ \\
\SRLLLLA    & $0.070$ &  $9.79$ & $ { 8.28 }^{ +3.58 }_{ -2.30 }$ \\
\SRLLLLB    & $0.028$ &  $3.87$ & $ { 4.29 }^{ +1.56 }_{ -0.86 }$ \\
\SRLLLLG    & $0.046$ &  $6.33$ & $ { 3.78 }^{ +1.59 }_{ -0.66 }$ \\
\SRLLLTA    & $0.046$ &  $6.37$ & $ { 7.46 }^{ +2.92 }_{ -2.04 }$ \\
\SRLLLTB    & $0.032$ &  $4.47$ & $ { 4.22 }^{ +1.63 }_{ -1.04 }$ \\
\SRLLLTC    & $0.033$ &  $4.56$ & $ { 4.59 }^{ +1.77 }_{ -1.22 }$ \\
\SRLLTTA    & $0.061$ &  $8.45$ & $ { 7.45 }^{ +2.36 }_{ -1.24 }$ \\
\SRLLTTB    & $0.041$ &  $5.63$ & $ { 3.53 }^{ +1.06 }_{ -0.15 }$ \\
\SRLLTTC    & $0.030$ &  $4.17$ & $ { 3.16 }^{ +1.20 }_{ -0.16 }$ \\
\SRLLLLL    & $0.129$ &  $17.88$~~ & $ { 9.88 }^{ +4.08 }_{ -2.44 }$ \\
\bottomrule
\end{tabular}
}
\end{table}
 
Exclusion limits at $95\%$ CL are calculated for the SUSY models and shown in Figure~\ref{fig:ExclLimits}.
A statistical combination is performed using the results from a number of disjoint signal regions targeting the SUSY model, and signal contamination in the CRs used to constrain $ZZ$ and $\ttbar Z$ background is accounted for.
Where two (or more) signal regions overlap, the signal region with the better (best) expected exclusion is used in the combination: specifically either \SRLLLLC\ or \SRLLLLD\ or \SRLLLLE\ or \SRLLLLF, and either \SRLLLLA\ or \SRLLLLB, and either \SRLLLTA\ or \SRLLLTB, and either \SRLLTTA\ or \SRLLTTB.
The four-lepton signal regions are more sensitive than SR5L in all signal models considered and so SR5L is not used to calculate model-dependent exclusion limits.
 
For the exclusion limits, the observed and expected 95\% CL limits take into account the theoretical and experimental uncertainties in the SM background and the experimental uncertainties in the signal.
For all expected and observed exclusion limit contours,
the $\pm1\sigma_\mathrm{exp}$ uncertainty band shows the impact on the expected limit of the systematic and statistical uncertainties included in the fit.
The $\pm1\sigma_\mathrm{theory}^\mathrm{SUSY}$ uncertainty lines around the observed limit illustrate the change in the observed limit as the nominal signal cross-section is
scaled up and down by the theoretical cross-section uncertainty.
Experimental uncertainties affecting irreducible backgrounds and signal are treated as correlated between regions and processes, while uncertainties associated with the estimate of the reducible background are correlated between regions only.
Theoretical uncertainties in the irreducible background and signal are treated as correlated between regions, while statistical uncertainties from MC simulation and data in the CR are treated as uncorrelated between regions and processes.

The exclusion contours for the higgsino GGM models are shown in Figure~\ref{fig:GGMexclLimits}, where \SRLLLLE\ and \SRLLLLF\ dominate the exclusion for low and high higgsino masses, respectively.
The analysis is more sensitive to scenarios with higher $\mathcal{B}(\ninoone \ra Z + \gravino)$ as they produce more four-lepton events. For scenarios with $\mathcal{B}(\ninoone \ra h + \gravino) = 100\%$, final states with lower lepton multiplicity are more sensitive~\cite{SUSY-2017-02}.
Higgsino masses up to $540\,\GeV$ are excluded for $\mathcal{B}(\ninoone \ra Z + \gravino) = 100\%$, while the exclusion is weaker for scenarios with $\mathcal{B}(\ninoone \ra Z + \gravino) < 100\%$.
These results significantly improve upon limits in Ref.~\cite{SUSY-2016-21} by around $200$--$260\,\GeV$, and in particular, set stronger limits for scenarios with $\mathcal{B}(\ninoone \ra Z + \gravino) < 100\%$.

Figure~\ref{fig:ExclLimits} also shows the exclusion contours for the RPV models considered here, where the limits extend to high NLSP masses due to the high lepton multiplicity in these scenarios ($\ninoone\ra \ell\ell\nu$ with 100\% branching ratio) and the high efficiency of the $\meff$ selections.
The exclusion in the $\LLE 12k$ models is dominated by the combination of \SRLLLLB\ and \SRLLLLG\, while the combination of \SRLLLTB\ and \SRLLTTB\ is important for the $\LLE i33$ models.
The sensitivity is reduced for large mass splittings between the NLSP and the $\ninoone$, where the decay products are strongly boosted.
These results improve on the limits set in similar models in Ref.~\cite{SUSY-2016-21} by around $100$--$350 \,\GeV$.
 
The exclusion contours for the RPV wino NLSP $\LLE 12k$ models are shown in Figure~\ref{fig:winoRPVexclLimits}, where
$\chinoonepm/\ninotwo$ masses up to $\sim 1.6\,\TeV$ are excluded for $m(\ninoone)>800\,\GeV$.
Figure~\ref{fig:winoRPVexclLimits} also shows exclusion contours for the RPV wino NLSP $\LLE i33$ models,
where $\chinoonepm/\ninotwo$ masses up to $\sim 1.13\,\TeV$ are excluded for $ 500\,\GeV < m(\ninoone) < 700\,\GeV$.
Figure~\ref{fig:LHSlepRPVexclLimits} shows exclusion contours for the RPV $\sleptonL/\snu$ NLSP model,
where left-handed slepton/sneutrino masses are excluded up to $\sim 1.2\,\TeV$ for $m(\ninoone) \simeq 800\,\GeV$ for $\LLE 12k$ models,
and up to $0.87\,\TeV$ for $m(\ninoone)\simeq500\,\GeV$ for $\LLE i33$ models.
Finally, the exclusion contours for the RPV $\gluino$ NLSP model are shown in Figure~\ref{fig:gluinoRPVexclLimits},
where gluino masses are excluded up to $\sim 2.45\,\TeV$ for $m(\ninoone)>1\,\TeV$ for $\LLE 12k$ models, increasing to $\sim 2.55\,\TeV$ for $m(\ninoone)>2.4\,\TeV$.
For $\LLE i33$ models, gluino masses are excluded up to $\sim 1.78\,\TeV$ for $m(\ninoone)>500\,\GeV$, increasing to $\sim 1.97\,\TeV$ for $m(\ninoone)>1.8\,\TeV$.

\begin{figure}[h]
\centering
\subfloat[][Higgsino GGM]{\includegraphics[width=0.49\textwidth]{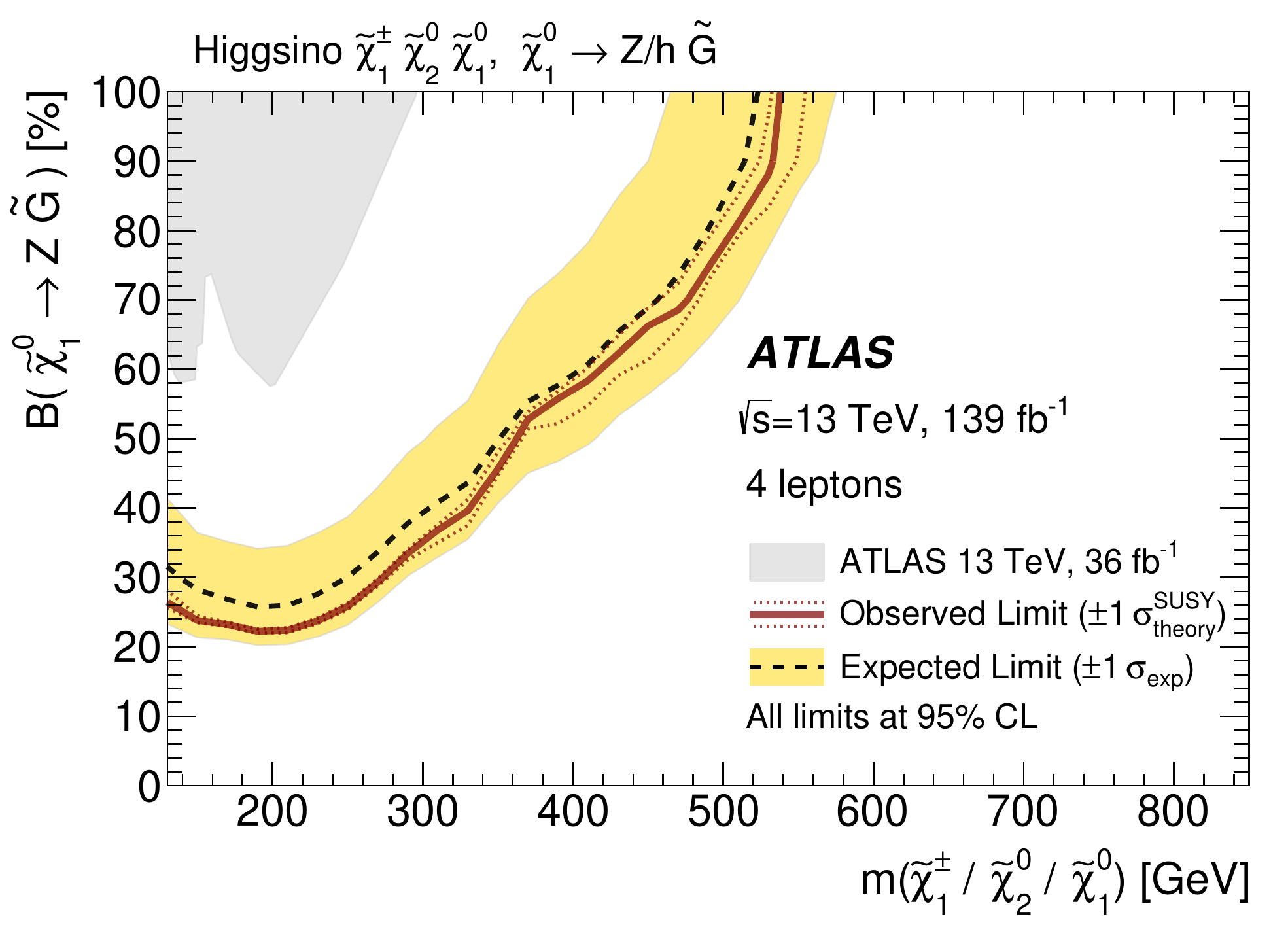}\label{fig:GGMexclLimits}}
\subfloat[][RPV wino NLSP]{\includegraphics[width=0.49\textwidth]{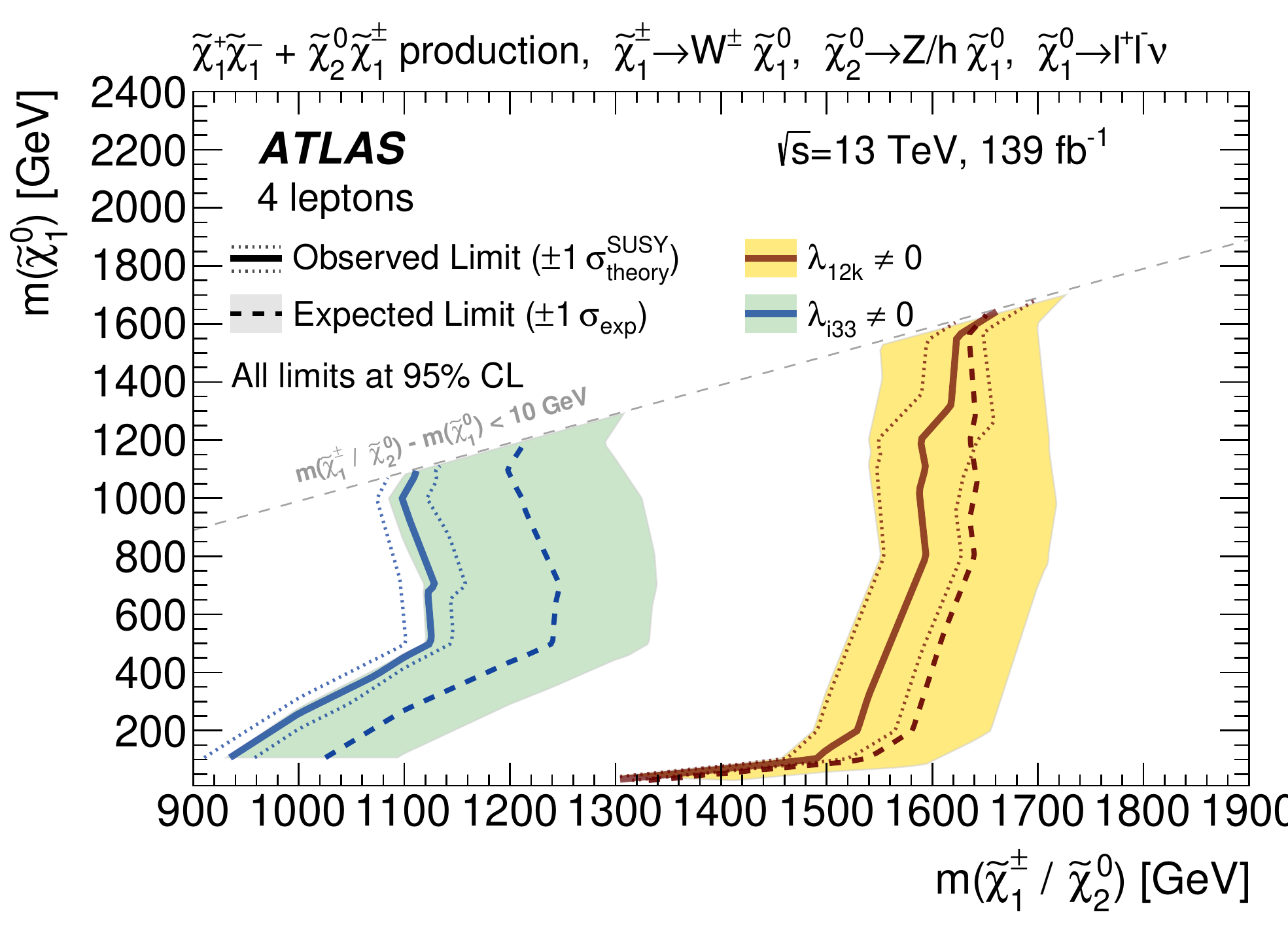}\label{fig:winoRPVexclLimits}}\\
\subfloat[][RPV $\sleptonL/\snu$ NLSP]{\includegraphics[width=0.49\textwidth]{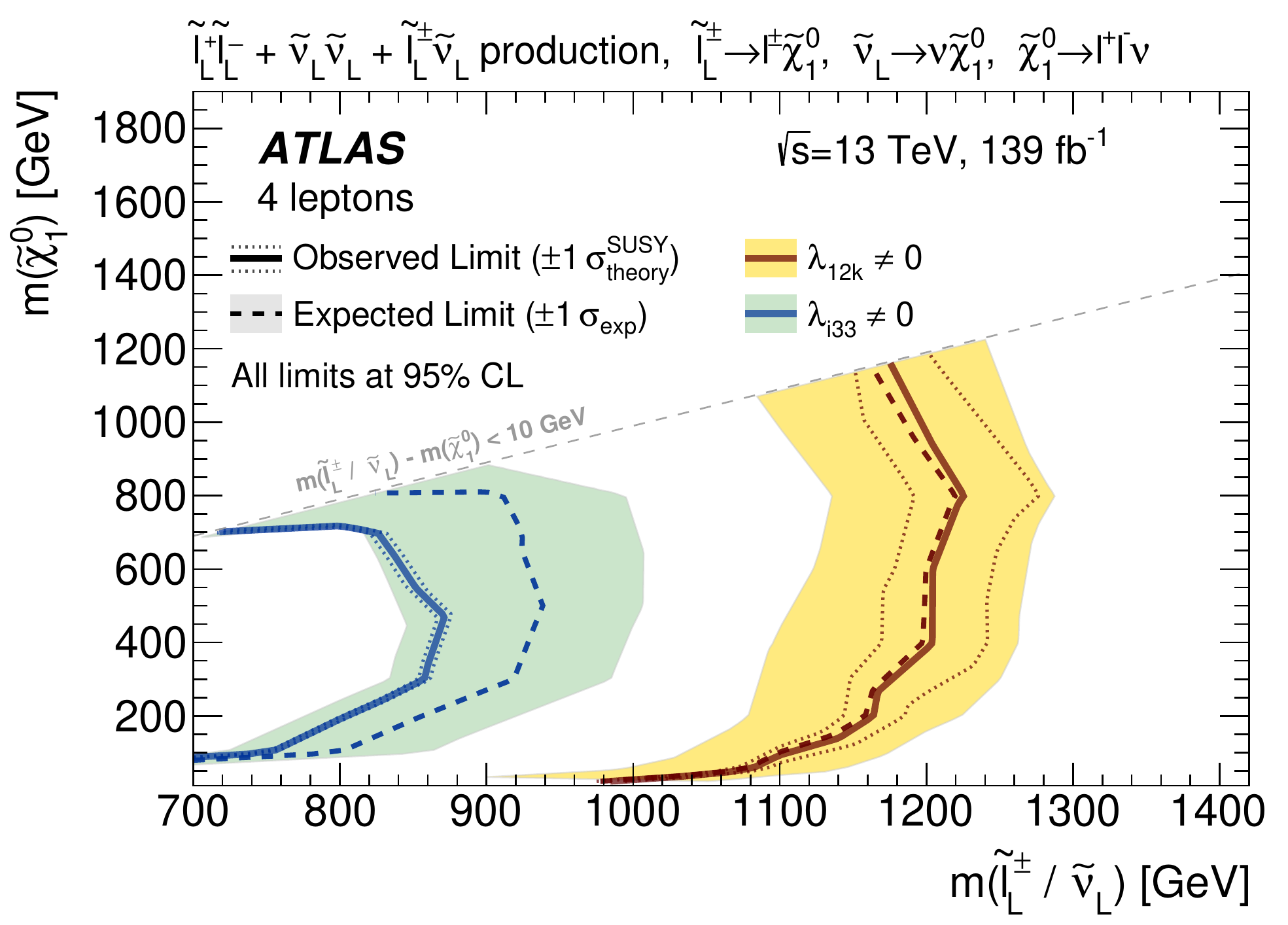}\label{fig:LHSlepRPVexclLimits}}
\subfloat[][RPV $\gluino$ NLSP]{\includegraphics[width=0.49\textwidth]{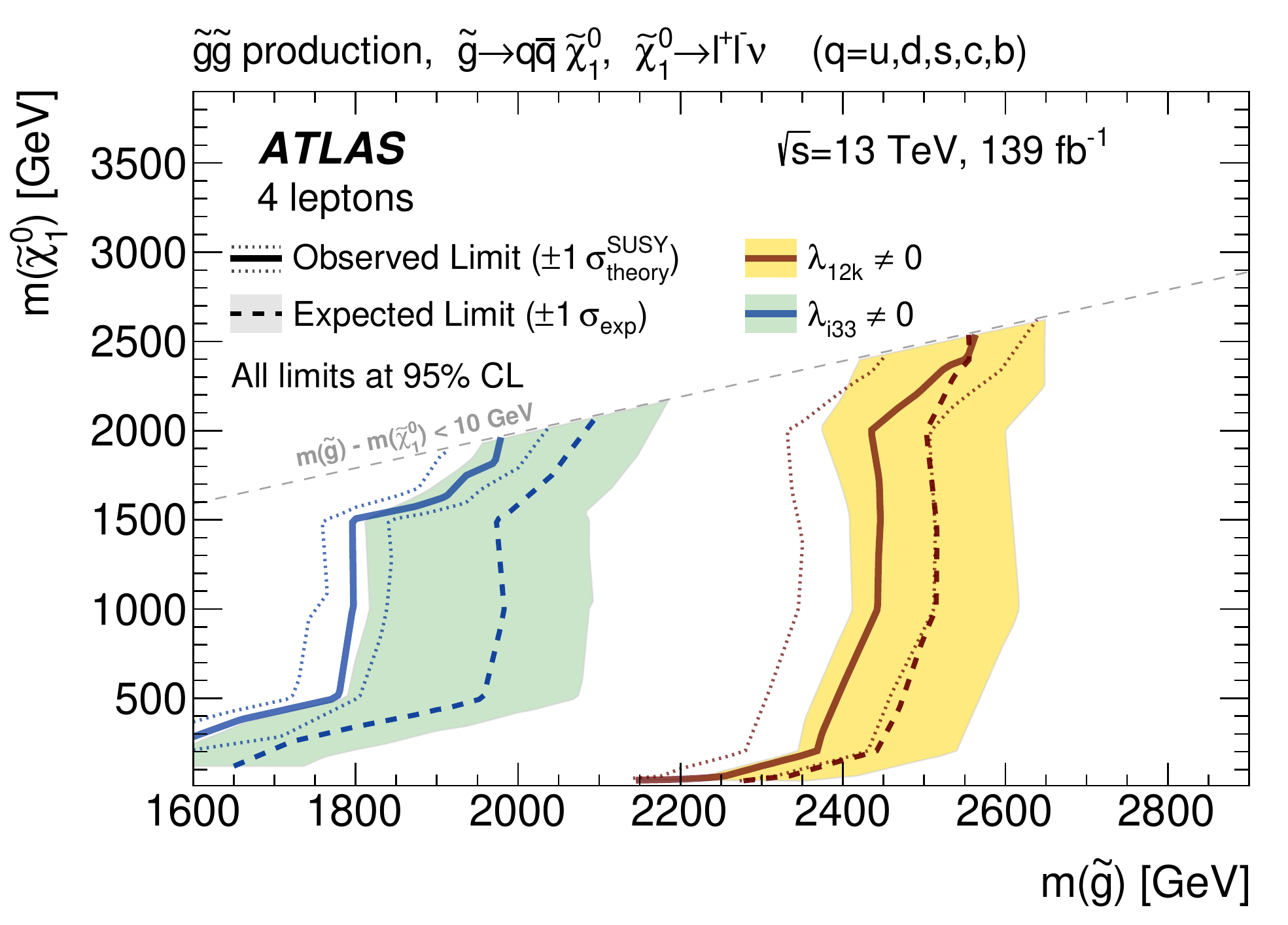}\label{fig:gluinoRPVexclLimits}}\\
\caption{
Expected (dashed) and observed (solid) 95\% CL exclusion limits on
(a) the higgsino GGM models, and
(b) wino NLSP,
(c) $\sleptonL/\snu$ NLSP, and
(d) gluino NLSP pair production
with RPV $\ninoone$ decays via $\lambda_{12k}$, or $\lambda_{i33}$ where $i,k \in{1,2}$.
The limits are set using a statistical combination of disjoint signal regions.
Where two (or more) signal regions overlap, the signal region contributing its observed $\mathrm{CL}_{\mathrm{s}}$ value to the combination is the one with the better (best) expected $\mathrm{CL}_{\mathrm{s}}$ value.
\label{fig:ExclLimits}
}
\end{figure}

\FloatBarrier
\section{Conclusion}\label{sec:conc}
A search for new physics in final states with four or more leptons (electrons, muons or hadronically decaying $\tau$-leptons) is presented,
using $139\,\ifb$ of $\sqrt{s}=13\,\TeV$ $pp$ collision data collected by the ATLAS detector at the LHC from 2015 to 2018.
No significant excess over the expected background is observed.
Observed 95\% CL limits on the visible cross-section are placed in each signal region, and the null result is interpreted in
simplified higgsino GGM models, where higgsino $\chinoonepm/\ninotwo/\ninoone$ masses up to $540\,\GeV$ are excluded in scenarios with a 100\% branching ratio for $\ninoone$ decay into a $Z$ boson and a gravitino.
This significantly improves upon previous GGM higgsino mass limits by around $200$--$260\,\GeV$.
The results are also interpreted in simplified models of NLSP pair production with RPV LSP decays,
where wino $\chinoonepm/\ninotwo$,
$\sleptonL/\snu$,
and $\gluino$ masses up to
$1.6\,\TeV$,
$1.2\,\TeV$,
and $2.5\,\TeV$ are excluded, respectively, improving upon previous limits set in similar models by around $100$--$350 \,\GeV$.
 
\FloatBarrier
\section*{Acknowledgements}

We thank CERN for the very successful operation of the LHC, as well as the
support staff from our institutions without whom ATLAS could not be
operated efficiently.
 
We acknowledge the support of ANPCyT, Argentina; YerPhI, Armenia; ARC, Australia; BMWFW and FWF, Austria; ANAS, Azerbaijan; SSTC, Belarus; CNPq and FAPESP, Brazil; NSERC, NRC and CFI, Canada; CERN; ANID, Chile; CAS, MOST and NSFC, China; Minciencias, Colombia; MSMT CR, MPO CR and VSC CR, Czech Republic; DNRF and DNSRC, Denmark; IN2P3-CNRS and CEA-DRF/IRFU, France; SRNSFG, Georgia; BMBF, HGF and MPG, Germany; GSRT, Greece; RGC and Hong Kong SAR, China; ISF and Benoziyo Center, Israel; INFN, Italy; MEXT and JSPS, Japan; CNRST, Morocco; NWO, Netherlands; RCN, Norway; MNiSW and NCN, Poland; FCT, Portugal; MNE/IFA, Romania; JINR; MES of Russia and NRC KI, Russian Federation; MESTD, Serbia; MSSR, Slovakia; ARRS and MIZ\v{S}, Slovenia; DST/NRF, South Africa; MICINN, Spain; SRC and Wallenberg Foundation, Sweden; SERI, SNSF and Cantons of Bern and Geneva, Switzerland; MOST, Taiwan; TAEK, Turkey; STFC, United Kingdom; DOE and NSF, United States of America. In addition, individual groups and members have received support from BCKDF, CANARIE, Compute Canada, CRC and IVADO, Canada; Beijing Municipal Science \& Technology Commission, China; COST, ERC, ERDF, Horizon 2020 and Marie Sk{\l}odowska-Curie Actions, European Union; Investissements d'Avenir Labex, Investissements d'Avenir Idex and ANR, France; DFG and AvH Foundation, Germany; Herakleitos, Thales and Aristeia programmes co-financed by EU-ESF and the Greek NSRF, Greece; BSF-NSF and GIF, Israel; La Caixa Banking Foundation, CERCA Programme Generalitat de Catalunya and PROMETEO and GenT Programmes Generalitat Valenciana, Spain; G\"{o}ran Gustafssons Stiftelse, Sweden; The Royal Society and Leverhulme Trust, United Kingdom.
 
The crucial computing support from all WLCG partners is acknowledged gratefully, in particular from CERN, the ATLAS Tier-1 facilities at TRIUMF (Canada), NDGF (Denmark, Norway, Sweden), CC-IN2P3 (France), KIT/GridKA (Germany), INFN-CNAF (Italy), NL-T1 (Netherlands), PIC (Spain), ASGC (Taiwan), RAL (UK) and BNL (USA), the Tier-2 facilities worldwide and large non-WLCG resource providers. Major contributors of computing resources are listed in Ref.~\cite{ATL-SOFT-PUB-2020-001}.
 

\printbibliography
\clearpage 
 
\begin{flushleft}
\hypersetup{urlcolor=black}
{\Large The ATLAS Collaboration}

\bigskip

\AtlasOrcid[0000-0002-6665-4934]{G.~Aad}$^\textrm{\scriptsize 102}$,    
\AtlasOrcid[0000-0002-5888-2734]{B.~Abbott}$^\textrm{\scriptsize 128}$,    
\AtlasOrcid[0000-0002-7248-3203]{D.C.~Abbott}$^\textrm{\scriptsize 103}$,    
\AtlasOrcid[0000-0002-2788-3822]{A.~Abed~Abud}$^\textrm{\scriptsize 36}$,    
\AtlasOrcid[0000-0002-1002-1652]{K.~Abeling}$^\textrm{\scriptsize 53}$,    
\AtlasOrcid[0000-0002-2987-4006]{D.K.~Abhayasinghe}$^\textrm{\scriptsize 94}$,    
\AtlasOrcid[0000-0002-8496-9294]{S.H.~Abidi}$^\textrm{\scriptsize 167}$,    
\AtlasOrcid[0000-0002-8279-9324]{O.S.~AbouZeid}$^\textrm{\scriptsize 40}$,    
\AtlasOrcid{N.L.~Abraham}$^\textrm{\scriptsize 156}$,    
\AtlasOrcid[0000-0001-5329-6640]{H.~Abramowicz}$^\textrm{\scriptsize 161}$
% ...
\bigskip
\\

% ...
$^{36}$CERN, Geneva; Switzerland.\\
% ...
$^{40}$Niels Bohr Institute, University of Copenhagen, Copenhagen; Denmark.\\
% ...
$^{53}$II. Physikalisches Institut, Georg-August-Universit\"{a}t G\"ottingen, G\"ottingen; Germany.\\
% ...
$^{94}$Department of Physics, Royal Holloway University of London, Egham; United Kingdom.\\
% ...
$^{102}$CPPM, Aix-Marseille Universit\'e, CNRS/IN2P3, Marseille; France.\\
$^{103}$Department of Physics, University of Massachusetts, Amherst MA; United States of America.\\
% ...
$^{128}$Homer L. Dodge Department of Physics and Astronomy, University of Oklahoma, Norman OK; United States of America.\\
% ...
$^{156}$Department of Physics and Astronomy, University of Sussex, Brighton; United Kingdom.\\
% ...
$^{161}$Raymond and Beverly Sackler School of Physics and Astronomy, Tel Aviv University, Tel Aviv; Israel.\\
% ...
$^{167}$Department of Physics, University of Toronto, Toronto ON; Canada.\\
% ...

\end{flushleft}


\end{document}